\documentclass{aa}

\usepackage[T1]{fontenc} 
\usepackage{siunitx}
\usepackage{amsmath}
\usepackage{amssymb}
\usepackage{graphicx}
\usepackage{longtable}   
\usepackage[para]{footmisc}
\usepackage{lscape}
\usepackage[export]{adjustbox}
\usepackage{subcaption}
\usepackage{perpage}
\usepackage{alphalph}
\usepackage{hyperref}
\usepackage[capitalise]{cleveref} 
\usepackage{float}
\usepackage{xcolor}
\usepackage{caption}

\newcommand{\kms}{km.s$^{\rm -1}$}

\newcommand{\Mjup}{M$_{\rm Jup}$}

\newcommand{\vsini}{$v\sin{i}$}
\newcommand{\Msun}{M$_{\sun}$}

\newcommand{\Mstar}{M$_{\rm \star}$}
\newcommand{\bv}{$B-V$}
\newcommand{\msini}{$m_{\rm p}\sin{i}$}
\newcommand{\rhk}{log$R'_{\rm HK}$}
\newcommand{\safir}{S{\small AFIR}}
\newcommand{\sophie}{S{\small OPHIE}}
\newcommand{\harps}{H{\small ARPS}}
\newcommand{\hipp}{H{\small IPPARCOS}}

\newcommand{\elodie}{E{\small LODIE}}

\newcommand{\sphere}{S{\small PHERE}}

\DeclareSIUnit\year{yr}
\DeclareSIUnit\arcsecond{arcsec}
\DeclareSIUnit\MJ{M_{Jup}}
\DeclareSIUnit\au{au}
\DeclareSIUnit\jdb{d}
\DeclareSIUnit\day{days}
\DeclareSIUnit\hour{h}
\DeclareSIUnit\night{nights}
\DeclareSIUnit\arcsec{as}
\DeclareSIUnit\msun{M_{\odot}}
\DeclareSIUnit\parsec{pc}
\makeatletter
\newcommand\footnoteref[1]{\protected@xdef\@thefnmark{\ref{#1}}\@footnotemark}
\makeatother

\begin{document} 

   \title{A SOPHIE RV search for giant planets around young nearby stars (YNS)} 
 \subtitle{A combination with the HARPS YNS survey}

   \author{A. Grandjean
          \inst{1}
          \and
          A.-M. Lagrange\inst{1,2,3}
        \and
        N. Meunier\inst{1}
        \and
        P. Rubini\inst{4}
        \and
        S. Desidera \inst{5}
        \and
        F. Galland \inst{1}
        \and
        S. Borgniet\inst{6}
        \and
        N. Zicher  \inst{7}
        \and
        S. Messina \inst{8}
        \and
        G. Chauvin\inst{1}
        \and
        M. Sterzik \inst{9}
        \and
        B. Pantoja \inst{10}
          }

   \institute{
Univ. Grenoble Alpes, CNRS, IPAG, 38000 Grenoble, France
 \\
\email{antoine.grandjean1@univ-grenoble-alpes.fr}
\and 
LESIA, Observatoire de Paris, Université PSL, CNRS, Sorbonne Université, Université de Paris, 5 place Jules Janssen, 92195 Meudon, France
\and
IMCCE – Observatoire de Paris, 77 Avenue Denfert-Rochereau, 75014 Paris, France
\and
Pixyl, 5 Avenue du Grand Sablon, 38700 La Tronche, France
\and
INAF-Osservatorio Astronomico di Padova, Vicolo dell’Osservatorio 5, Padova, Italy, 35122-I
\and
CNRS Lesia (UMR8109) - Observatoire de Paris, Paris, France
\and
Oxford Astrophysics, Department of Physics, Denys Wilkinson Building,UK
\and
INAF–Osservatorio Astrofisico di Catania, via Santa Sofia, 78 Catania, Italy 
\and
European Southern Observatory, Karl-Schwarzschild-Str 1, D-85748 Garching, Germany
\and
Departamento de Astronomía, Universidad de Chile, Camino al Observatorio, Cerro Calán, Santiago, Chile
             }

   \date{Received 14 October 2020 / Accepted 1 March 2021}

 
  \abstract
   {The search of close ($a\lesssim \SI{5}{\au}$) giant planet (GP) companions with radial velocity (RV) around young stars and the estimate of their occurrence rates  is important to constrain the migration timescales. Furthermore, this search will allow   the giant planet occurrence rates to be computed at all separations via  the combination with direct imaging techniques.
The RV search around young stars is a challenge as they are generally faster rotators than older stars of similar spectral types and they exhibit signatures of  magnetic activity (spots) or pulsation in their RV time series.
Specific analyses are necessary to characterize, and possibly correct for, this activity.

 }
   {Our aim is to search for planets around young nearby stars and to estimate the GP occurrence rates for periods up to \SI{1000}{\day}.}
   {We used the \sophie \ spectrograph on the \SI{1.93}{\meter} telescope at the Haute-Provence Observatory to observe $63 \ A-M$ young ($< \SI{400}{\mega\year}$) stars. 
We used our  Spectroscopic data via Analysis of the Fourier Interspectrum Radial velocities (\safir) software  to compute the RVs and other spectroscopic observables. 
We then combined this survey with the HARPS YNS survey to compute the companion occurrence rates on a total of $120$ young $A-M$ stars.}
   {We report one new trend compatible with a planetary companion on HD 109647. We also report  HD 105693 and  HD 112097  as binaries,  and we confirm the binarity of HD 2454, HD13531, HD 17250 A,  HD 28945, HD  39587, HD 131156, HD 142229,  HD 186704 A, and HD 195943. We constrained for the first time the orbital parameters of HD 195943 B. We refute the HD 13507 single brown dwarf (BD) companion solution and propose a double BD companion solution.
Two GPs were previously reported from this survey  in the HD 113337 system.  
Based on our sample of $120$ young stars, we obtain a GP occurrence rate of  $1_{-0.3}^{+2.2} \ \%$ for  periods lower than  $\SI{1000}{\day}$, and we obtain an upper limit on BD occurrence rate  of $0.9_{-0.9}^{+2}\ \%$ in the same period range. We report a possible lack of close ($P\in[1;1000]\ \si{\day}$) GPs around young FK stars compared to their older counterparts, with a confidence level of $90\%$.
 }
   {}

   \maketitle
%

\section{Introduction}

More than four thousand exoplanets and brown dwarfs (BDs) have been detected and most of them have been found by transit or radial velocity (RV) techniques\footnote{\url{exoplanet.eu}}. The occurrence rates of these planets are well established for main sequence (MS) and evolved stars, as are some relations between their occurrence rates and their host star characteristics.

The MS late-type stars are the most studied in RV \citep{cumming,Mayor_occ,fernandes,Fischer,Santos}, together with the evolved stars \citep{Bowler_2009,Johnson,Jones}.
On the other side of the stellar mass range, early-type  MS stars are usually avoided in RV surveys because their optical spectra present fewer spectral lines than late-type stars and because they generally present higher projected rotation velocities (\vsini) than   late-type stars.
One previous survey was carried out on peculiar AF-type MS dwarfs by \cite{Hartmann}. A large RV survey of AF-type MS stars was carried out by \cite{Simon_X} with the \harps\footnote{High Accuracy Radial velocity Planet Searcher} \ and \sophie\footnote{Spectrographe pour l’Observation des Phénomènes des Intérieurs stellaires et des Exoplanètes} \ instruments. This survey, whose sample selection criterion was not based on age, contained targets as young as $\sim \SI{100}{\mega\year}$.

These surveys on  MS  and evolved stars permit us to identify two correlations between giant planet (GP) occurrence rates and host stellar characteristics.
First, there is a positive correlation with the star metallicity for MS FGK stars \citep{Fischer,Santos}.
Then, there is a positive correlation with host star mass. This correlation was observed for evolved stars in RV \citep{Bowler_2009,Johnson,Jones} and for wide orbit planets around young stars in direct imaging \citep{Lannier_Massive,Baron_2019}. 
These observations are consistent with the predictions of core accretion models  \citep{Kennedy}.

Young stars, however, were poorly studied in RV due to their high stellar induced jitter (spots, plages, convection, and pulsations), which can reach amplitudes of up to a few $\SI{1}{\kilo\meter\per\second}$ \citep{Lagrange,Grandjean_HARPS}, larger than the planet's induced signal. Moreover, young stars are generally faster rotators than their older counterparts \citep{Stauffer_2016, Rebull_2016, Gallet_2015}. Consequently, several false positive detections were reported around young stars in the past \citep{Huelamo,Figueira,Soto}.

One of the remaining questions about planet formation and early evolution is their migration timescales. These timescales can be constrained by the study of young stars. GP formation models predict a formation at a few $\si{\au}$ \citep{Pollack}, yet migration through disk--planet interactions \citep{Migration-disk} or gravitational interaction with a third body can allow the planet to finally orbit close to the star \citep{High_excentricity_migration}. 
Massive hot Jupiters (HJs; $m_p\sin{i} \in [1,13] \ \si{\MJ}$, $P \in [1,10] \ \si{\day}$) are common among exoplanets orbiting solar to late-type MS stars,  representing one  detected planet out of five \citep{Wright_2012}, yet their occurrence rate is low ($\sim 1 \ \%$; \cite{cumming,Wright_2012}). 
While previous RV surveys on young stars ($<\SI{300}{\mega\year}$) showed no evidence of the presence of young HJs \citep{Esposito,Paulson,Grandjean_HARPS}, three HJs were recently discovered around such young stars with RV \citep{Johns_Krull_2016} and with RV derived from spectropolarimetry \citep{Donati,Yu}. Young HJs are also known from transit  \citep{Cameron,Tanimoto,Mann,David,Rizzuto}, and from both transit and RV  \citep{Deleuil,Alsubai}.
In addition, no BDs with periods shorter than $\SI{10}{\day}$ were discovered with RV around young MS stars, although one was discovered from transit \citep{James}.
The occurrence rates of HJs and short period BDs still need to be constrained at young ages.

We carried out three RV surveys on young stars from A to M types with the final aim of coupling RV data with direct imaging (DI) data, which will allow  the computation of detection limits  for each target at all separations and then of  GP and BD occurrence rates for all separations.
The first survey was performed with the High Accuracy Radial velocity Planet Searcher  (\harps) \citep{HARPS} on young nearby stars (hereafter \harps\ YNS survey) and its results are presented in \cite{Grandjean_HARPS}. The second survey was performed with the \sophie \ \citep{SOPHIE} spectrographs on similar young nearby stars. Finally the third survey was performed with \harps \ on Sco-Cen stars.

We present in this paper the results of our \sophie \ YNS survey and its combination with the \harps \ YNS survey.
We describe our \sophie \ survey sample in \cref{survey_description} and we describe the GP, BD, or stellar companion detections along with their characterizations  in \cref{comp}.
We present the \sophie \ and \harps \ combined samples and its statistical analysis, including the occurrence rate computation for GPs and BDs, in \cref{HS}.
Finally, we present our conclusion in \cref{conc}.

\section{Description of the SOPHIE survey}
\label{survey_description}

\subsection{Sample}
\label{SOPHIE_sample}
The initial sample of our SOPHIE YNS survey included $63$ stars;   most of the targets are part of the \sphere\footnote{Spectro-Polarimetric High-contrast Exoplanet REsearch} \ GTO  SHINE\footnote{The SpHere INfrared survey for Exoplanets} survey sample \citep{Chauvin_shine}. The targets were selected according to their declination ($\in[+0:+80] \ \si{degree}$), brightness ($V < 10$),   age as found in the literature ($\lesssim \SI{300}{\mega\year}$ for most of them; see   \cref{tab_carac_s}), and  distance ($< \SI{80}{pc}$) as determined from  \hipp \ parallax \citep{hipp2}.  The Gaia mission refined the target distances, and one target (HD 48299) is now outside the distance criterion ($d=95.3\pm0.4 \ \si{pc}$ \cite{DR2A1}).
These criteria ensure  the best detection limits for both the \sophie \ RV and SPHERE DI surveys at, respectively, small (typically $2-5 \ \si{\au}$), and large (further than typically $\SI{5}{\au}$) separations.
The declination criterion was chosen to ensure that the stars' declinations are close to the Haute-Provence Observatory's latitude ($+\ang{43;55;54}$). This ensures a low airmass during most of the observation time, which permits spectra to be obtained with a good  signal-to-noise ratio (S/N). This criterion also ensures that most of the targets can be studied with the VLT/SPHERE instrument as the VLT\footnote{Very Large Telescope} can point up to a declination of $+\SI{46}{\degree}$.
The V-band apparent magnitude criterion was chosen to ensure that the stars are bright enough to obtain spectra with a good  S/N.
The age and distance criteria were chosen to obtain the best detection limits from direct imaging;  young planets are still warm from their formation, and are thus brighter than old ones. This lowers the contrast between them and their host stars. Moreover, nearby stars are better suited for direct imaging. 
Binary stars with an angular separation on the sky lower than $\SI{2}{\arcsecond}$  were not selected to avoid contamination in the spectra from the companion.

Our \sophie \ observations permitted us to measure the projected rotational velocity  (\vsini) of the stars  in our sample (see \Cref{observable}) and nine of them  presented a \vsini \ too high ($>\SI{300}{\kilo\meter\per\second}$) to allow RV measurement: HD 56537, HD 87696, HD 97603, HD 116842, HD 126248, HD 159651, HD 177178, HD 203280, and HD 222439. These stars were excluded from our analysis.
The ages of the stars  in the survey were re-evaluated by the community during the execution of the survey and in some cases the new estimated age was older than our initial age criterion. However, for most of these cases the new estimated age was poorly constrained and the associated uncertainties were huge (up to several $\si{\giga\year}$), which does not exclude a low age for these stars. We then chose to keep the stars for which the lower limit on the age is under  $< \SI{500}{\mega\year}$.
As a consequence, five stars were excluded from our analysis due to their old age: HD 2454 \citep{Gomez}, HD 48299 \citep{Casagrande}, HD 89449 \citep{Gaspar}, HD 148387 \citep{Baines}, and HD 166435 \citep{Gomez}. Nevertheless, we discuss the RV trend we observe on HD 2454 in \cref{2454}. 
After these different exclusions,  $49$ targets were available for our analysis.

\begin{figure*}[ht!]
  \centering
\begin{subfigure}[t]{0.23\textwidth}
\includegraphics[width=1\hsize]{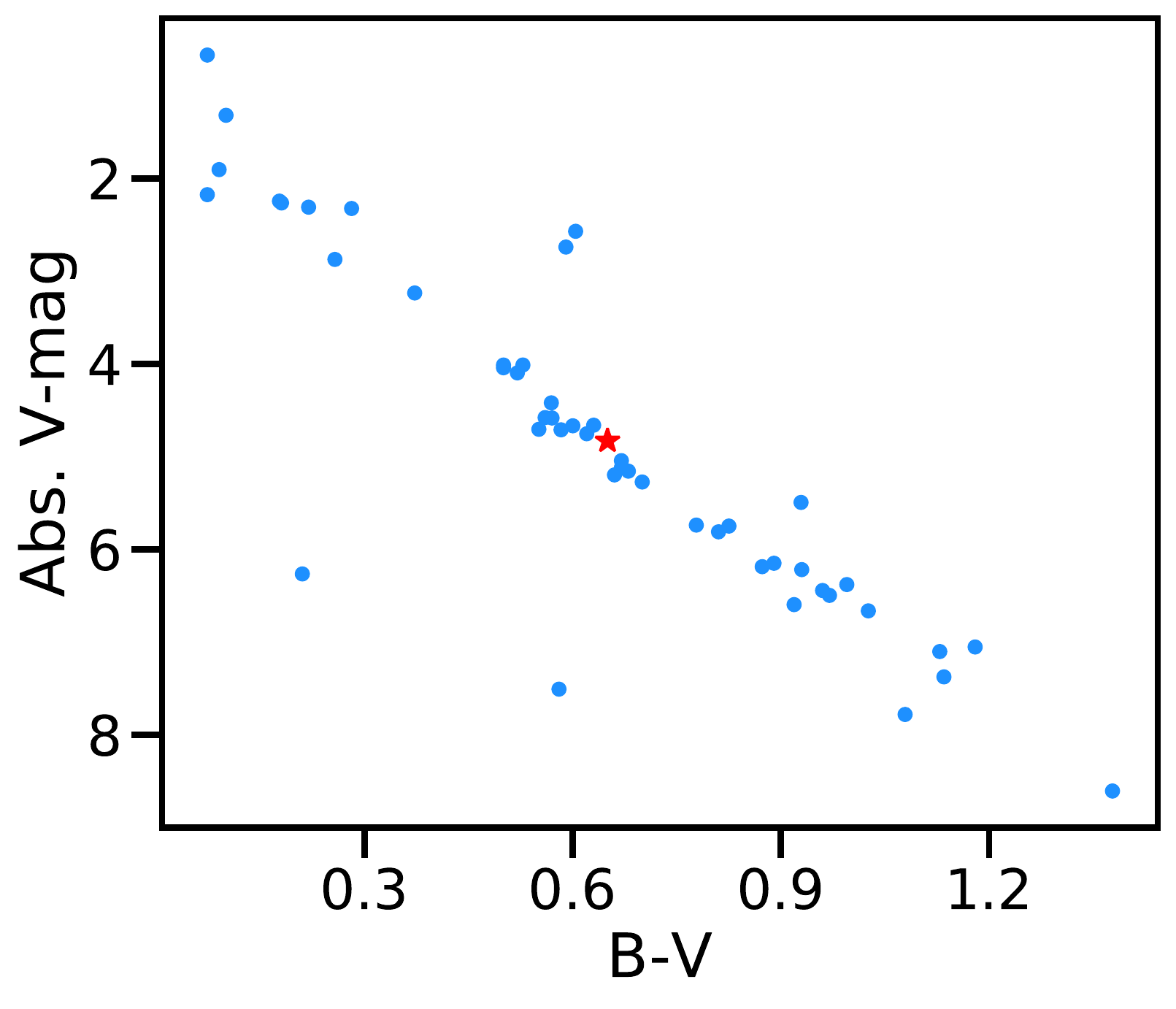}
\caption{\label{HR}}
\end{subfigure}
\begin{subfigure}[t]{0.24\textwidth}
\includegraphics[width=1\hsize]{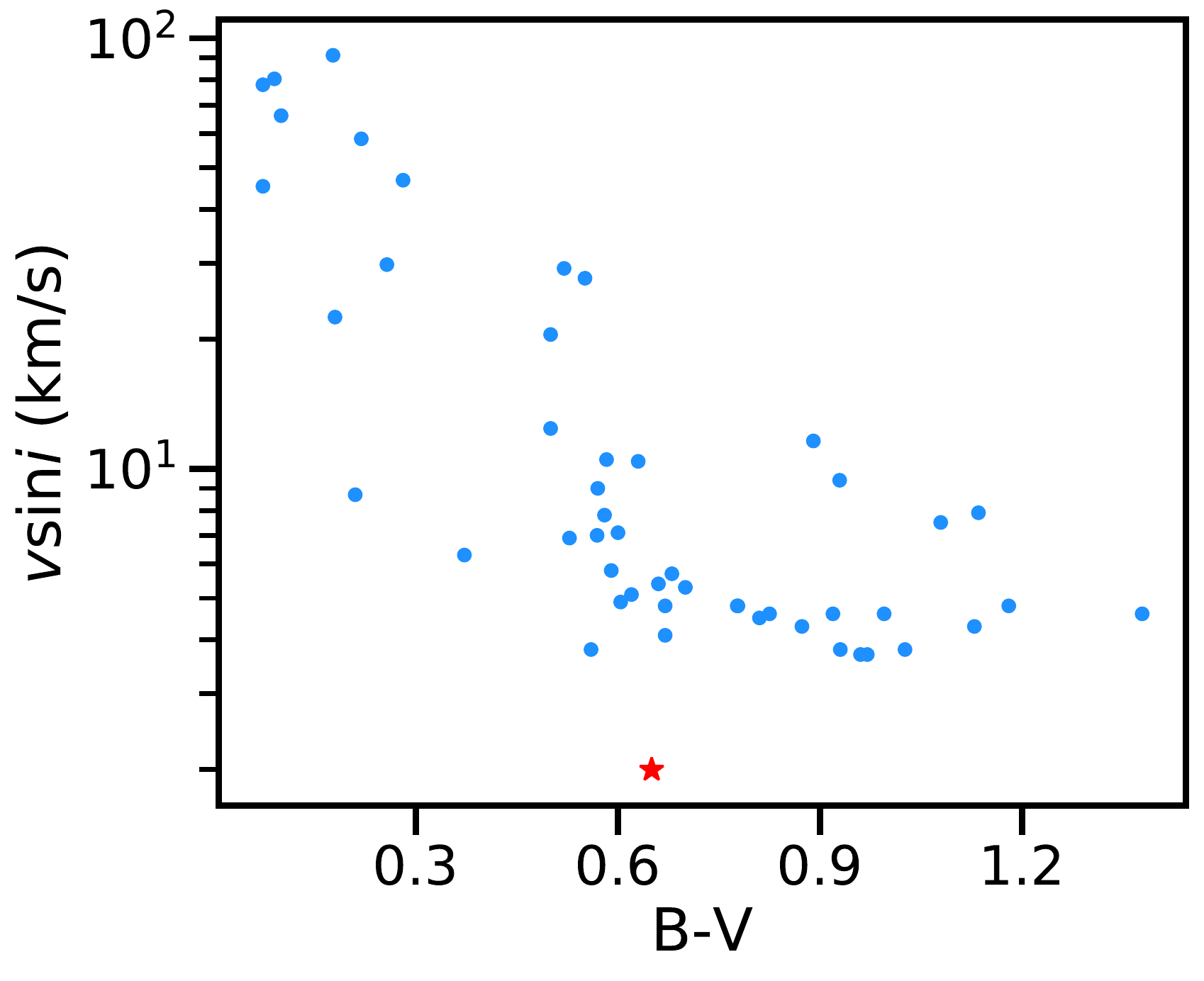}
\caption{\label{vsini}}
\end{subfigure}
\begin{subfigure}[t]{0.24\textwidth}
\includegraphics[width=1\hsize]{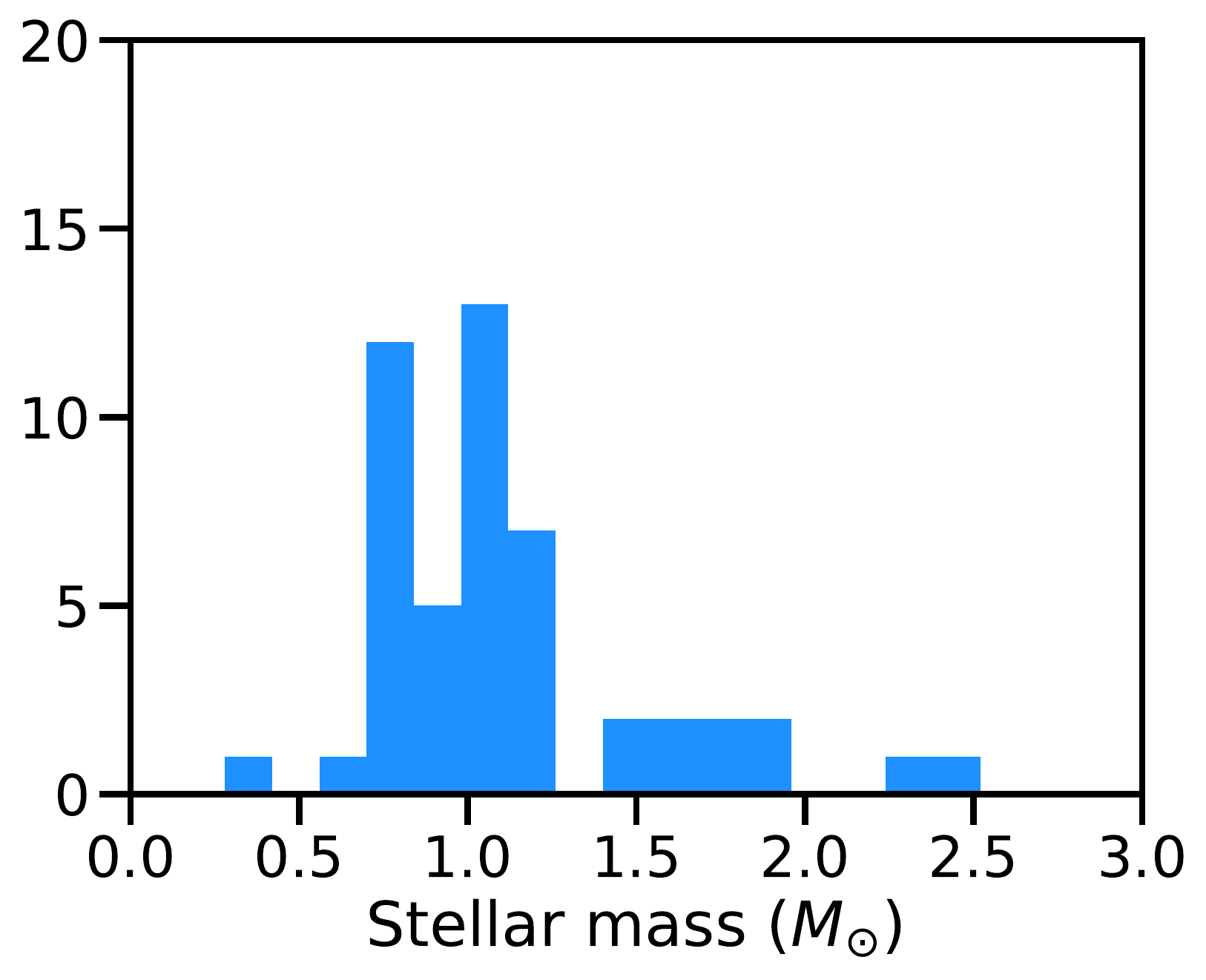}
\caption{\label{mass}}
\end{subfigure}
\begin{subfigure}[t]{0.24\textwidth}
\includegraphics[width=1\hsize]{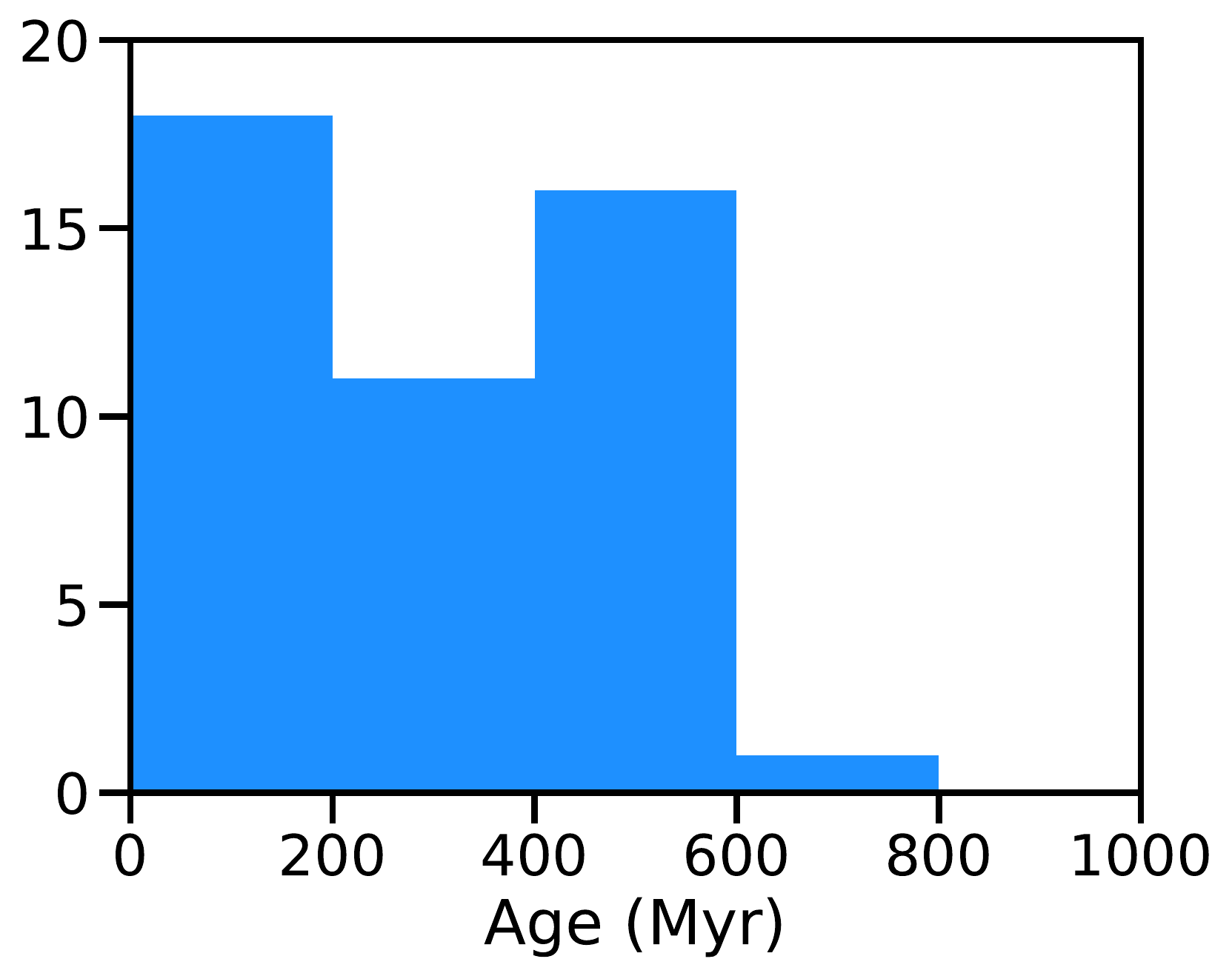}
\caption{\label{age}}
\end{subfigure}
\caption{Main physical properties of our \sophie \ sample.
 \subref{HR})  Absolute $V$-magnitude vs \bv. Each blue dot corresponds to one target.
 The Sun is displayed (red star) for comparison. 
\subref{vsini}) \vsini~vs \bv~distribution.
\subref{mass})  Mass histogram (in \Msun).
\subref{age}) Age histogram.}
       \label{survey_carac_1_s}
\end{figure*}

The spectral type of out targets ranges from A1V to K0V (\Cref{survey_carac_1_s}).  Their projected rotational velocities  (\vsini) range from $1$ to $\SI{90}{\kilo\meter\per\second}$, with a median of  $\SI{6.9}{\kilo\meter\per\second}$ (see  \cref{observable}).
 Their V-band magnitudes range between  $2.1$ and $10.1$, with a median of $7.2$.  
Their masses are between $0.42$ and $\SI{2.52}{\msun}$, with a median of  $\SI{1.07}{\msun}$  (see \Cref{age_masse} for mass determination).
 Our sample includes $13$ targets between A0 and F5V ($B-V \in [-0.05:0.52[$), $35$ between F6 and K5   ($B-V \in [-0.05:1.33[$), and $1$ between K6 and M5 ($B-V \geq 1.33$).
This survey has nine targets in common with the  \harps \ YNS survey: HD 25457, HD 26923, HD 41593, HD 89449, HD 90905, HD 171488, HD 186704 A, HD 206860, and HD 218396.

We present the main characteristics  of our star sample in \Cref{survey_carac_1_s} and \cref{tab_carac_s}.

\begin{figure*}[ht!]
  \centering
\begin{subfigure}[t]{0.315\textwidth}
\includegraphics[width=1\hsize]{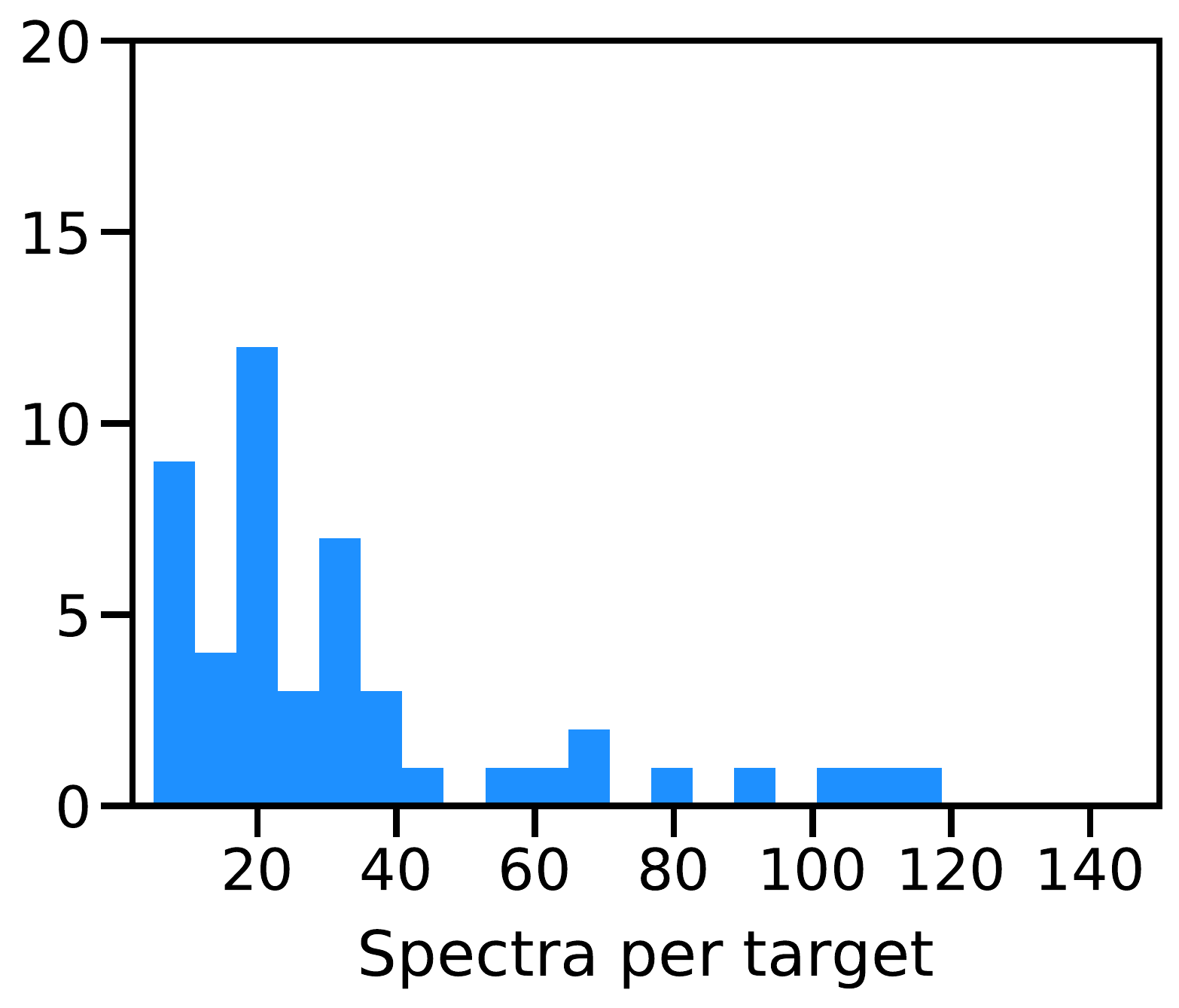}
\caption{\label{Nmes}}
\end{subfigure}
\begin{subfigure}[t]{0.32\textwidth}
\includegraphics[width=1\hsize]{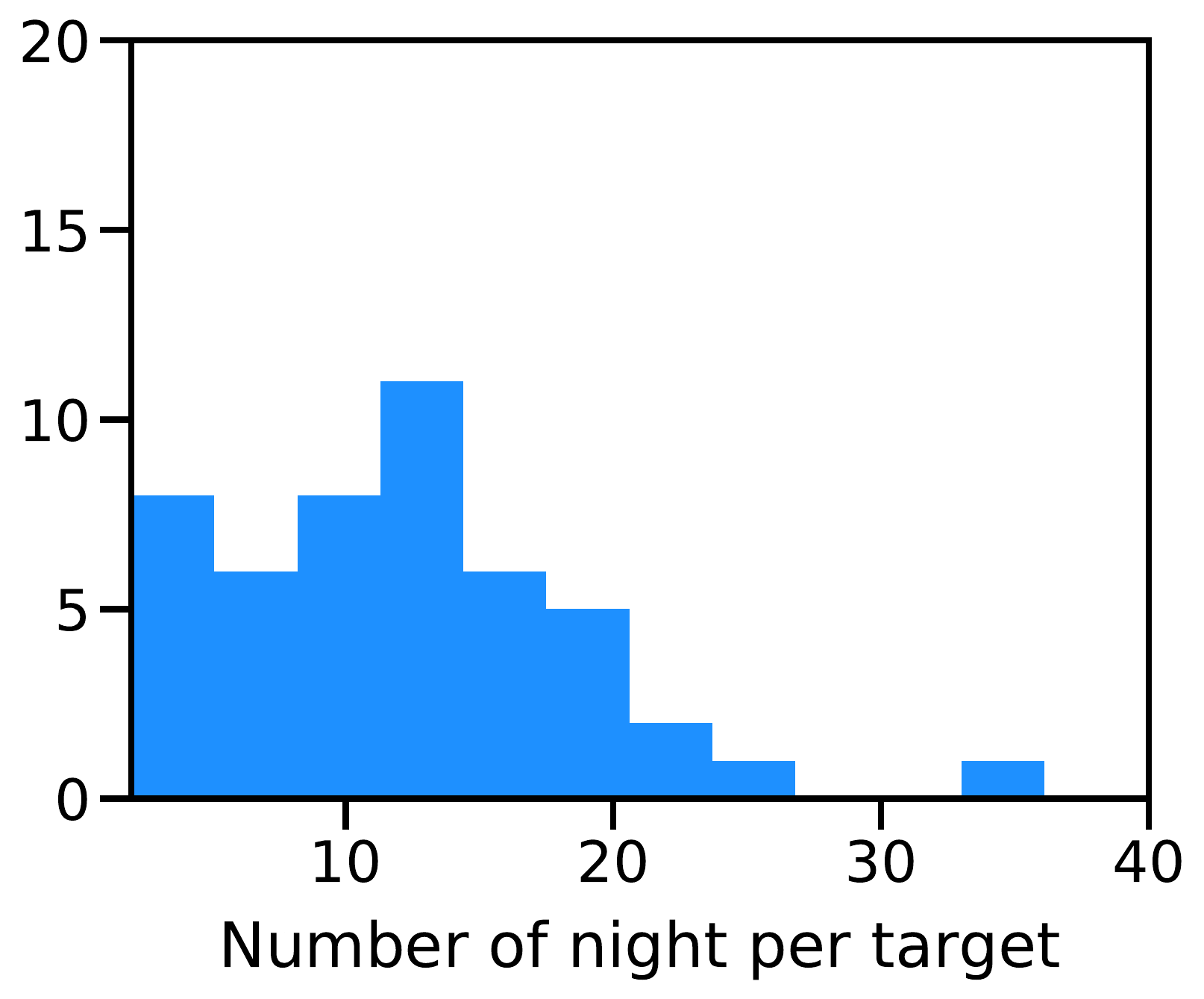}
\caption{\label{Nb_day}}
\end{subfigure}
\begin{subfigure}[t]{0.325\textwidth}
\includegraphics[width=1\hsize]{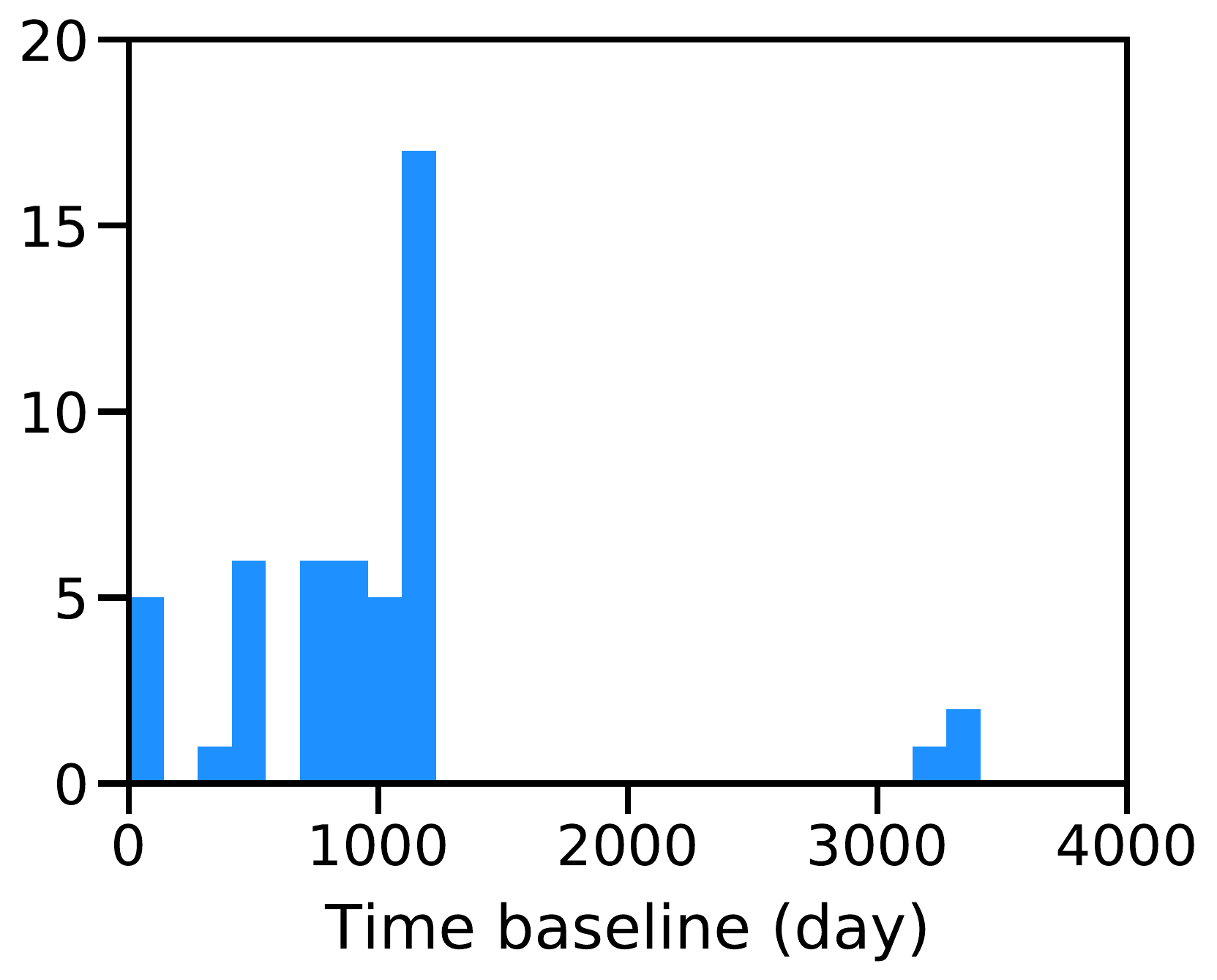}
\caption{\label{time_bsl}}
\end{subfigure}
\caption{\sophie \ observation summary.
 \subref{Nmes}) Histogram of the number of spectra per target. HD 113337 is not displayed (304 spectra).
 \subref{Nb_day}) Histogram of the  number of nights per target. HD 113337 is not displayed (157 nights).
\subref{time_bsl}) Histogram of the time baselines.} 
       \label{survey_carac_2_s}
\end{figure*}

\subsection{Observations}

The \sophie \ YNS survey observations were  performed between 2013 and 2016.
Some stars were previously observed as part of previous surveys made by \cite{Simon_VIII,Simon_IX,Simon_X}. Their time baselines extend to 9 years.

We adopted the observational strategy presented in \cite{Simon_VIII}, which consists in recording two spectra per visit and  observing each target on several consecutive nights. This strategy permits us to sample the short-term jitter of late-type
stars. For early-type stars the strategy consists in obtaining long sequences of observations ($>\SI{1.5}{h}$), in order to sample the pulsations of these stars. The number of spectra per target and the number of visits per nights of these stars  is then higher  than for  the late-type stars.
The median time baseline of the sample is \SI{975}{\day} (mean time baseline of  \SI{975}{\day}), with a  median number of spectra per target of $22$ ($38$ on average), spaced on a median number of $\SI{12}{\night}$  ($15$ on average; see \Cref{survey_carac_2_s}).
Details can be found in \cref{tab_result_s}.

\subsection{Observables}

\label{observable}

We used the  Spectroscopic data via Analysis of the Fourier Interspectrum Radial velocities (\safir) software to  derive the observables  from the \sophie \ spectra: RVs  and whenever possible the cross-correlation function (CCF), the bisector velocity span (BVS),  the star $V\sin{i}$ (from  the full width at half maximum of the CCF), and the \rhk.  
\safir \ builds a reference spectrum from the median of all spectra available on a given star, and computes the relative RV in the Fourier plane.
The computed RVs are then only relative to the  reference spectrum,  hence  we do not compute RVs in heliocentric or barycentric referentials.
The efficiency of this method was demonstrated in the search for  low-mass companions  around AF-type MS stars \citep{Galland_05,Galland_GP}.

To filter the bad spectra we used the selection criteria used for the  \harps \ YNS survey \citep{Grandjean_HARPS}: $S/N_{\SI{550}{\nano\meter}}$\footnote{Signal-to-noise ratio per pixel between $554.758$ and \SI{555.299}{\nano\meter}, estimated with a sampling of $1$ pixel every $\SI{3}{\pico\meter}$.} $\in [80;380]$, $sec \ z < 3$, and $\chi^2 < 10$ \citep{Galland_05}. For faint stars ($V > 8.8 \ mag$) we included spectra with $S/N_{\SI{550}{\nano\meter}}$ down to $30$ as it was the best compromise to include enough spectra to perform our analysis without degrading the RV uncertainties.
To determine the main source of RV variability of each star (magnetic activity, pulsations, or  companions), we used  the correlation between BVS and RV  in addition to the shape of the bisectors \citep{Lagrange_2009,Lagrange,Simon_IX}. 
\label{safir}

\section{Detected companions in the \sophie \ survey}
\label{comp} 

Among the $49$ stars used in our analysis, two planets with periods lower than $\SI{1000}{\day}$ were discovered in the HD 113337 system \citep{Simon_VIII,Simon_X}. In addition, $13$ stars present RVs that are dominated by the signal of a companion. We present their characterizations below.

\subsection{RV long-term trends, and stellar binaries}
\label{trend_bin}

We present here the stars that exhibit a  long-term trend, as well as the stars that exhibit a binary signal in their RVs (single-lined spectroscopic  binary, SB1) or in their CCF (double-lined spectroscopic  binary, SB2).
When possible, we characterized the  companion that induces the SB1 binary signal with \emph{yorbit} \citep{Segransan}. These characterized SB1 binaries are presented in \cref{binary}, and their companions' parameters are summarized in \cref{tab_bin}\footnote{We provide the uncertainties as given by  \emph{yorbit}, but it should be noted that they are often underestimated.}. In addition, we present the stars with a long-term trend in   \cref{trend}.
For these long-term trends, we estimated the minimum mass needed for a companion to produce the trend. It corresponds to the minimum mass needed for a companion on a circular orbit with a period equal to the time baseline to produce RV variations with an amplitude equal to the drift amplitude.  It represents the limit below which a companion on a circular orbit cannot explain the observed total amplitude in the RVs. A companion whose period is  equal  to the time baseline would   produce a signal with visible curvature in our data. The actual period, and thus the actual  mass of the companion, is therefore significantly greater than this limit. On the other hand, for a fixed period a companion on an eccentric orbit would produce an identical amplitude with a lower mass, yet with a more visible curvature. A longer period and therefore a higher mass would thus be necessary to produce only a linear trend in the observational window.  We can therefore assume that our limit closely  corresponds to the limit below which a companion cannot explain the observed total amplitude in the RVs. We present the lower limits of the masses in \cref{tab_trend}.
We also verified that the slope induced by the secular drift of these stars is negligible towards the slope of the observed trend. These trends thus cannot be attributed to  secular drift.

\begin{table}[h!]
\center
\begin{tabular}[h!]{|l|c|c|c|}
System& $P$ & $e$ & $M_p\sin{i}$  \\ 
& $(\si{\year})$ && $(\si{\MJ})$  \\ \hline
HD 39587& $\sim14$ & $\sim0.4$ & $\sim150$ \\ 
HD 195943& $3.73\pm0.03$ & $0.093 \pm 0.004$ & $561\pm4$  \\ 
\end{tabular}
\caption{Orbital parameters of the characterized binaries.}
\label{tab_bin}
\end{table}

\begin{table}[h!]
\center
\begin{tabular}[h!]{|l|c|c|}
Star & $P (\si{\year})$ & $M_c (\si{\MJ})$  \\ \hline
HD 2454 & $>2.2$ & $>16$  \\ 
HD 17520 & $>0.01$ & $>7$  \\ 
HD 109647& $>3.1$ & $>2$  \\ 
HD 112097 & $>0.01$  & $>52$  \\ 
\end{tabular}
\caption{Lower limits on periods and masses for the companions that lead trends. We present these limits for the companions whose  periods and masses were not estimated in previous studies.}
\label{tab_trend}
\end{table}

\subsubsection{HD 2454}

\label{2454}

HD 2454 is an F5V-type star reported as a spectroscopic binary by \cite{Escorza}. This star is known to present magnetic activity. \cite{Rutten} measured a rotation period of $\SI{7.8}{\day}$ from the Ca II variations, while \cite{Olspert} measured its harmonic at $3.47\pm0.01 \si{\day}$.
 We observe a trend with a slope of $\SI{222}{\meter\per\second\per\year}$ over baseline of $\SI{805}{\day}$ with a sign of curvature, in addition to a short term jitter. The minimum mass needed for a companion to induce this trend is  $\SI{16}{\MJ}$. We attribute this trend to the known binary companion.
We performed a polynomial fit on the RVs to remove the companion signal (see \Cref{trend}). We chose a third-degree model as it presented the best reduced $\chi^2$. The residuals have  a standard deviation of $\SI{11}{\meter\per\second}$. The (BVS, RV residuals) diagram is vertically spread. However, the star presents a low sign of activity with a $<$\rhk$>$ \ of $-4.89$ (with a standard deviation of $0.08$). In addition, the periodicities  present  in the residuals between $4$ and $\SI{10}{\day}$ (with a maximum at $\sim \SI{5}{\day}$)  are also present in the BVS, while they are not present in the time window periodogram. This indicates that the RV jitter could also come from stellar activity. According to the star spectral type (F5V) and its relatively old age ($800\pm300 \ \si{\mega\year}$; \cite{Gomez}), it is difficult to determine if the RV jitter is dominated by magnetic activity (spots) or pulsations.
Due to the weak correlation between the BVS and the RV residuals, we chose to not correct the RV residuals for this correlation.

\begin{figure}[h]
  \centering
\includegraphics[width=0.9\hsize]{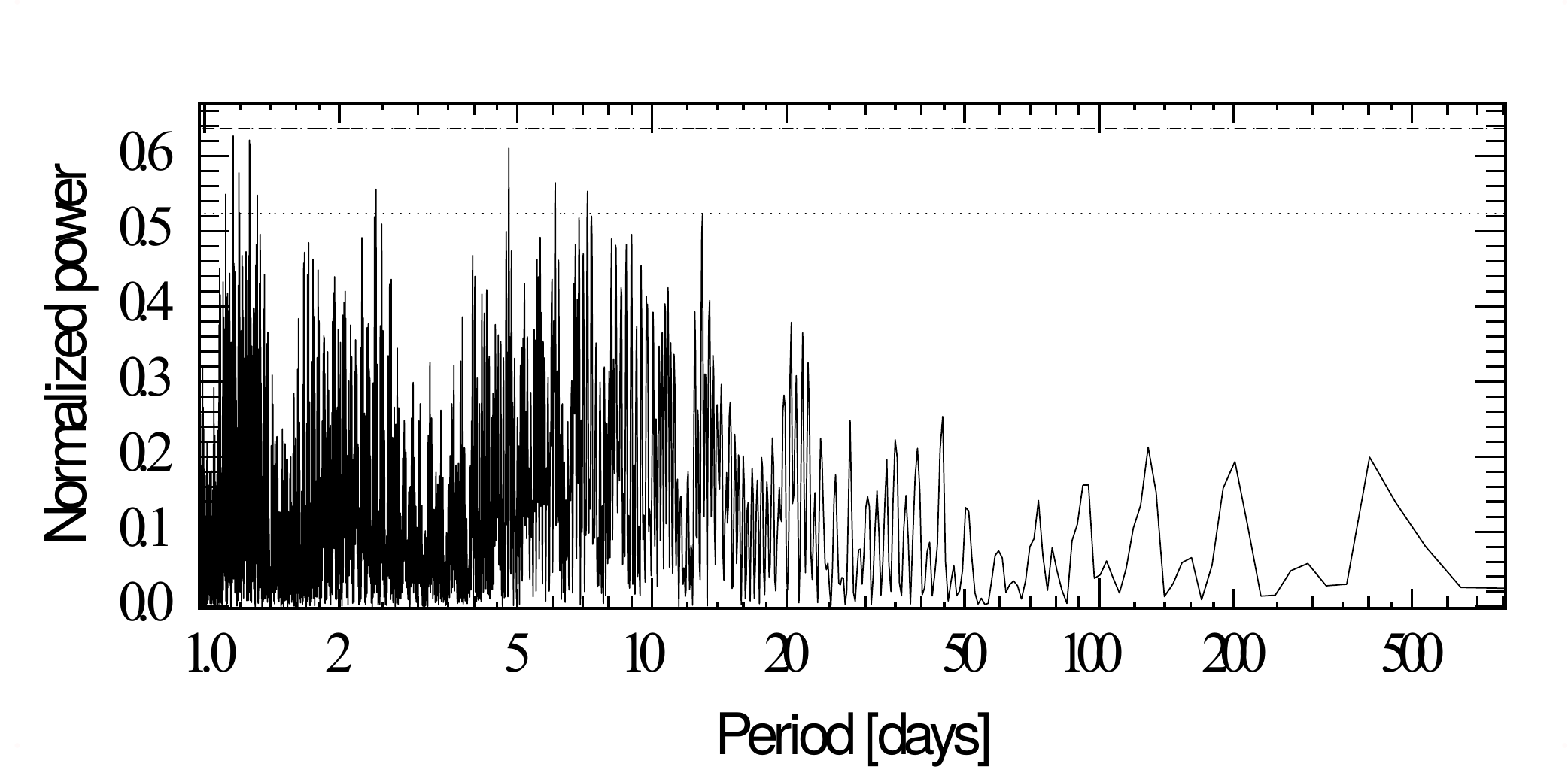}
\includegraphics[width=0.9\hsize]{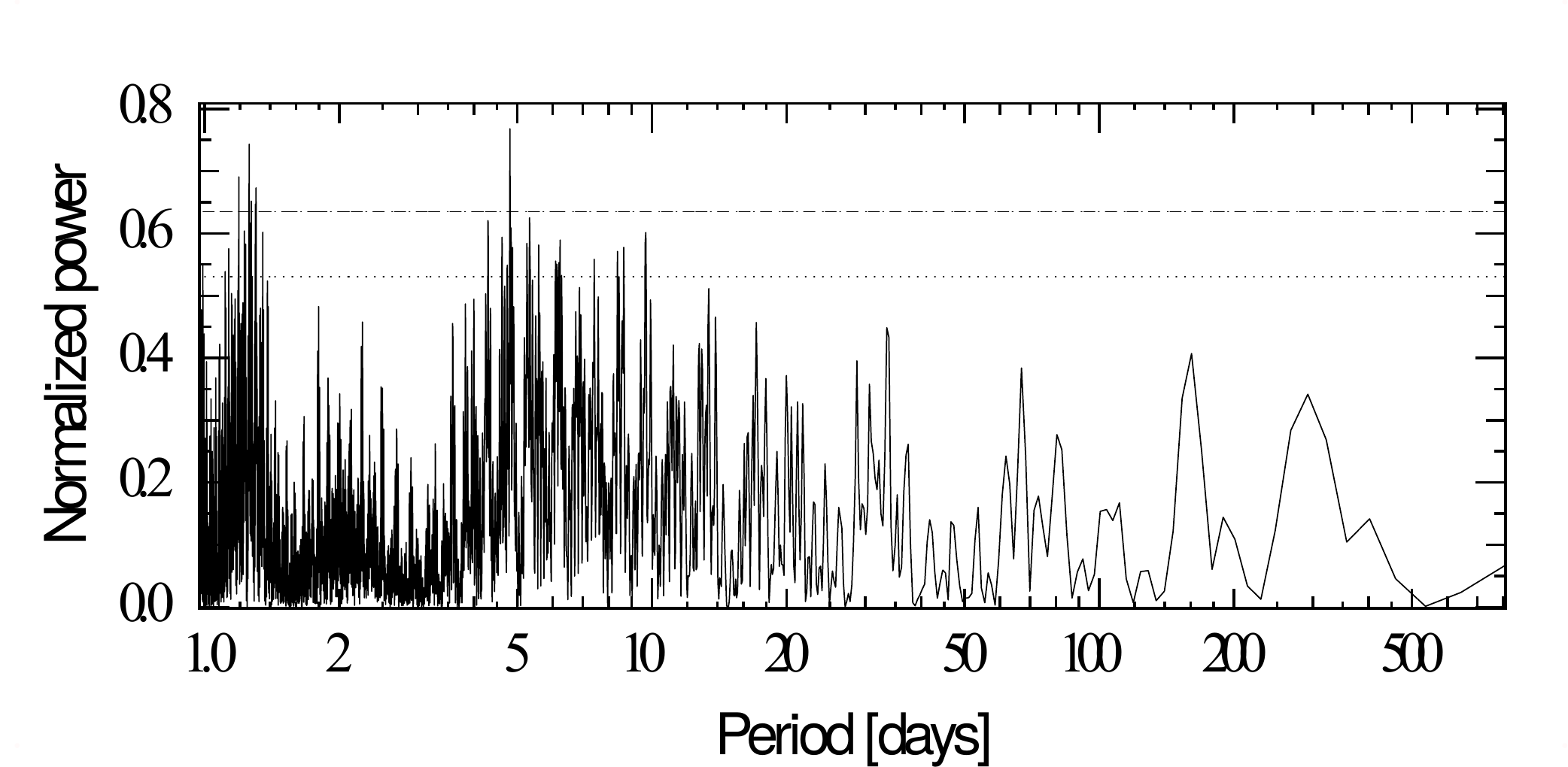}
\includegraphics[width=0.9\hsize]{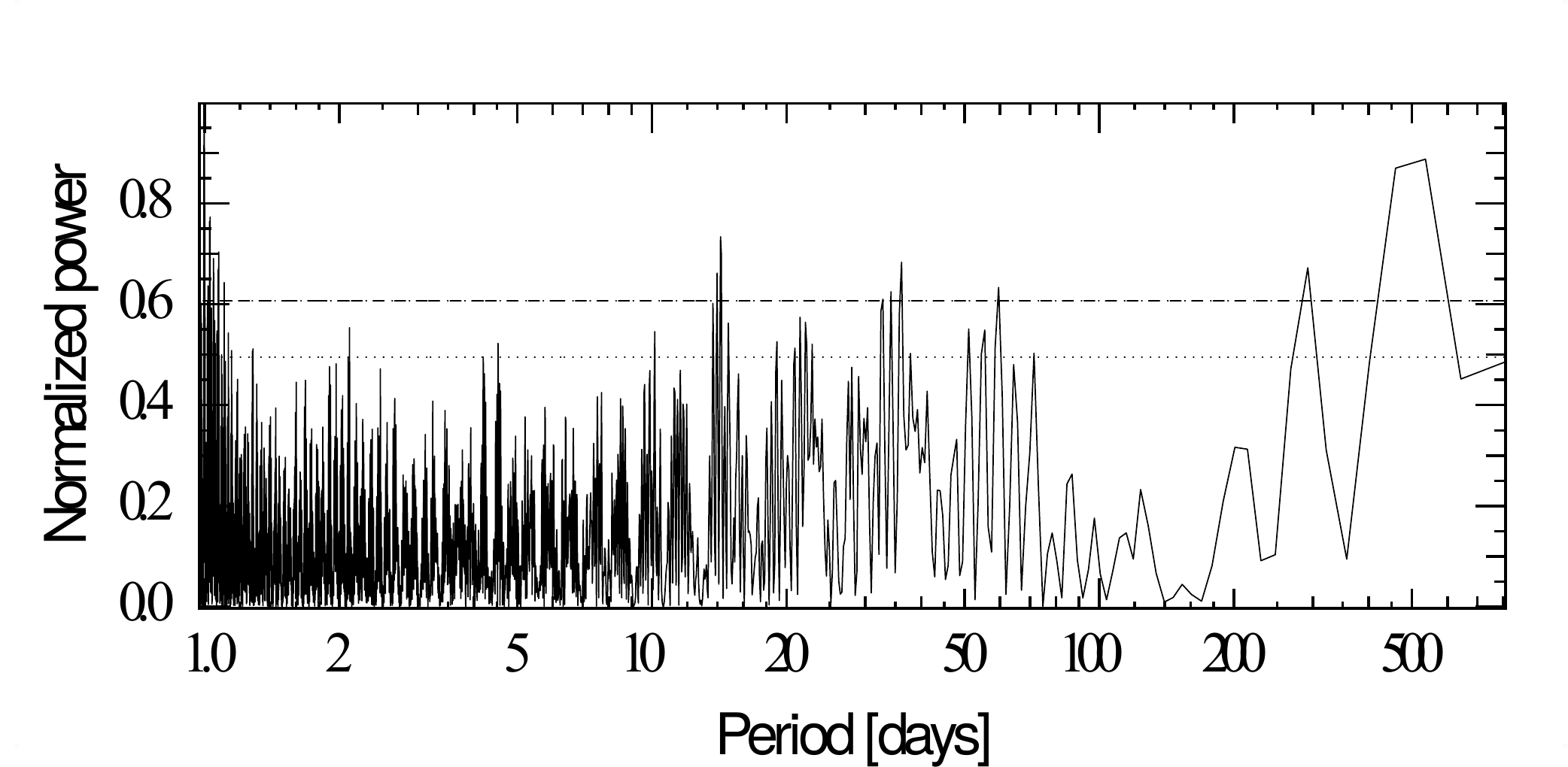}
\caption{HD 2454 periodograms. From \emph{top} to \emph{bottom} : periodogram of the RV residuals of the third-degree polynomial fit, BVS periodogram, time window periodogram. The $10 \%$ false alarm probability (FAP) is shown as  a dotted line and the $1 \%$ FAP  as a dashed line.}
       \label{2454}
\end{figure}

\subsubsection{HD 13507}

\begin{figure}[h]
  \centering
\includegraphics[width=0.9\hsize]{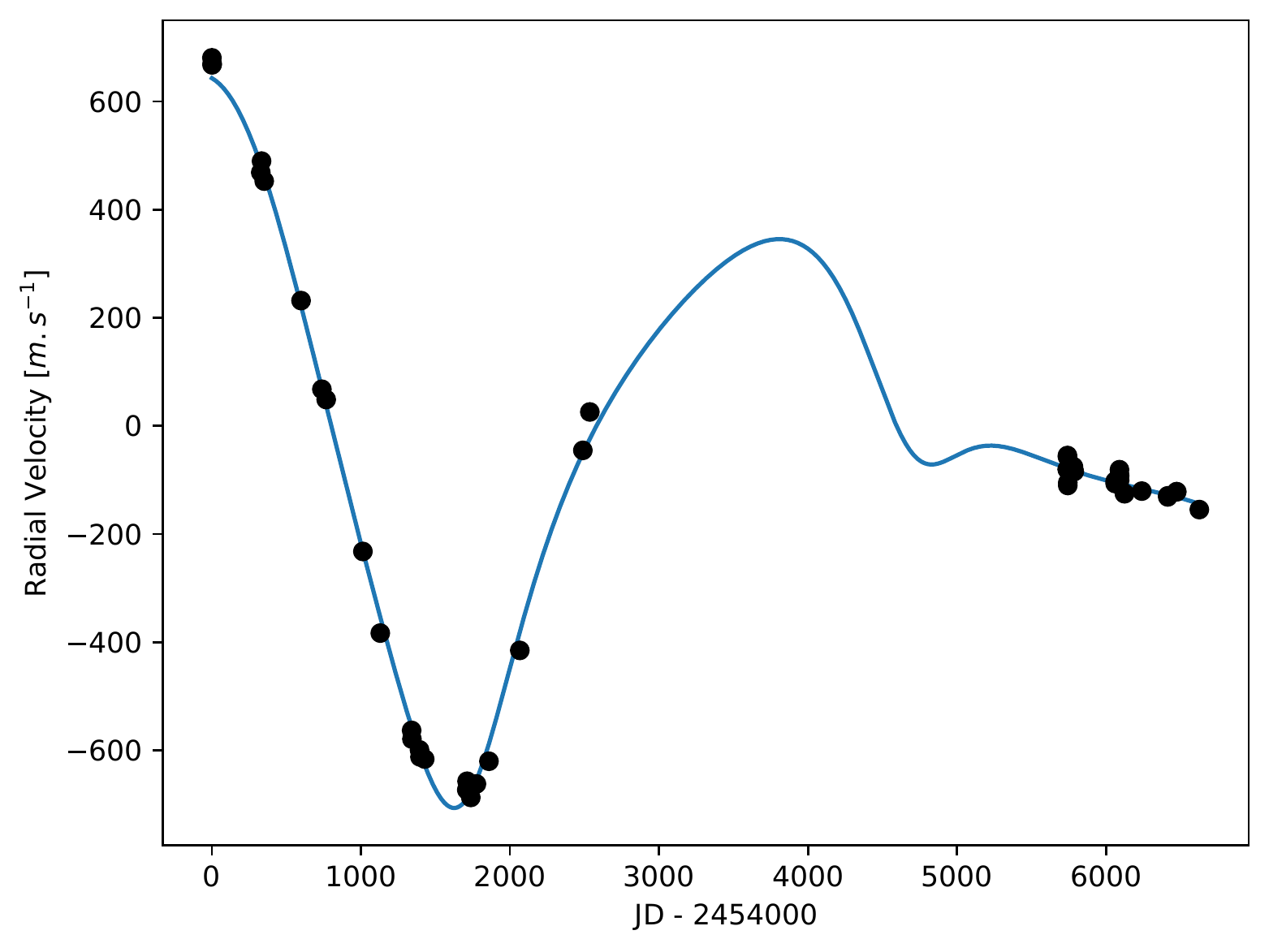}
\caption{HD 13507 RVs fit using two Keplerians, performed with Dpass \citep{Beta_pic_c} on the combination of the \cite{Perrier}  ELODIE RV dataset  and   our \sophie \ RV dataset.} 
       \label{fig_13507}
\end{figure}

HD 13507 is a G5V-type  star, with a $7.45-day$ rotation period \cite{Wright_prot}.  \cite{Perrier} reported a $\SI{52}{\MJ}$ companion with a $\sim 3000-day$ period and an eccentricity of $0.14$ from \elodie \ RVs.
\cite{Wilson_16} refined the orbital solution using the \cite{Perrier} RVs together with  \elodie \  additional RVs. They obtained a period of $4890^{+209}_{-109} \ \si{\day}$ and a $M_p\sin{i}$ of $67^{+8}_{-9} \ \si{\MJ}$.
 We observe a trend in our \sophie \ RVs with a slope of $\SI{26}{\meter\per\second\per\year}$ over a baseline of $\SI{886}{\day}$.  In order to constrain the BD companion parameters, we combined our \sophie \ RV dataset with the ELODIE RV dataset of  \cite{Perrier} and we fit them together. For this fit we used our Dpass  tool, which is based on an evolutionary algorithm \citep{Beta_pic_c}, and we took into account an offset between the two datasets.
Our fit does not confirm the \cite{Perrier} and \cite{Wilson_16} solutions, and it clearly shows that the system is not single. We then considered a two-planet system\footnote{No constraint was put on the two companion properties, except on the eccentricities that were constrained to be lower than 0.3. Both eccentricities were regularized under the arbitrary assumption that eccentricities follow a normal law (0, 0.1).}, and found a possible solution including two BD companions of lower masses, $0.02$ and $\SI{0.03}{\msun}$  ($\SI{21}{\MJ}$ and $\SI{31}{\MJ}$, respectively ) orbiting at $4.2$ ($\Leftrightarrow$ \SI{8.3}{\year}) and $\SI{5.4}{au}$ ($\Leftrightarrow$ \SI{12.1}{\year}), and with low eccentricities, $0.27$ and  $0.2$, respectively ({cf.}  \cref{fig_13507}). Other degenerate solutions may exist, which prevents us from giving   a proper estimate of the uncertainties on these parameters.
We note that such companions could be detected in high-contrast imaging, and that the orbits are close to 3:2 resonance. We also note that this solution should be taken with caution, given the limited number of data points available. 

\subsection{HD 13531}

HD 13531 is a G7V-type star, with a $7.49-day$ rotation period \citep{Wright_prot},   that presents an IR excess \citep{McDonald}. The radius of the corresponding warm disk was estimated at $\SI{1.78}{\au}$ \citep{Gaspar}. We observe a trend in the RVs with a slope of $\SI{37}{\meter\per\second\per\year}$ over a baseline of $\SI{882}{\day}$. The minimum mass needed for a companion to induce this trend is  $\SI{3}{\MJ}$. \cite{Metchev} discovered a low-mass star candidate companion by direct imaging with a semimajor axis of $\SI{18.6}{\au}$ (\SI{81}{\year}) and a mass estimated at $\SI{0.19}{\msun}$. For circular and edge-on orbits the total amplitude in RV of this companion is   $\SI{2.65}{\kilo\meter\per\second}$, and the mean annual variation is   $\SI{65}{\meter\per\second\per\year}$, which is on the order of magnitude of the slope of the trend we observe.
We thus attribute   this RV trend to this companion. 
The residuals of a linear regression show a jitter with an amplitude of  $\SI{11}{\meter\per\second}$ (see \Cref{trend}). These residuals show a significant correlation between the BVS and the RVs ($Pearson=-0.54$, $p_{value}=0.1 \%$), which indicate that the jitter is due to stellar activity (spots).

\subsubsection{HD 17250}

HD 17250 is known as a hierarchical multiple system composed of four stars \citep{Tokovinin_I}. The main star, HD 17250 A, presents an IR excess \citep{McDonald} and a spectroscopic binary companion  \citep{Tokovinin_I}.  
We observe a trend of $\SI{1145}{\meter\per\second}$ over $\SI{5}{\day}$. According to the amplitude of the signal on a such short timescale, it is unlikely that the signal is produced by a companion, other than the one reported by \citet{Tokovinin_I}, as it would require   the latter to produce a signal of even greater amplitude over a longer timescale.
We thus attribute  this signal to the spectroscopic companion reported by  \cite{Tokovinin}. 
Our data are too sparse to study the residuals of a linear regression on the RVs (see \cref{trend}).

\subsubsection{HD 28495}

\begin{figure}[h]
  \centering
\includegraphics[width=0.7\hsize]{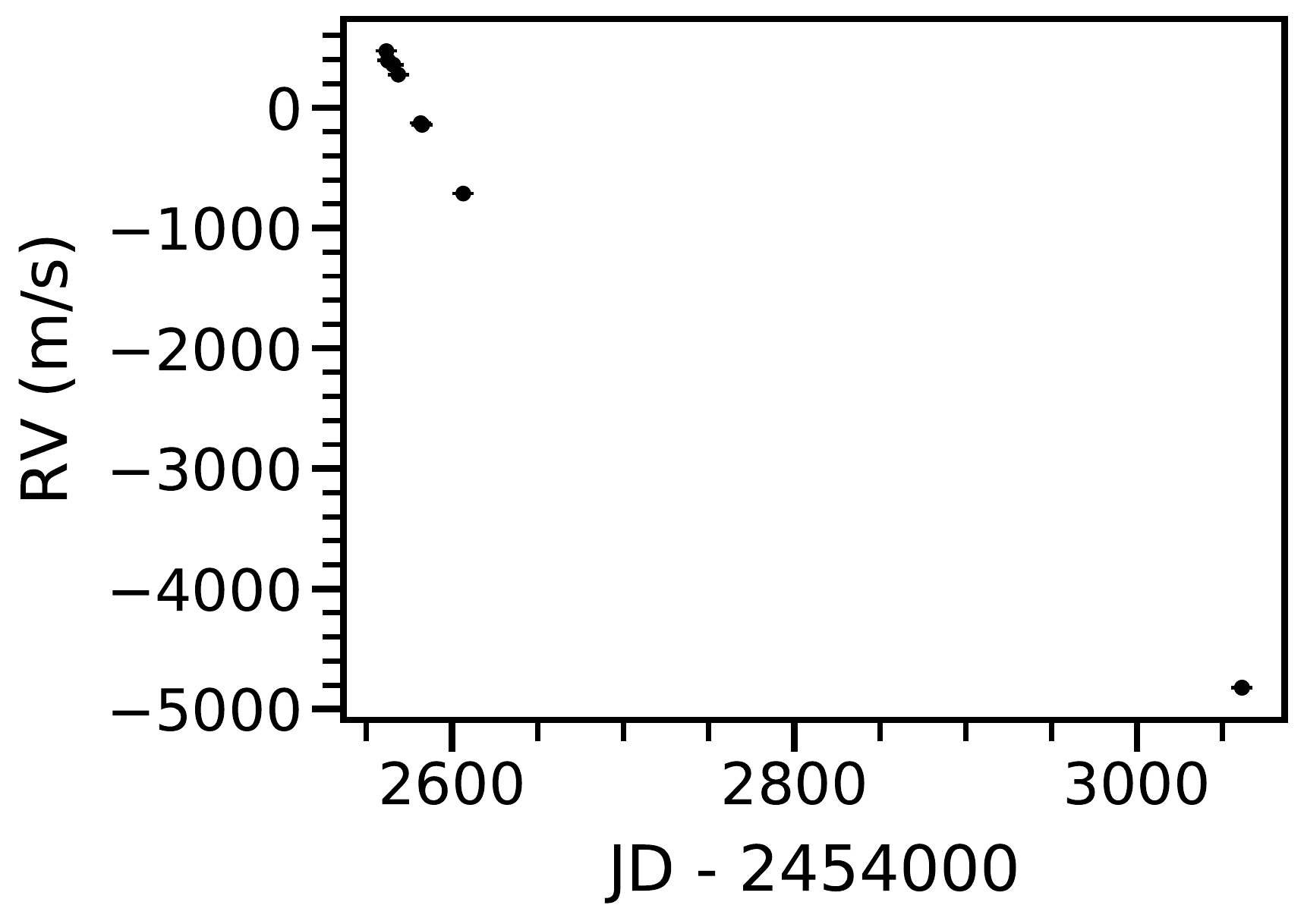}
\caption{HD 28495 RV time variations} 
       \label{28495}
\end{figure}

HD 28945 is a G0V-type star that was reported as a spectroscopic binary by \cite{Nordstrom_181}, then as an astrometric binary from Hipparcos proper motion \citep{Makarov,Frankowski}.
We observe a variation in the RVs with an amplitude of $\SI{5.3}{\kilo\meter\per\second}$ over $\SI{500}{\day}$ with a sign of curvature.  However, this sign of curvature relies on only two data points (see \Cref{28495}). We attribute these RV variations to the known stellar companion, but more data points are needed to characterize the system.

\subsubsection{HD 39587}

\begin{figure}[h]
  \centering
\includegraphics[width=1\hsize]{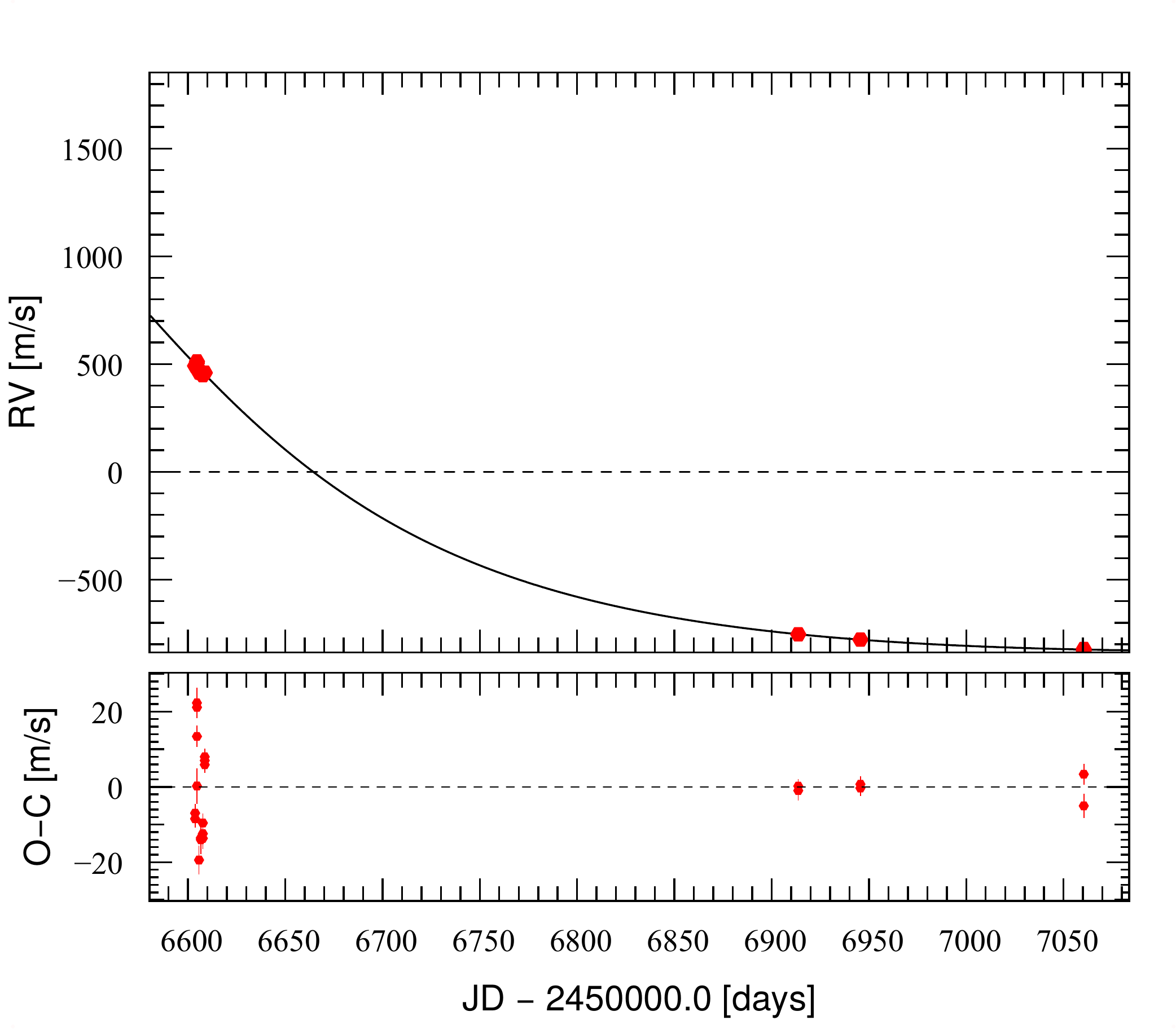}
\caption{HD 39587 RVs best fit with one Keplerian and its residuals.} 
       \label{39587}
\end{figure}

HD 39587 ($\chi^1$ Orionis) has been  known as a binary since 1978 from absolute astrometry; it is  composed of a solar-like G0V star and a low-mass star companion on a close and nearly edge-on orbit ($i = \SI{95}{\degree}$) \citep{Lippincott_1978}. 
The orbital parameters of the stellar companion were refined by combining radial velocity and absolute astrometry by \citet{han2002precise}. They obtained a period of $5156.7 \pm 2.5 \si{\day}$, an eccentricity of $0.451 \pm 0.003$, and a mass of $0.15 \pm 0.02 \si{\msun}$ for the companion. This companion was then directly imaged in 2002 by \cite{Konig}.
We observe the signal of the stellar companion in our RVs. We use \emph{yorbit} to fit the RVs with one Keplerian model. For this fit we adopt a mass of $\SI{1.07}{\msun}$ for the primary \citep{Tokovinin}.
We obtain a period of $5600 \pm 3400 \si{\day}$ ($\SI{6.5}{\au}$, $\SI{750}{\milli\arcsec}$), an eccentricity of $0.89 \pm 0.15$. However, the $M_p\sin{i}$ is not constrained in this fit. We present this fit in \Cref{39587} and the (BVS, RV) diagram of its residuals in \Cref{binary}).
Our solution is consistent at $1 \sigma$  with previous literature values in terms of period, but not in terms of eccentricity.
Our solution is more eccentric, which leads to a lower mass for the companion. We see a strong correlation between the BVS and the RVs of the spectra taken after 2014, which indicates that they might be altered by the activity of the star (spots). This can explain the strong difference in eccentricity between our values and those found in    the literature.

\begin{figure}[h]
  \centering
\includegraphics[width=0.9\hsize]{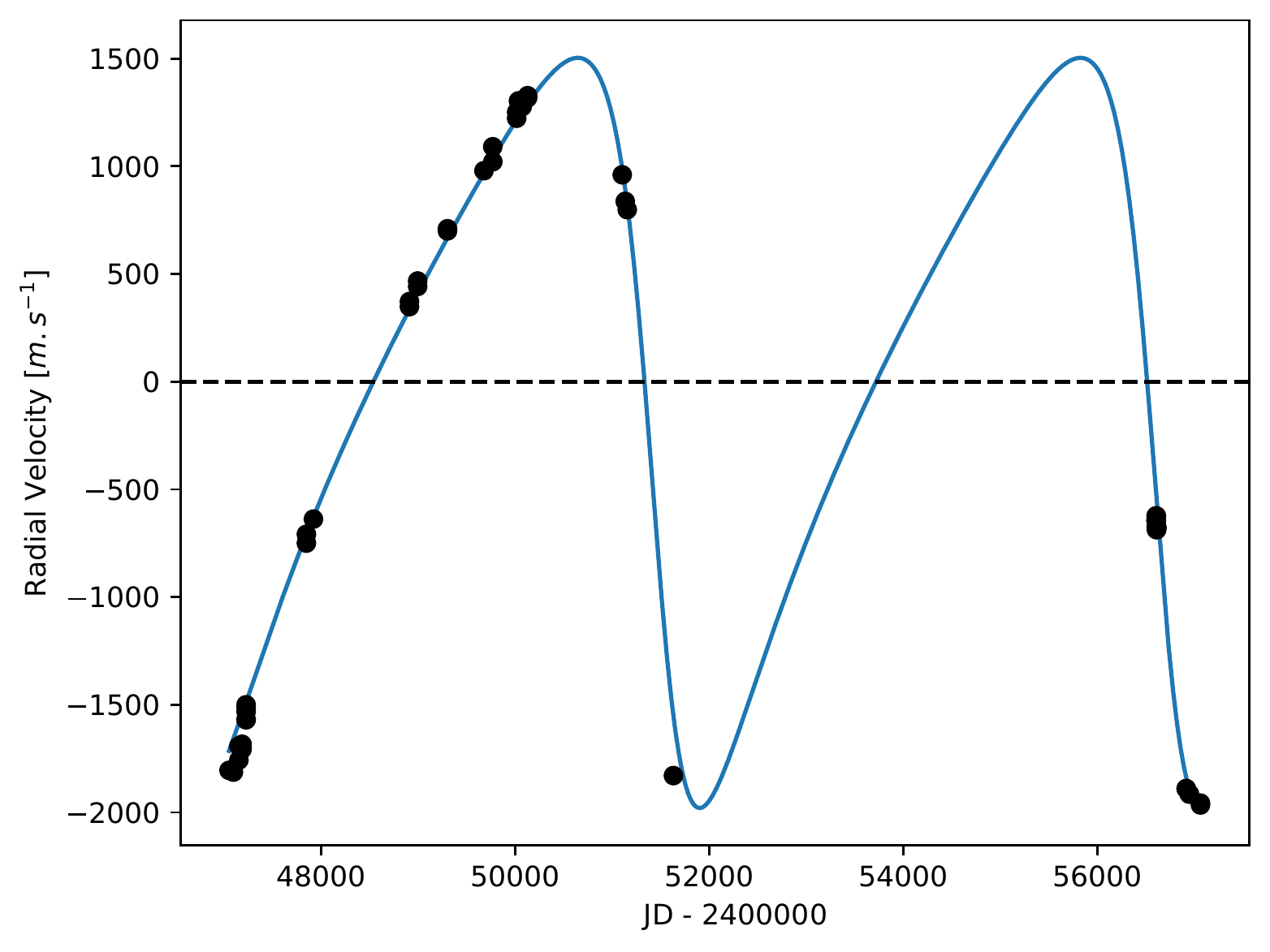}
\caption{HD 39587 RVs fit using one Keplerian, performed with Dpass \citep{Beta_pic_c} on the combination of the \cite{han2002precise} Lick Observatory RV dataset  and   our \sophie \ RV dataset.} 
       \label{39587_join}
\end{figure}

We then combined our \sophie \ RV dataset with the Lick Observatory RV dataset of  \cite{han2002precise} in order to perform a joined fit  with our Dpass  tool \citep{Beta_pic_c} using a singular Keplerian model. We also adopted a mass of  $\SI{1.07}{\msun}$ \citep{Tokovinin} for the primary  and we took into account an offset between the two datasets.  We added quadratically $\SI{10}{\meter\per\second}$  to the error bars of our  \sophie \ RV dataset to take into account that the stellar origin jitter is badly sampled after 2014.  We present this fit in \Cref{39587_join}. We obtain a mass of $ \SI{0.15}{\msun}$, an eccentricity of $0.44$, and a period of $\SI{5180}{\day}$ for the companion, which is similar to the estimation of \cite{han2002precise} on these parameters.  However, as other degenerate solutions may exist in our fit we cannot make a proper estimation of the uncertainties on the value of the parameters we estimate, which prevent us from making a better comparison between our results and the results of \cite{han2002precise}.

\subsubsection{HD 105963}

HD 105963 is a known wide-orbit binary \citep{Lepine} with a $\ang{;;13.5}$ separation \citep{Durkan_2016}.
We observe a double component in the CCF. The field of view of the  \sophie\ fiber is   $\ang{;;3}$ \citep{SOPHIE+}, so this component cannot come from the wide separation companion.
We thus report  HD 105693 A as a SB2.

\subsubsection{HD 109647}

HD 109647 is a K0V-type star that presents an infrared excess \citep{McDonald}. We observe a trend in the RVs with a slope of $\SI{17}{\meter\per\second\per\year}$ over a baseline of $\SI{1147}{\day}$. The minimum mass needed for a companion to induce this trend is $\SI{2}{\MJ}$. 
The residuals of a linear regression show a jitter with an amplitude of  $\SI{22}{\meter\per\second}$ (see \Cref{trend}). These residuals show a significant correlation between the BVS and the RVs ($Pearson=-0.6$, $p_{value}=0.2 \%$), which indicates that the jitter is due to stellar activity (spots).
The age of the system is $\SI{412}{\mega\year}$ \citep{Stone}, which  does not allow us to confirm this companion by direct imaging with the current instrumentation.

\subsubsection{HD 112097}

HD 112097 is an  A7III-type star. We observe a trend of $\SI{8}{\kilo\meter\per\second}$ over $\SI{3}{\day}$. The minimum mass needed for a companion to induce this trend is  $\SI{52}{\MJ}$. We report thus HD 112097 as a spectroscopic binary.  Our data are too sparse to study the residuals of a linear regression (see \Cref{trend}).

\subsubsection{HD 131156}

HD 131156 ($\xi \ Boo$) is a G7V-type star known to be a visual binary since 1950 \citep{Muller}.
\citet{Stone} measured a separation of $\ang{;;4.94}$ ($\SI{33}{\au}$) and estimated the mass of the two components as $m1 = \SI{1}{\msun}$ and $m2 = \SI{0.7}{\msun}$.
We observe a trend with a slope of $\SI{35}{\meter\per\second\per\year}$ over a baseline of $\SI{1154}{\day}$.
The maximum annual RV variation that HD 131156 B can apply to its host star is $\SI{76}{\meter\per\second\per\year}$, which is greater than the slope of the trend we observe. 
We thus attribute  the RV trend we observe to HD 131156 B.
The residuals of a linear regression  show a jitter with an amplitude of  $\SI{30}{\meter\per\second}$ (see \Cref{trend}). These residuals show a significant correlation between the BVS and the RVs ($Pearson=-0.88$, $p_{value}< \num{4e-12}\%$), which indicates that the jitter is due to stellar activity (spots).

\subsubsection{HD 142229}

HD 142229 is a G5V-type star known to present an IR excess \citep{McDonald} and a RV trend \citep{Nidever}.
\cite{Gaspar} estimated the warm disk radius at $\SI{1.94}{\au}$ from $\SI{24}{\micro\meter}$ Spitzer data. 
From additional RV data \cite{Patel_2007} estimated a period greater than $\SI{16.4}{\year}$ and an $M_p\sin{i}$ greater than $\SI{150}{\MJ}$ for the companion.
We observe a trend with a slope of $\SI{71}{\meter\per\second\per\year}$ over a baseline of $\SI{1111}{\day}$. This trend is compatible with the solution obtained by  \cite{Patel_2007}, we thus attribute this trend to HD 142229 B. 
The residuals of a linear regression  present a standard deviation of $\SI{21}{\meter\per\second}$ with a weak correlation between BVS and RV (see \Cref{trend}). According to the star spectral type we attribute this jitter to spots.

\subsubsection{HD 186704 A}

HD 186704 is a known binary system with a companion at \SI{10}{\arcsecond} \citep{Zuckerman_13}.
\cite{Nidever} reported a trend in the RVs of $88 \pm 8$ $\si{\meter\per\second\per\jdb}$ with a negative curvature based on four observations spaced over \SI{70}{\day} for HD 186704 AB.
\cite{Tremko} observed a change in the RVs of \SI{4200}{\meter\per\second} in \SI{8682}{\day} on HD 186704 A.
Finally, \cite{Tokovinin} reported HD 186704 A as hosting a spectroscopic binary (SB) companion with a $3990-day$ period.
We reported  in \cite{Grandjean_HARPS} a trend in the  \harps \ RVs with a slope of  $\SI{275}{\meter\per\second\per\year}$ over a baseline of  $\SI{450}{\day}$ baseline (data taken between 2014 and 2015). This trend was attributed to the known SB companion.
We observe in the \sophie \ RVs a trend with a slope of $\SI{268}{\meter\per\second\per\year}$ over a baseline of $\SI{1042}{\day}$ (data taken between 2013 and 2016).
It emphasizes the consistency between \sophie \ and \harps \ data. 
The residuals of a linear regression show a jitter with an amplitude of  $\SI{38}{\meter\per\second}$ (see \Cref{trend}). These residuals show a significant correlation between the BVS and the RVs ($Pearson=-0.64$, $p_{value} = 0.2\%$), 
which indicates, in addition to the star's relatively fast rotation ($P_{rot} = \SI{3.511}{\day}$; \cite{Kiraga}) and Ca II H and K activity ($<$\rhk$> =-4.32$), that the measured jitter is likely due to stellar activity (spots).

\subsubsection{HD 195943}

\begin{figure}[h]
  \centering
\includegraphics[width=1\hsize]{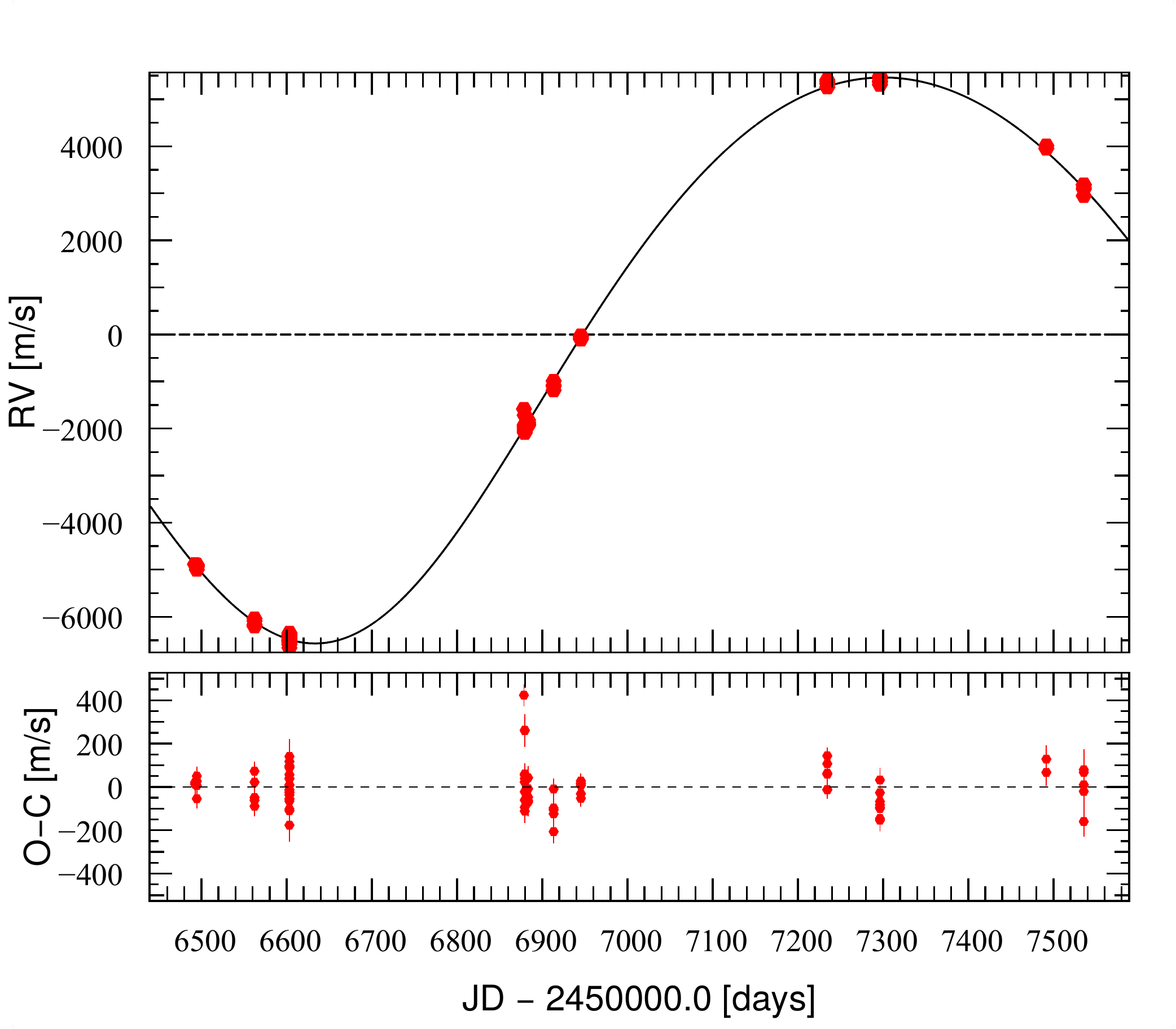}
\caption{HD 195943 RVs best fit with one Keplerian and its residuals.} 
       \label{195943}
\end{figure}

\begin{figure*}[t]
  \centering
\includegraphics[width=1\hsize]{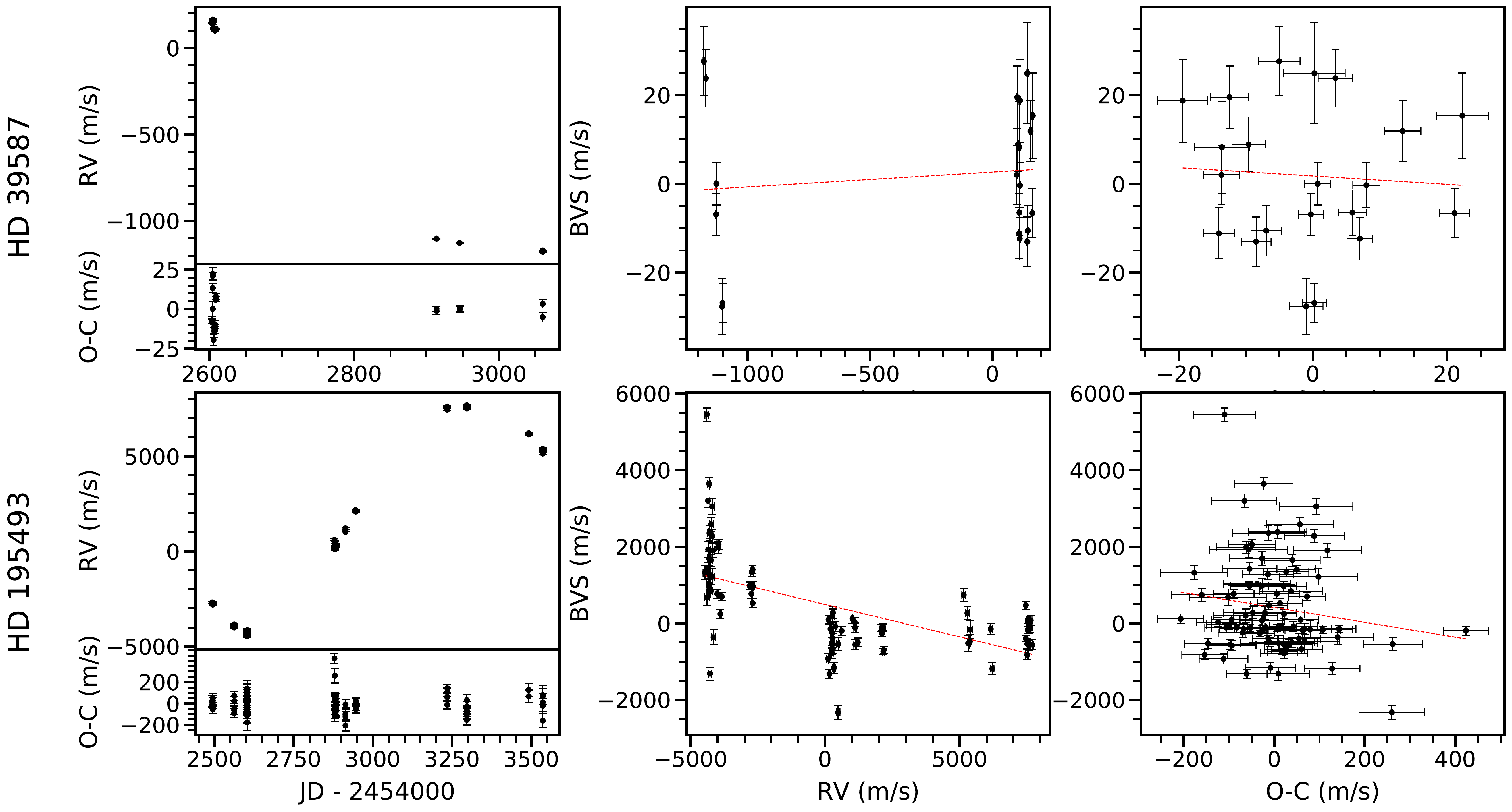}
\caption{Binary star RV analysis. \emph{First column}: RV time variations  (\emph{top}) and its Keplerian fit residuals. These fits are presented in \cref{39587} for HD 39587 and in \cref{195943} for HD 195943. \emph{Second column}: BVS vs RVs  (black) and its best linear model (red dashed line).  \emph{Third column}: BVS vs RV residuals  and its best linear model (red dashed line).} 
       \label{binary}
\end{figure*}

HD 195943 ($\eta \ Del$) is an A3IV-type star that was reported as a binary on the basis of Hipparcos proper motion \citep{Makarov,Frankowski}.
We get a good coverage of this companion signal in RV. Our best \emph{yorbit} fit gives a period of  $1363 \pm 11 \ {\si{\day}}$ ($\SI{3.73}{\year}$), an eccentricity of $0.093 \pm 0.004$, and a $M_p\sin{i}$ of $561\pm\SI{4}{\MJ}$ ($\SI{0.54}{\msun}$). 
We present our fit in \Cref{195943}. For this fit we assumed that the $\SI{2.25}{\msun}$ mass estimated  from an evolutionary model by \cite{Zorec} corresponds to the primary mass. However, this mass estimation might be affected by the binary companion. This will thus lead to an overestimation of the $M_p\sin{i}$ and an underestimation of the period in our fit. The residuals of the fit show a vertical spread of the (BVS, RV) diagram, and can thus be attributed to pulsations (see \Cref{binary}).

\subsubsection{HD 218738}

HD 218738 (KZ Andromedae) is known  to be the component of a common proper motion binary along with HD 218739. HD 218738 is also known as a double line binary with a period of $\SI{3.03}{\day}$ \citep{Bopp_1975, Fekel}.  We observe this double component in our spectra. However, our data are too sparse to characterize this binary.

\subsection{Giant planets : HD 113337}

HD 113337 is an F6V-type star that presents an IR excess \citep{Rhee}.  The corresponding debris disk was resolved by \cite{Su}. A first planetary companion was discovered from previous \sophie \ surveys, in addition to a long-term variation attributed to stellar activity \citep{Simon_VIII}. The additional data obtained during the \sophie \ YNS survey permited \cite{Simon_X} to discover that the long-term variations are due to a second companion.
The parameters of the two companions are  $P_1=323\pm 1 \si{\day}$, $M_{p1}\sin{i_1} = 3 \pm 0.3 \si{\MJ}$ and  $P_2=3265\pm 134 \si{\day}$, $M_{p2}\sin{i_2} = 6.9 \pm 0.6 \si{\MJ}$.

\subsection{Detected companion summary}

Over the $49$ stars of our analysis a system composed of two planets with $P<\SI{1000}{\day}$ was discovered \citep{Simon_VIII,Simon_X}. In addition, we observed ten  single-lined spectroscopic binary systems and two double-lined binary systems, eight of which were already reported in the literature, while two  were unknown (HD 105693 B, HD 112097 B). Finally, we report a long-term trend compatible with a planetary companion on HD 109647.

\subsection{Known giant planet non-detections}

Some of our targets are known to host confirmed or debated giant planets. We present here the non-detection of these companions.

\subsubsection{HD 128311}

\begin{figure*}[t]
  \centering
\includegraphics[width=0.4\hsize,valign=m]{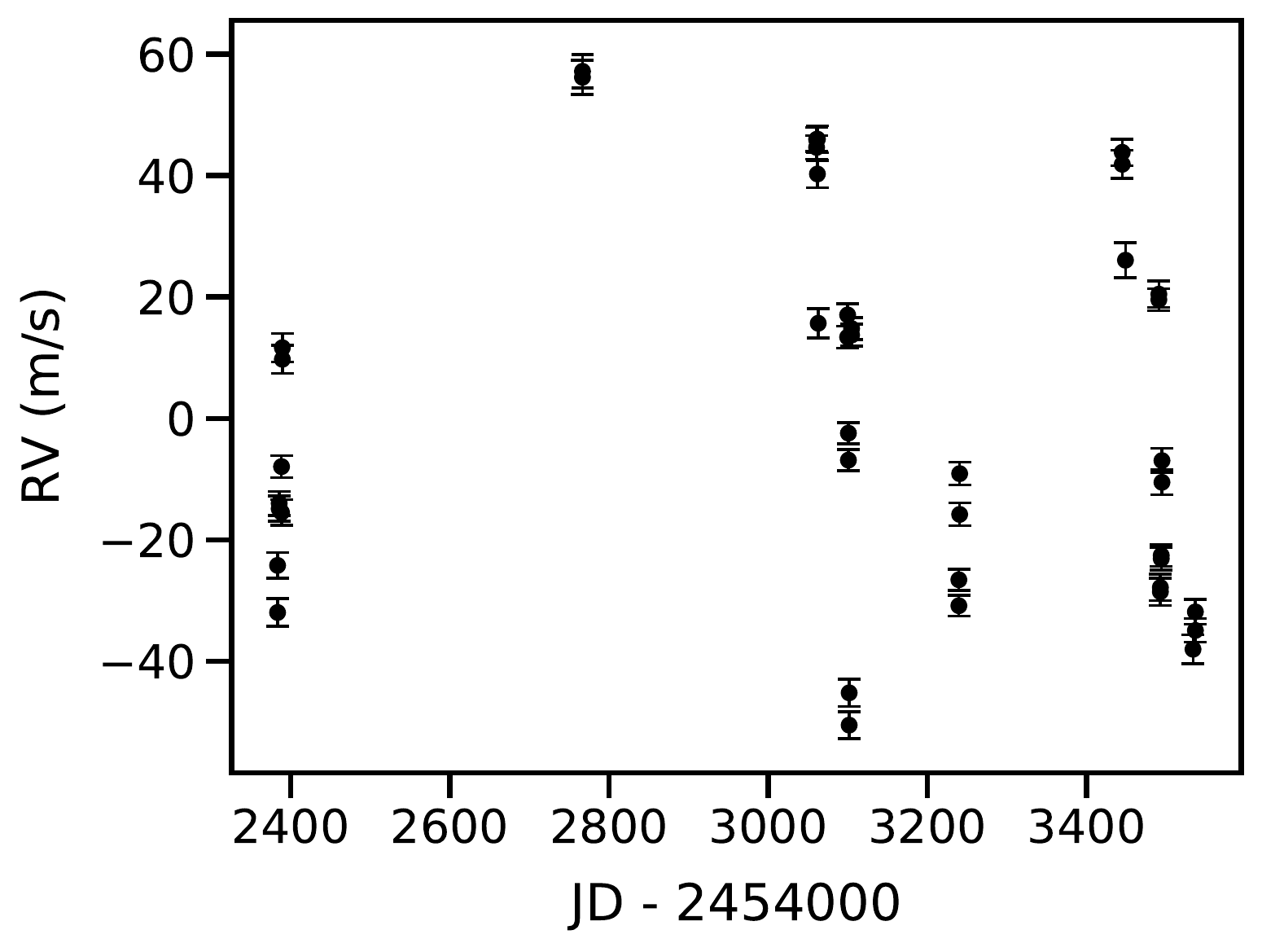}
\includegraphics[width=0.41\hsize,valign=m]{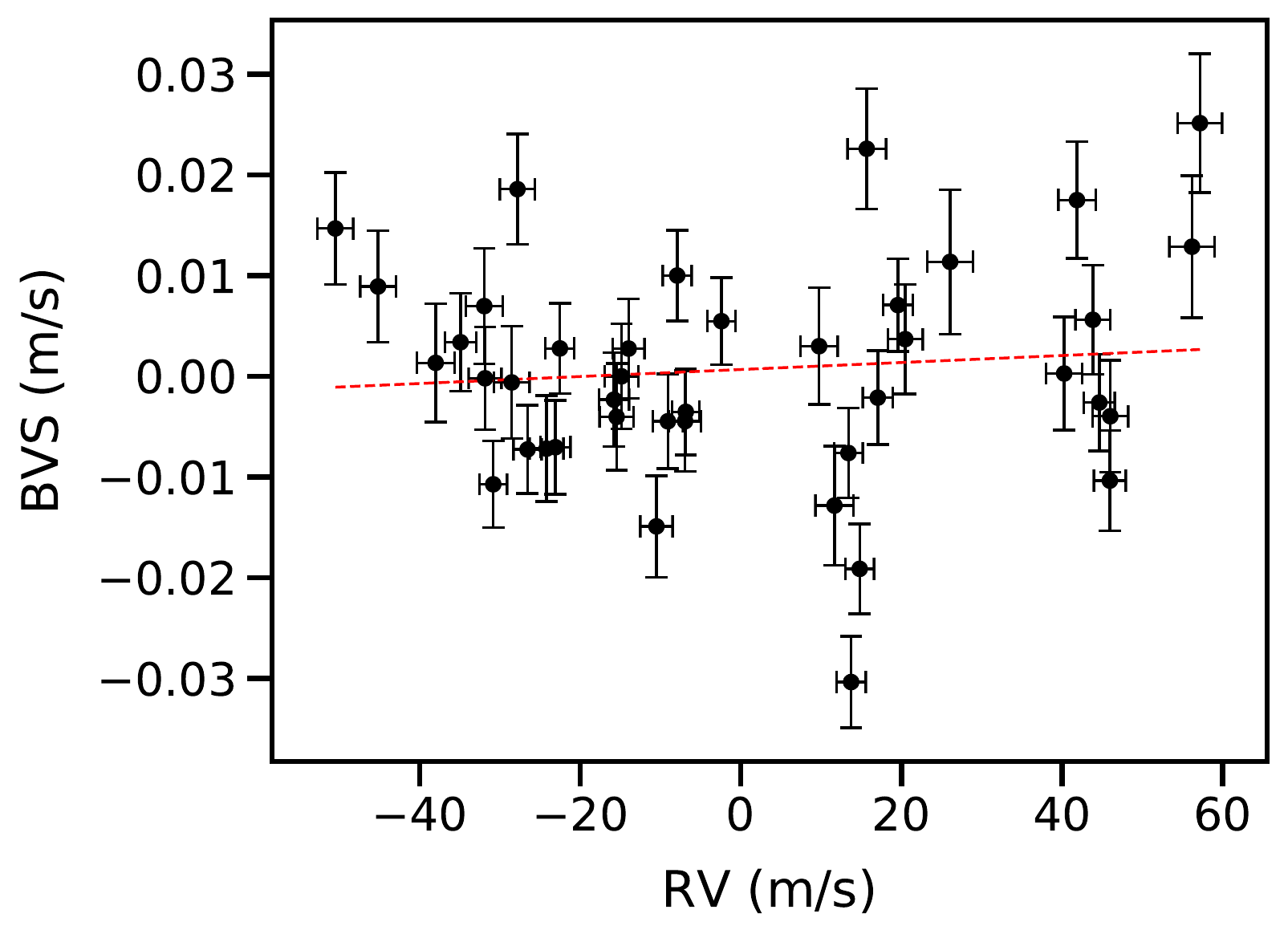}
\caption{HD 128311 RV variations (\emph{left}) and (BVS, RV) diagram (\emph{right}).} 
       \label{128311_a}
\end{figure*}

\begin{figure}[t!]
  \centering
\includegraphics[width=0.9\hsize]{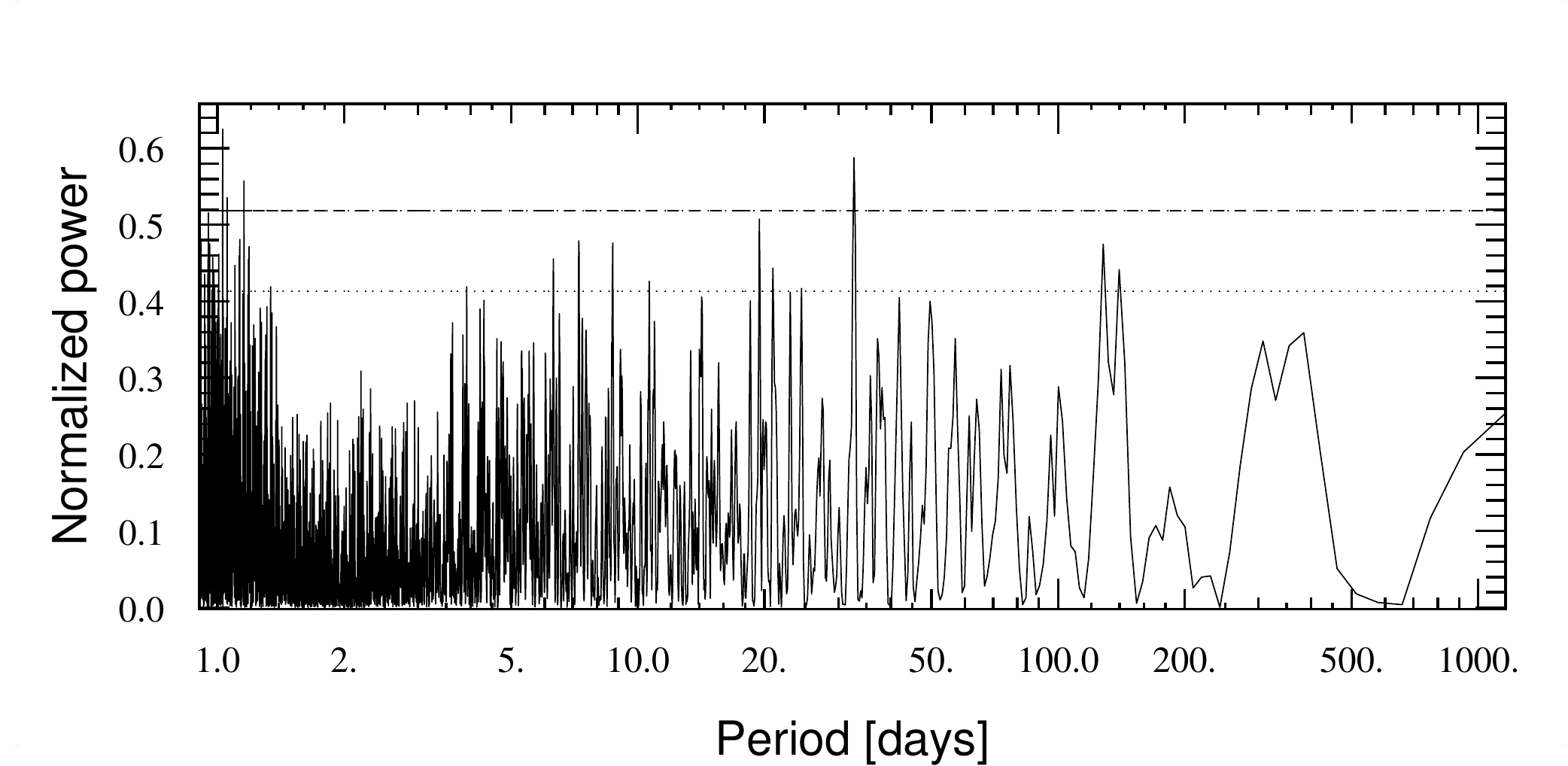}
\includegraphics[width=0.9\hsize]{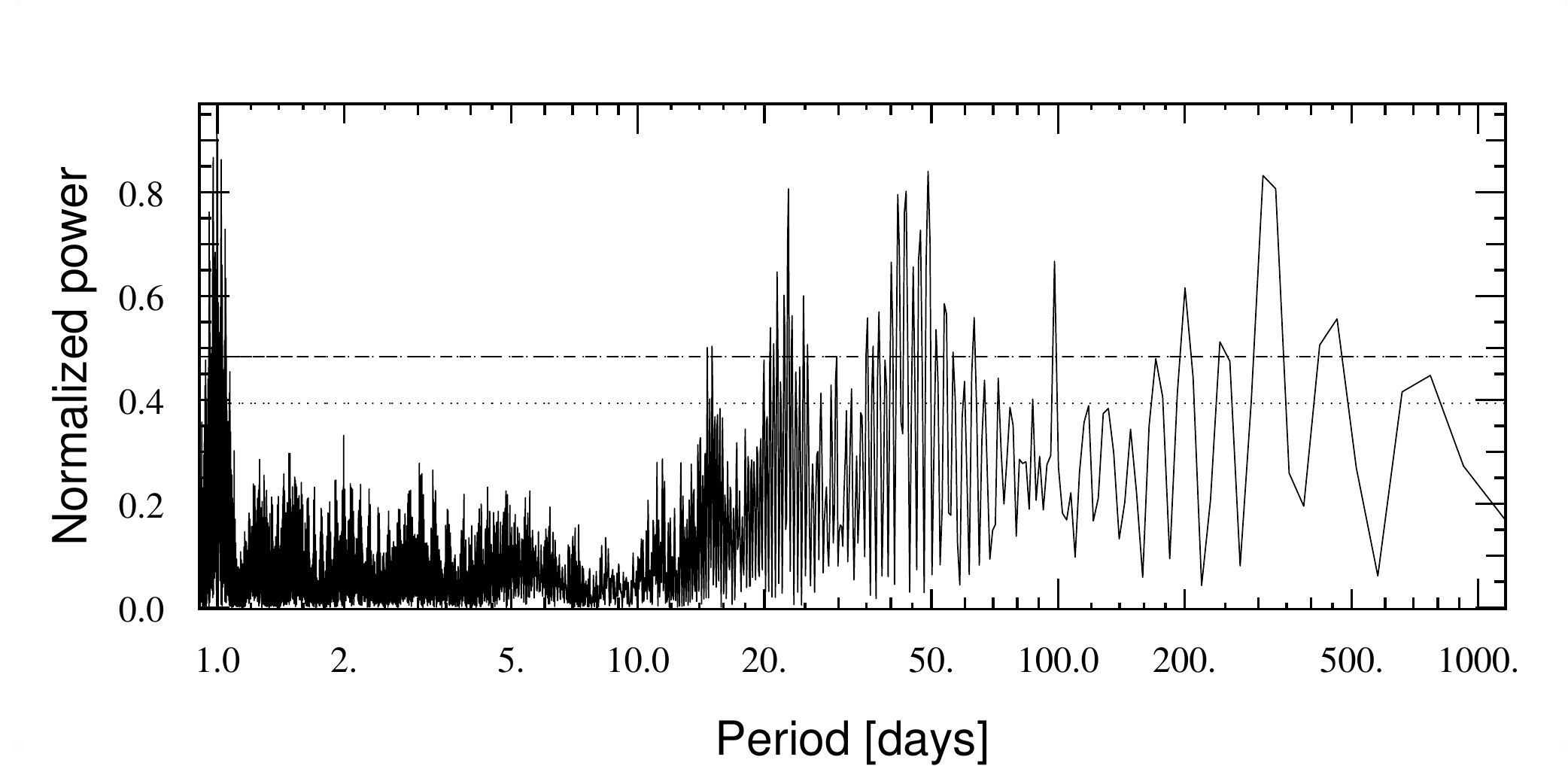}
\caption{HD 128311 periodograms \emph{Top}:  RVs periodogram. \emph{Bottom}: Time window periodogram. The $10 \%$ false alarm probability (FAP) is shown as a dotted line and the $1 \%$ FAP  as a dashed line. \label{128311_b}} 
\end{figure}

HD 128311 is a K0V-type star that is known to host two planets in 2:1 mean motion resonance. The first   was detected by \cite{Butler}, who estimated its period at $\SI{422}{\day}$ and its $M_p\sin{i}$ at $\SI{2.57}{\MJ}$. \cite{Vogt} confirmed this companion and discovered a second one. They estimated their periods to be $\SI{458.6}{\day}$ and  $\SI{928}{\day}$, and $M_p\sin{i}$ at $\SI{2.18}{\MJ}$ and  $\SI{3.21}{\MJ}$, respectively. They also found a periodicity in the photometry of the star with a period of $\SI{11.53}{\day}$.  \cite{Wittenmyer_2009} confirmed the two planets with the high-resolution spectrograph of the Hobby-Ebberly telescope, and they estimated their period at $454.2\pm1.6 \ \si{\day}$ and $923.8\pm5.3 \ \si{\day}$,  and their $M_p\sin{i}$ at $1.45\pm 0.13 \ \si{\MJ}$ and  $3.24\pm0.1 \ \si{\MJ}$,  respectively. They noted a strong periodicity in the residuals of their two-Keplerian fit, with a period  of $\SI{11.5}{\day}$, which corresponds to the rotation period of the star previously measured by \cite{Vogt}. Finally, \cite{McArthur_2014} combined Hubble Space Telescope astrometry and additional RVs from the Hobby-Ebberly telescope spectrograph to constrain the inclination of the system.
They find an inclination of $55.95 \pm \ 14.55 \si{\degree}$ and a true mass of $3.789^{+0.924}_{-0.432} \ \si{\MJ}$ for HD 128311 c.

Our data show parallel bisectors and a flat (BVS, RV) diagram ($Pearson=0.09$, $p_{value}=56 \%$, see \Cref{128311_a}), which indicates the presence of companion. However, we do not recover the two known companions in our data. The periodogram of our \sophie \ RVs show a strong signal at $\SI{30}{\day}$, a weak signal near $\SI{450}{\day}$, and a very weak signal near $\SI{900}{\day}$ (see \Cref{128311_b}). This  $30-day$ period could be a multiple of the rotation period seen in the photometry. We attribute  thus this period to magnetic activity.
The periodogram of the time window of our observations shows a strong signal near $\SI{450}{\day}$, which explains why we do not recover HD128311 b (see \Cref{128311_b}). Moreover, our time baseline is only    $\SI{1153}{\day}$ and our data are sparse ($41$ spectra), which explains why we do not recover HD 128311 c.

\subsubsection{BD+20 1790}

BD+20 1790 is a K5V star for which the presence of a giant planet was refuted. 
\cite{Obispo_10}  first reported a companion with a period of $\SI{7.78}{\day}$ from RV, along with a  $2.28-day$  period in the photometry of the star.
 The companion was then refuted by \cite{Figueira} as the star presented a strong BVS versus RV correlation, which indicated that the RV signal were dominated by the stellar activity (spots). \cite{Obispo_15} reanalyzed the RV data and found that the RV variations were composed of three signals: 
 the first   with a period of $\SI{2.8}{\day}$  that was linked to the photometric rotation period; the second   with a period of $\SI{4.36}{\day}$ that was linked to the synodic period of the star--planet system; and the third   with a period of $\SI{7.78 }{\day}$, which they attributed to the companion.
\cite{Gagn__2016} combined the \cite{Obispo_10} and \cite{Figueira} data with their CSHELL data, and did not identify any significant periodicity in them or in any combination of them.
Finally, \cite{Carleo} used multiband spectroscopy to show that the RV variations of BD+20 1790 are chromatic, ruling out the companion.

In our \sophie \ RVs we observe a $\sim \SI{1}{\kilo\meter\per\second}$ amplitude variation, which is consistent with the previous literature. The BVS are strongly correlated to the RVs, which is consistent with the  \cite{Figueira} analysis. However, we have too few  spectra  (ten) to permit the characterization of this stellar signal.

\begin{figure*}[t]
  \centering
\caption{Stars with trend RV analysis. \emph{First column},  \emph{top}: RV time variations (black) with the model of its linear regression (red line);  \emph{bottom}: Residuals of the linear regression.  \emph{Second column}: BVS vs RVs  (black) and its best linear model (red dashed line).  \emph{Third column}: BVS vs RV residuals  and its best linear model (red dashed line).} 
\includegraphics[width=1\hsize]{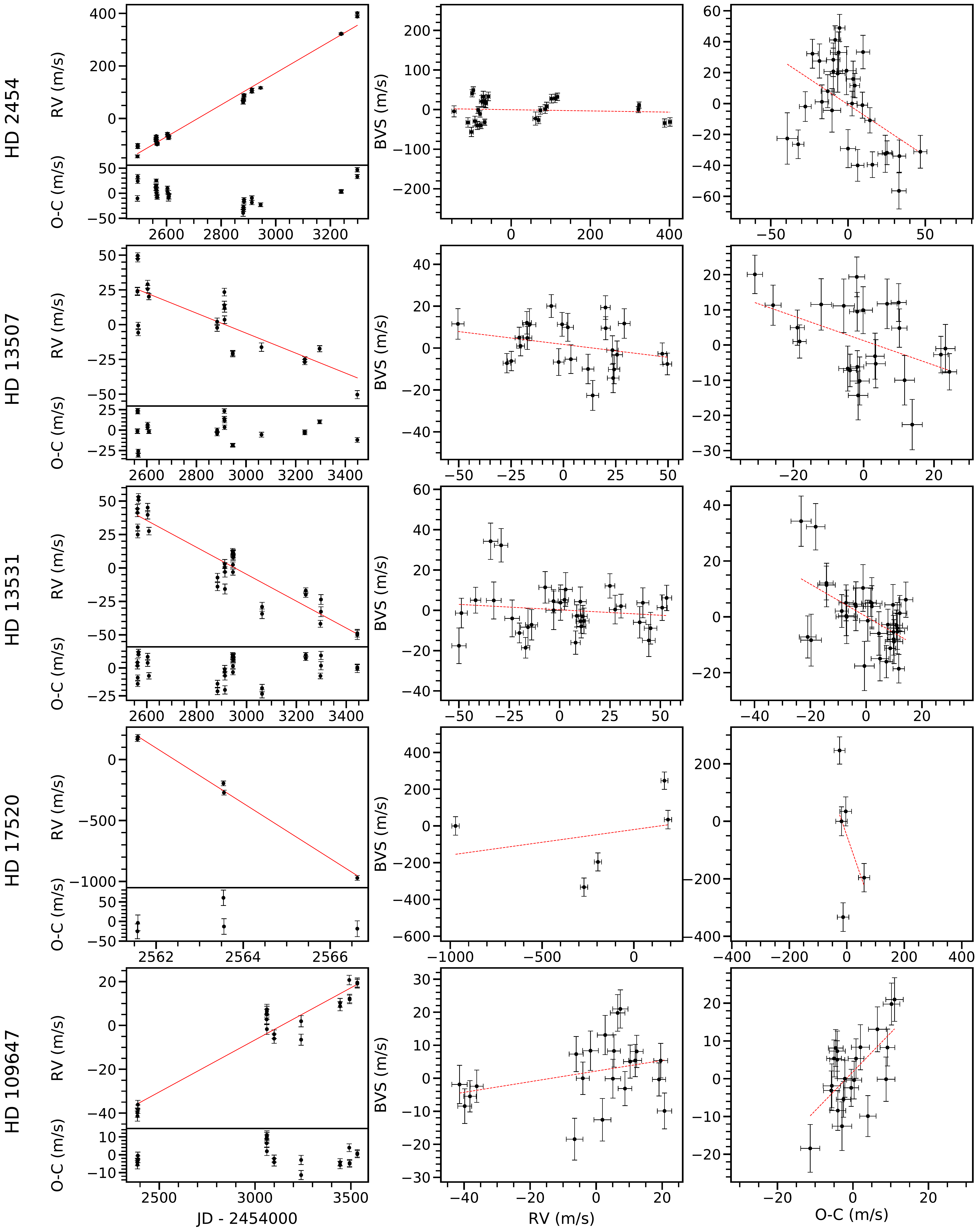}
       \label{trend}
\end{figure*}

\begin{figure*}[t]
 \ContinuedFloat
  \centering
\includegraphics[width=1\hsize]{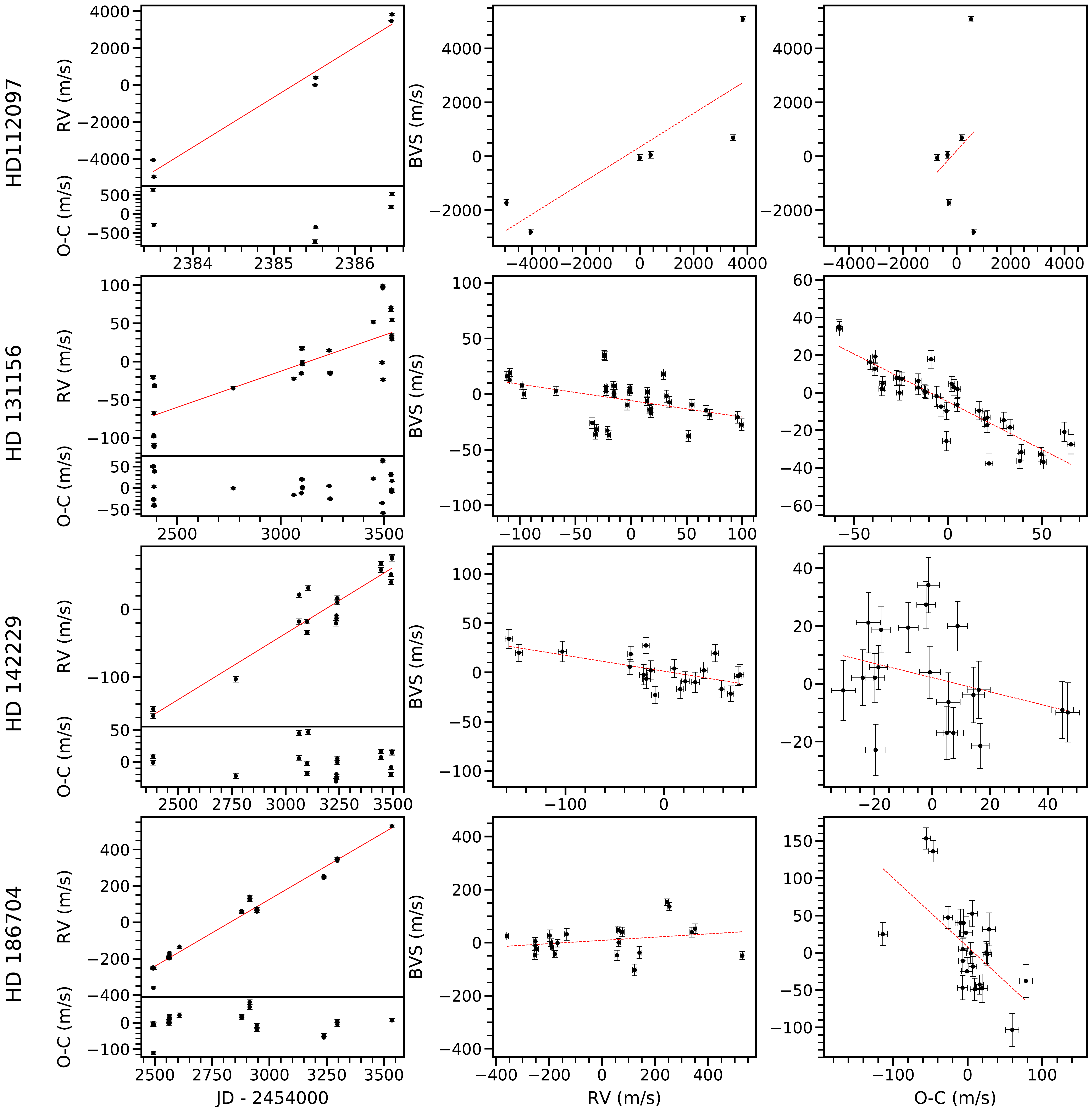}
\caption{Continued} 
       \label{trend}
\end{figure*}

\section{\sophie \ and \harps \ YNS combined survey analysis}
\label{HS}

In order to improve the statistics of GPs around young stars, we combine the \sophie \ YNS survey with the  \harps \ YNS survey (presented in \citealt{Grandjean_HARPS}). This combination is presented below.

\subsection{Sample}

The \harps \ survey consist of $89$ targets, $9$ of which are in common with the \sophie \ survey (see \cref{SOPHIE_sample}), leading to a total of 143 distinct targets in the combined survey.

In this combined survey, $19$ spectroscopic binary systems were highlighted  ($17$ SB1, $2$ SB2). Two planets with $P<\SI{1000}{\day}$ were discovered in the HD 113337 system \citep{Simon_VIII,Simon_X}. One long-period ($P>\SI{1000}{\day}$) sub-stellar candidate was discovered in the HD 206893 system \citep{Grandjean}. Finally, one star presents a trend compatible with a GP companion signal.

For the targets observed in both surveys we chose to use the instrument for which we have the lerger number of spectra and the longer time baseline (\sophie :  HD 89449, HD 171488, HD 186704 A , HD 206860. \harps : HD 25457, HD 26923, HD 41593, HD 90905).
For HD218396 the \sophie \ and \harps \ time baselines are similar. In this case we favored the number of spectra over the time baseline as it presents a better sampling of the jitter. We thus used the \sophie \  data for HD 218396 as it presents twice the number of  spectra even though the time baseline is $30\%$ shorter than for  \harps \ data.

We do not combine our HARPS and \sophie \ RV data for these targets, as it is not possible to determine an accurate value of the offset between the two datasets (as each dataset  is relative to its respective median spectra).
Moreover, our detection limits are based on RVs periodograms (see \cref{detlim}) and the uncertainty on the offset would lead to a biased combined periodogram, and thus to biased detection limits.

From the \harps \ YNS sample we excluded the targets that were excluded in Section 4.3 of \cite{Grandjean_HARPS}.
From the \sophie \  YNS survey we excluded the star excluded in \cref{SOPHIE_sample}. In addition, we excluded the binary stars for which the companion signal could not be fitted: HD 28495, HD 105963, and HD 218738.
This leads to a total of $120$  targets in the combined survey. All the following figures will present the  \harps \ targets in black and the \sophie \  targets in blue.

\begin{figure*}[ht!]
  \centering
\begin{subfigure}[t]{0.32\textwidth}
\includegraphics[width=1\hsize]{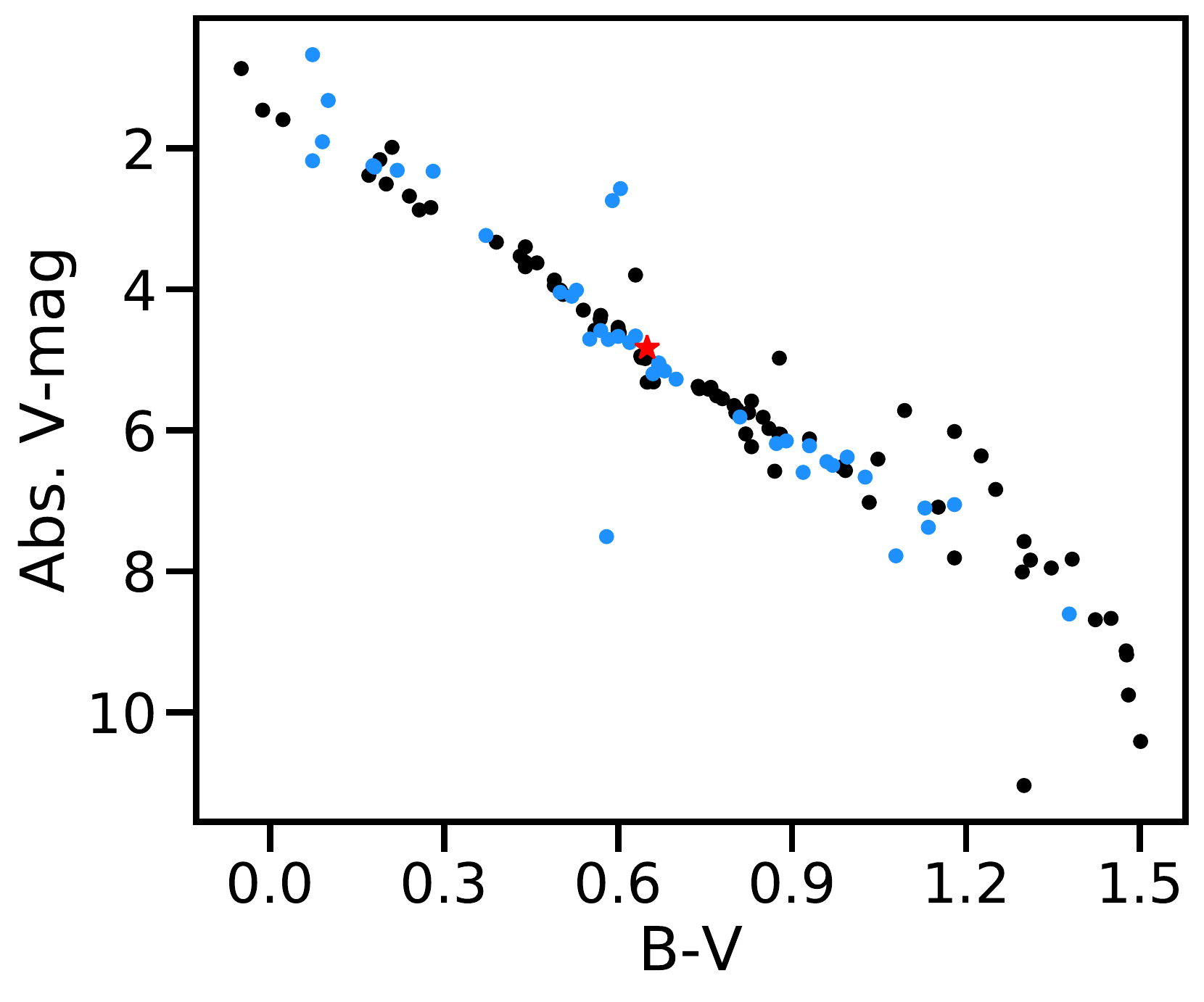}
\caption{\label{HS_HR}}
\end{subfigure}
\begin{subfigure}[t]{0.32\textwidth}
\includegraphics[width=1\hsize]{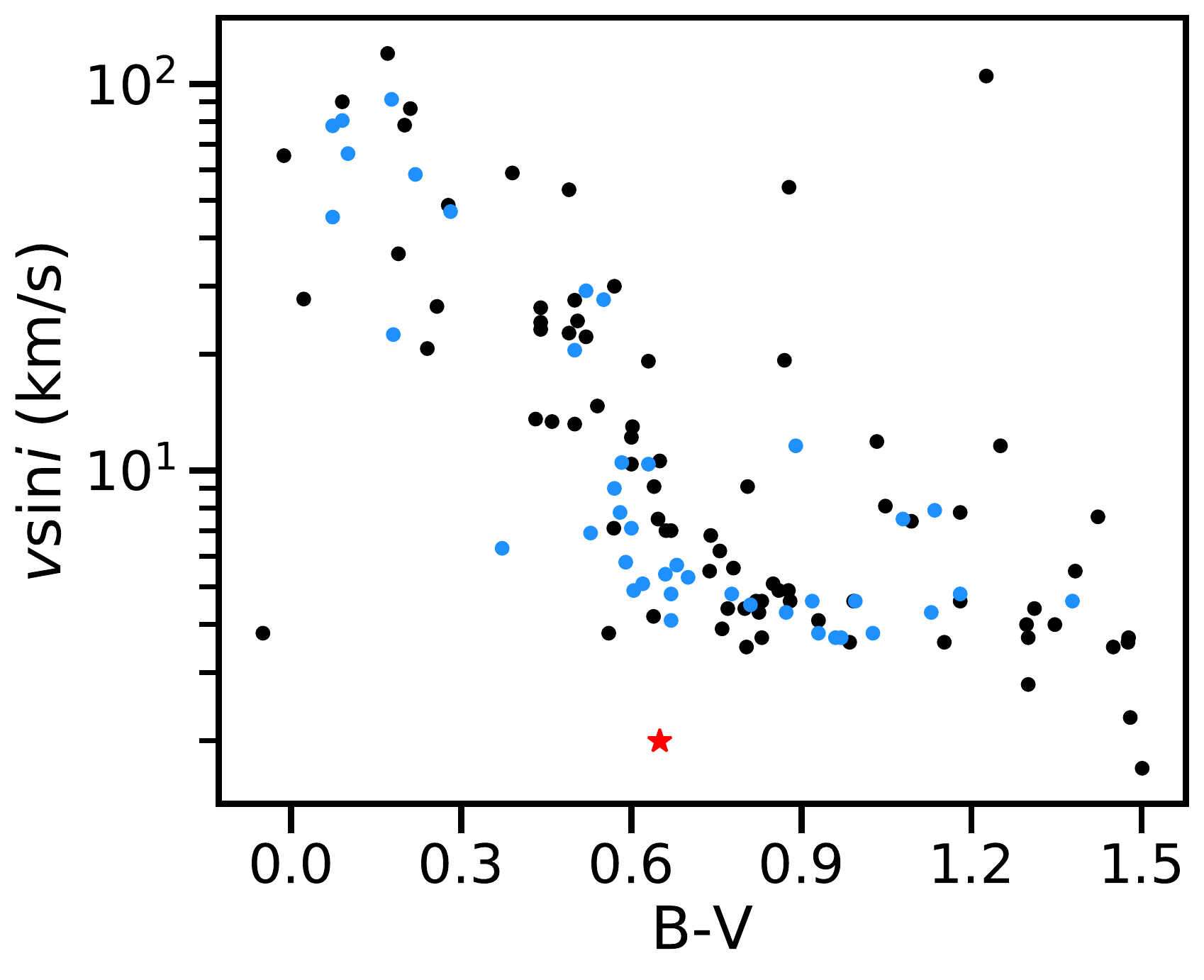}
\caption{\label{HS_vsini}}
\end{subfigure}
\begin{subfigure}[t]{0.32\textwidth}
\includegraphics[width=1\hsize]{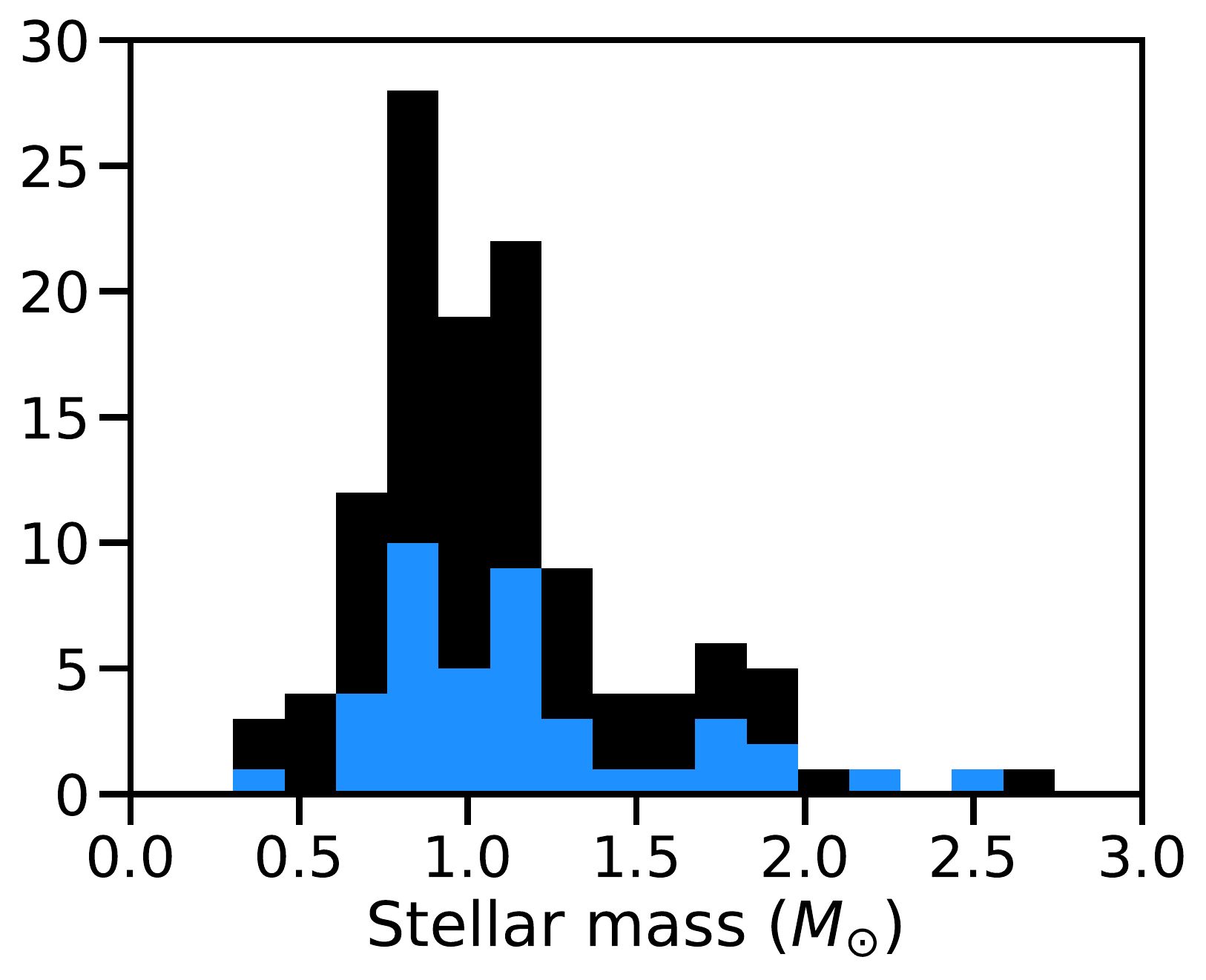}
\caption{\label{HS_mass}}
\end{subfigure}
\caption{Main physical properties of our combined  \harps \ and \sophie \  YNS sample. \harps \ targets related data  are  in black and \sophie \   related data are  in blue. The  \harps \ (black) histogram and the \sophie \ (blue) histogram are stacked.
 \subref{HS_HR})  Absolute $V$-magnitude vs \bv. Each dot corresponds to one target.
 The Sun is displayed (red star) for comparison. 
\subref{HS_vsini}) \vsini~vs \bv~distribution.
\subref{HS_mass})  Histogram of the star masses (in \Msun).}
       \label{survey_carac_1_hs}
\end{figure*}

The  targets of our final sample have spectral types that range from A0V to M5V (\Cref{survey_carac_1_hs}).  This sample includes $32$ targets between A0 and F5V ($B-V \in [-0.05:0.52[$, hereafter AF sub-sample), $79$ between F6 and K5   ($B-V \in [-0.05:1.33[$, hereafter FK sub-sample), and $9$ between K6 and M5 ($B-V \geq 1.33$,  hereafter M sub-sample).
Their projected rotational velocity  (\vsini) ranges from $1.7$ to $\SI{120}{\kilo\meter\per\second}$, with a median of  $\SI{7.1}{\kilo\meter\per\second}$.
Their V-band relative magnitude ranges between  $1.2$ and $10.1$, with a median of $7.6$.  
Their masses are between $0.42$ and $\SI{2.74}{\msun}$, with a median of  $\SI{1.0}{\msun}$ (see \Cref{age_masse} for masses determination). 
The AF sub-sample presents a median mass of $\SI{1.62}{\msun}$ with a standard deviation of  $\SI{0.38}{\msun}$, the FK  sub-sample presents a median mass of $\SI{0.93}{\msun}$ with a standard deviation of  $\SI{0.19}{\msun}$, and  the M sub-sample presents a median mass of $\SI{0.6}{\msun}$ with a standard deviation of  $\SI{0.08}{\msun}$.

\begin{figure*}[ht!]
  \centering
\includegraphics[width=0.32\hsize]{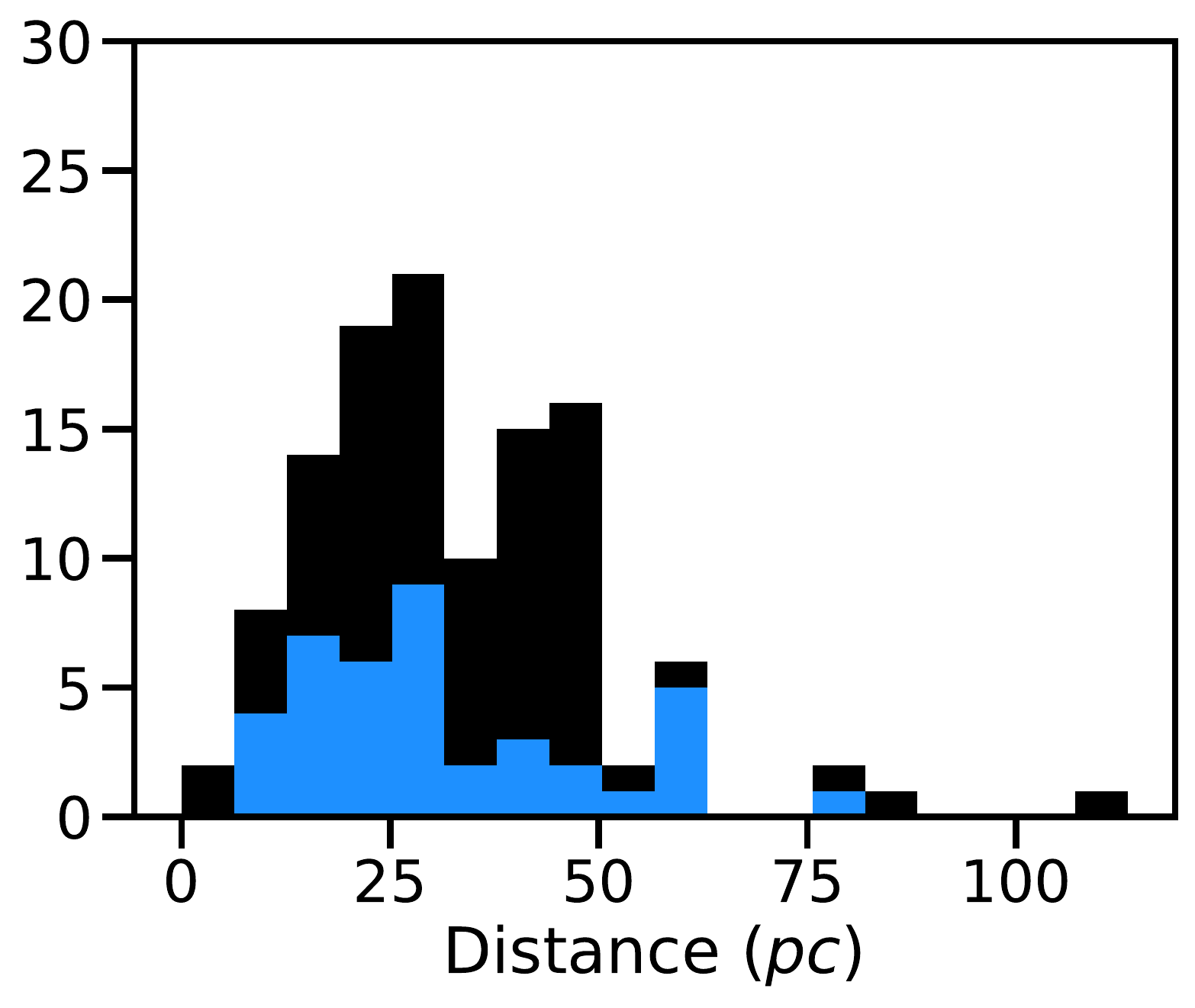}
\caption{Gaia DR2 \citep{DR2A1} distance histogram of our combined  \harps \ and \sophie \  YNS sample. The  \harps \ (black) histogram and the \sophie \ (blue) histogram are stacked. }
       \label{survey_dist_hs}
\end{figure*}

The distances of the stars in our sample range between $3$ and $\SI{113}{\parsec}$, with a median of $\SI{28}{\parsec}$ (see \cref{survey_dist_hs}, \cite{DR2A1}). 

\begin{figure*}[ht!]
  \centering
\begin{subfigure}[t]{0.32\textwidth}
\includegraphics[width=1\hsize]{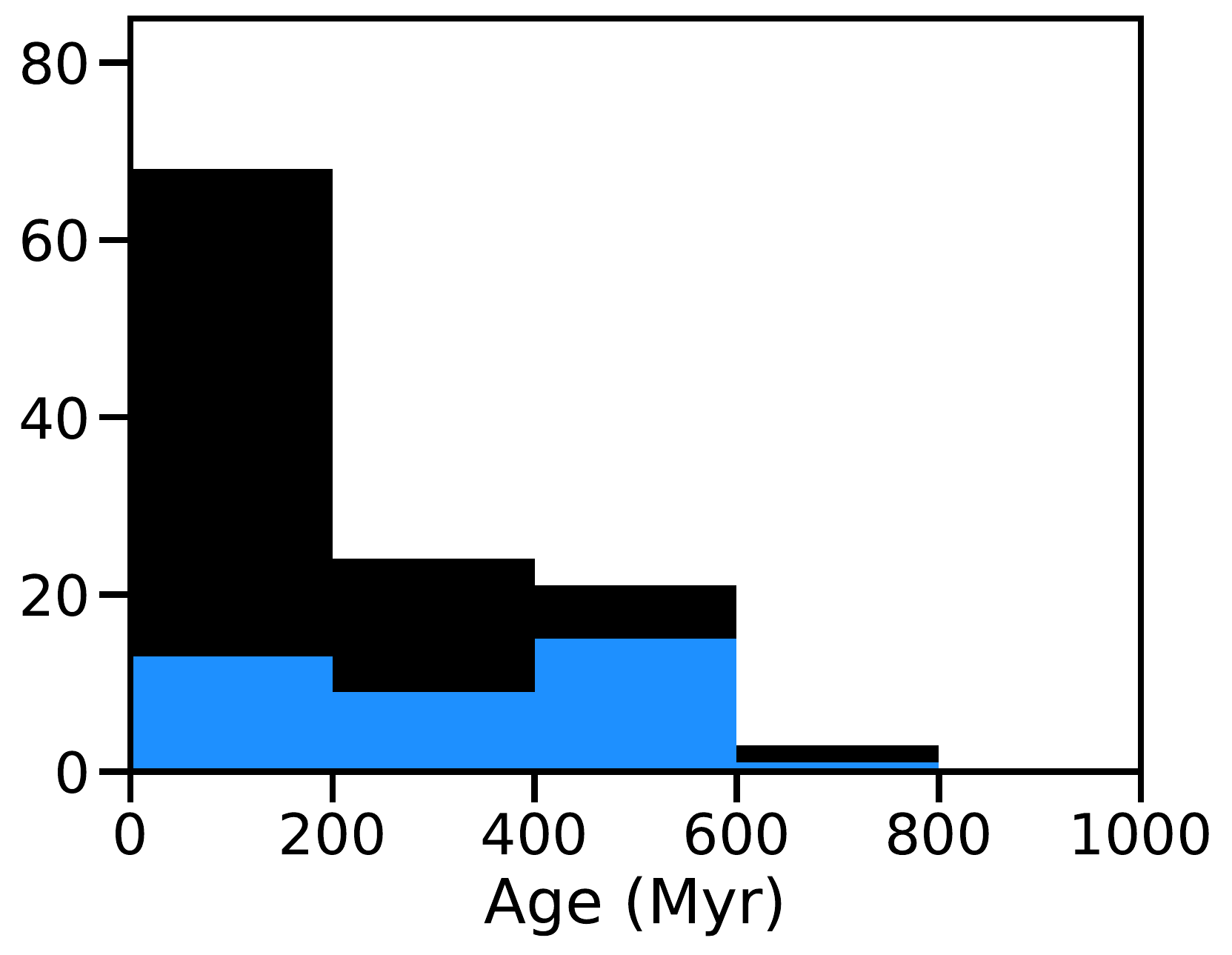}
\caption{\label{HS_age}}
\end{subfigure}
\begin{subfigure}[t]{0.32\textwidth}
\includegraphics[width=1\hsize]{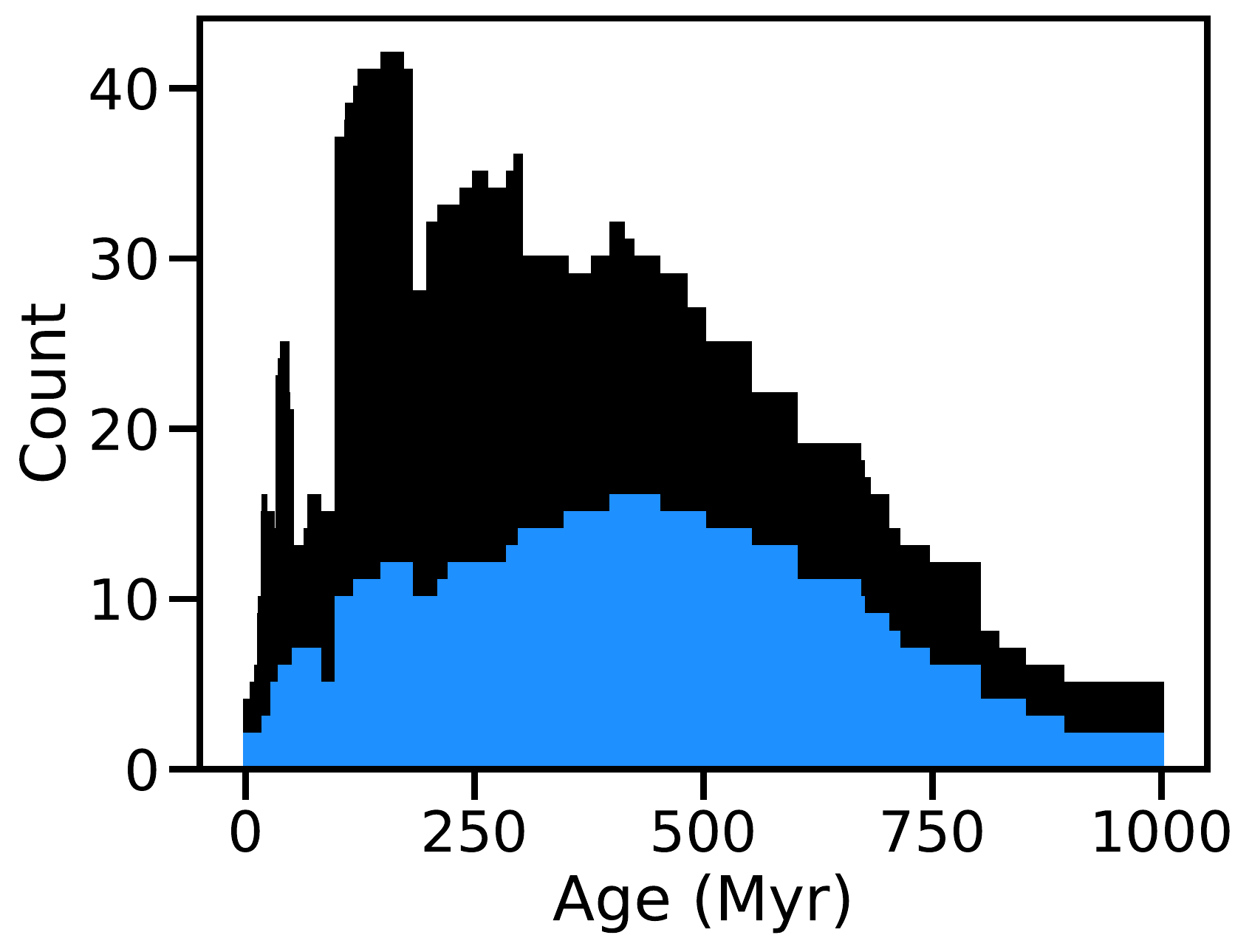}
\caption{\label{HS_age_distrib}}
\end{subfigure}
\caption{Age distribution of the combined \harps \ and \sophie \  YNS sample. 
\subref{HS_age}) Age histogram.
\subref{HS_age_distrib}) Histogram of the possible age. Each bin counts the number of stars for which the bin is within their age error bars.
The  \harps \ (black) histogram and the \sophie \  (blue) histogram are stacked.}
       \label{survey_age_hs}
\end{figure*}

The median age of the sample is  $\SI{149}{\mega\year}$ (see \Cref{age_masse} for age determination). The uncertainties on the ages range from several million  years to several hundred million years. We chose two ways to represent them.
First, we present a histogram of the ages in \Cref{HS_age}. We chose the  histogram bin to be larger than the median uncertainty on the age of the star in the survey. This ensures that the real ages of most of the stars in the survey are within the range where they were counted. 
Second, we considered an alternative way to present the age of the survey that we call the histogram of the possible age. It is a histogram where we count in each bin the number of stars for which the bin is within their errorbars.
 It represents the ranges of possible values for the ages of the stars. As the stars are counted several times, this histogram give only qualitative information.
We present the histogram of the possible age of the combined sample in \Cref{HS_age_distrib}. We can observe several peaks corresponding to different moving groups:  $\sim \SI{40}{\mega\year}$ (Tucana/Horlogium, Carina, Columba, Argus \citep{Bell}), $\sim 130-150 \ \si{\mega\year}$ (AB Doradus \citep{Bell}), and $ \SI{250}{\mega\year}$ (Hercules/Lyraes \citep{Eisenbeiss}).

\begin{figure*}[ht!]
  \centering
\begin{subfigure}[t]{0.32\textwidth}
\includegraphics[width=1\hsize]{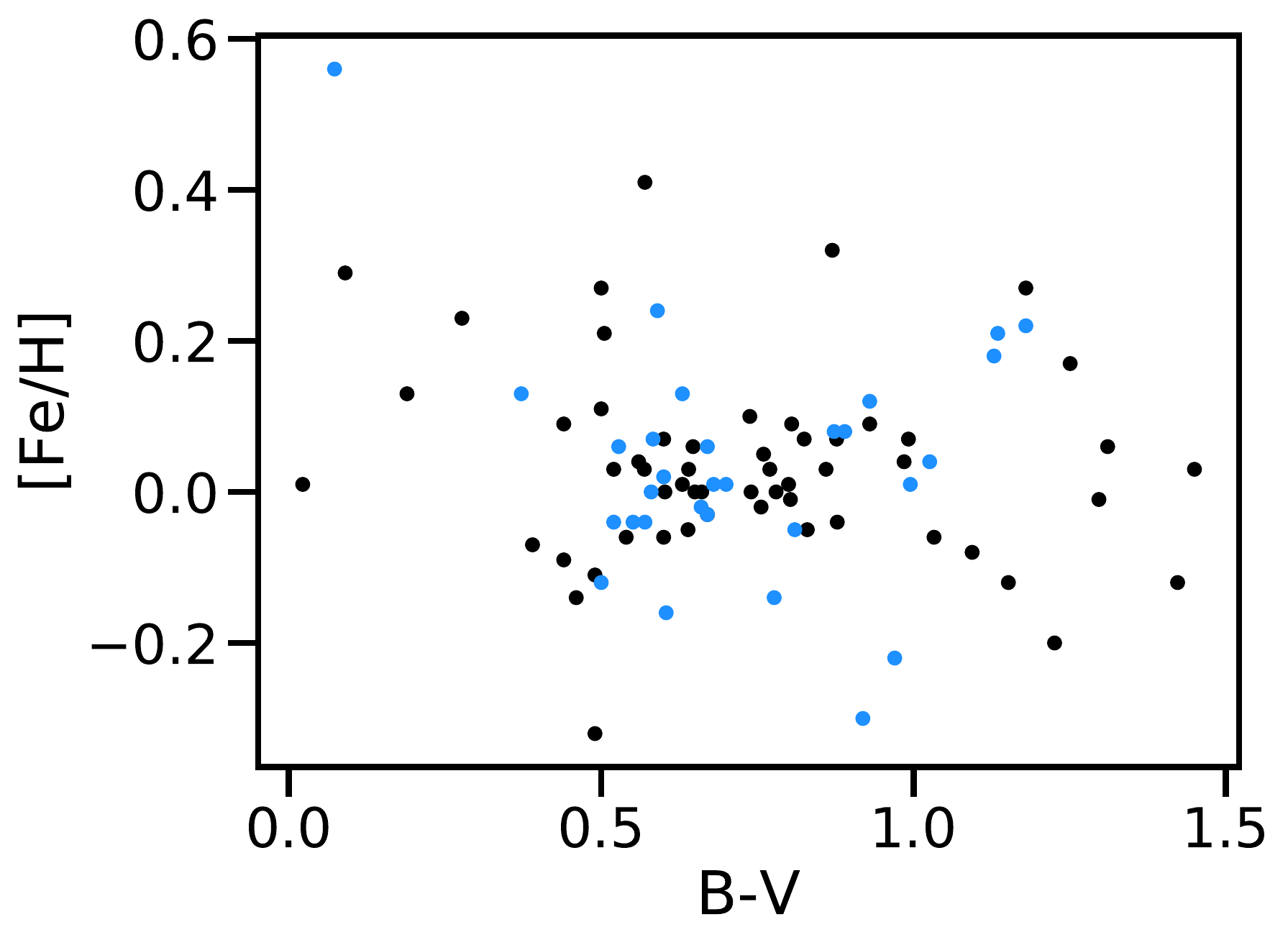}
\caption{\label{FEH_BV}}
\end{subfigure}
\begin{subfigure}[t]{0.32\textwidth}
\includegraphics[width=1\hsize]{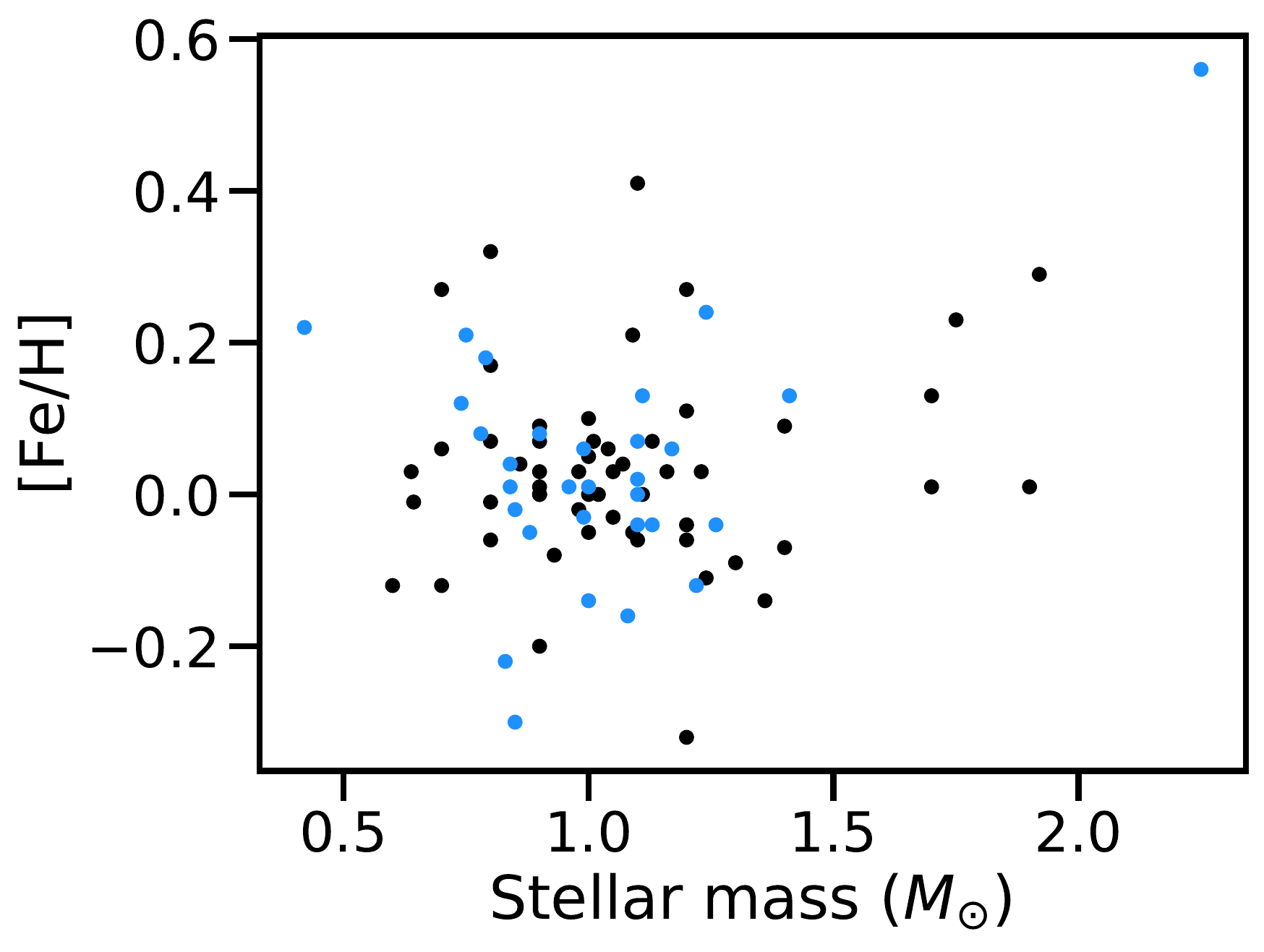}
\caption{\label{FEH_M}}
\end{subfigure}
\caption{Metallicity of the \harps \ and \sophie \  YNS sample against \bv \ (\subref{FEH_BV}) and against stellar mass (\subref{FEH_M}).
\harps \ targets related data  are  in black and \sophie \   related data are  in blue. }
       \label{FEH}
\end{figure*}

Metallicity measurements are only available for $86$ of our targets. 
Their metallicities  are close to the solar value, with a median of  $0.03 \ dex$ (mean of $0.03 \ dex$) and a standard deviation of $0.14 \ dex$.
We observe no statistically significant correlation between the metallicity and the \bv \ nor between the metallicity and the stellar mass in our sample. We present the metallicity of the combined sample in \Cref{FEH}.

\begin{figure*}[ht!]
  \centering
\begin{subfigure}[t]{0.32\textwidth}
\includegraphics[width=1\hsize]{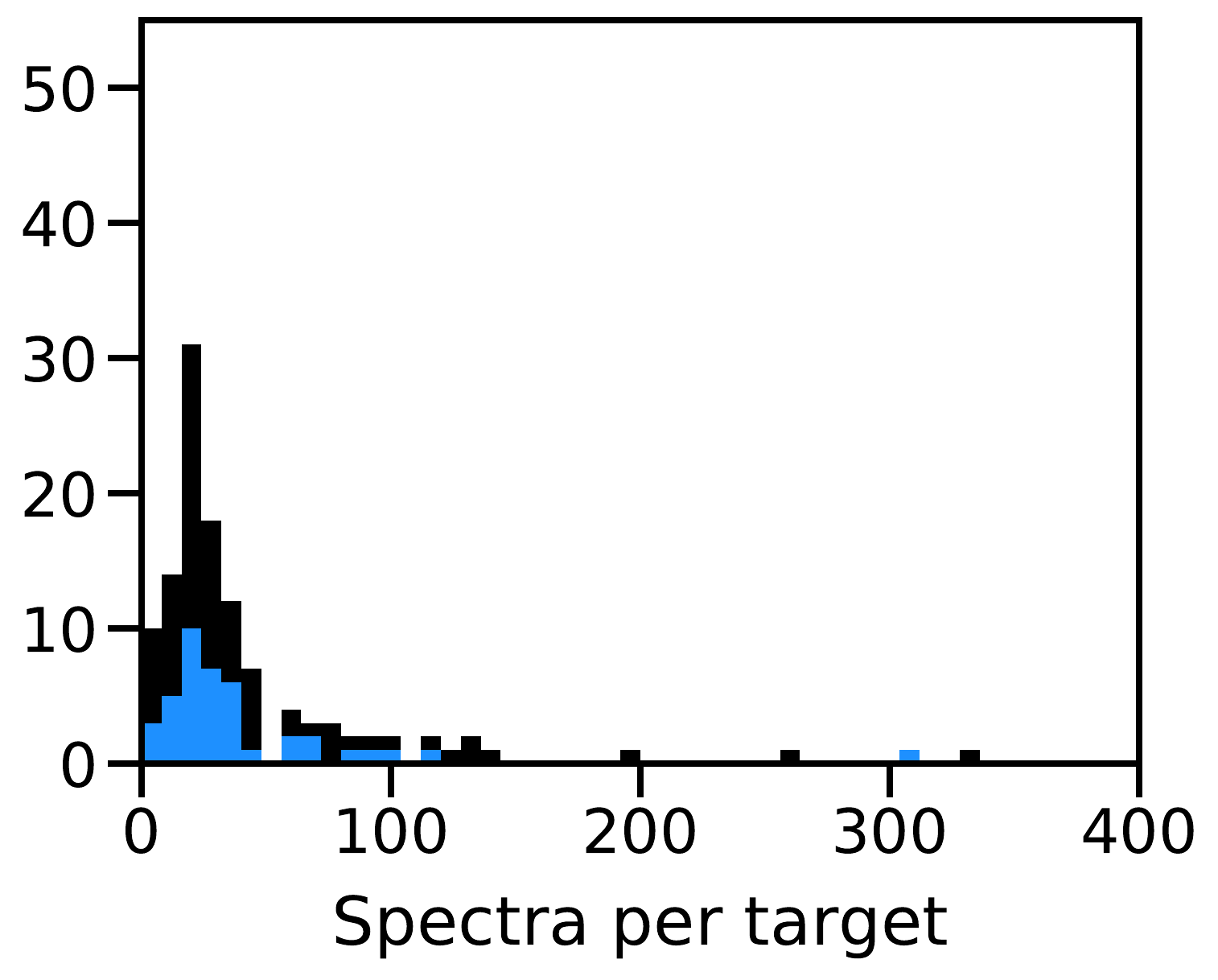}
\caption{\label{Nmes}}
\end{subfigure}
\begin{subfigure}[t]{0.32\textwidth}
\includegraphics[width=1\hsize]{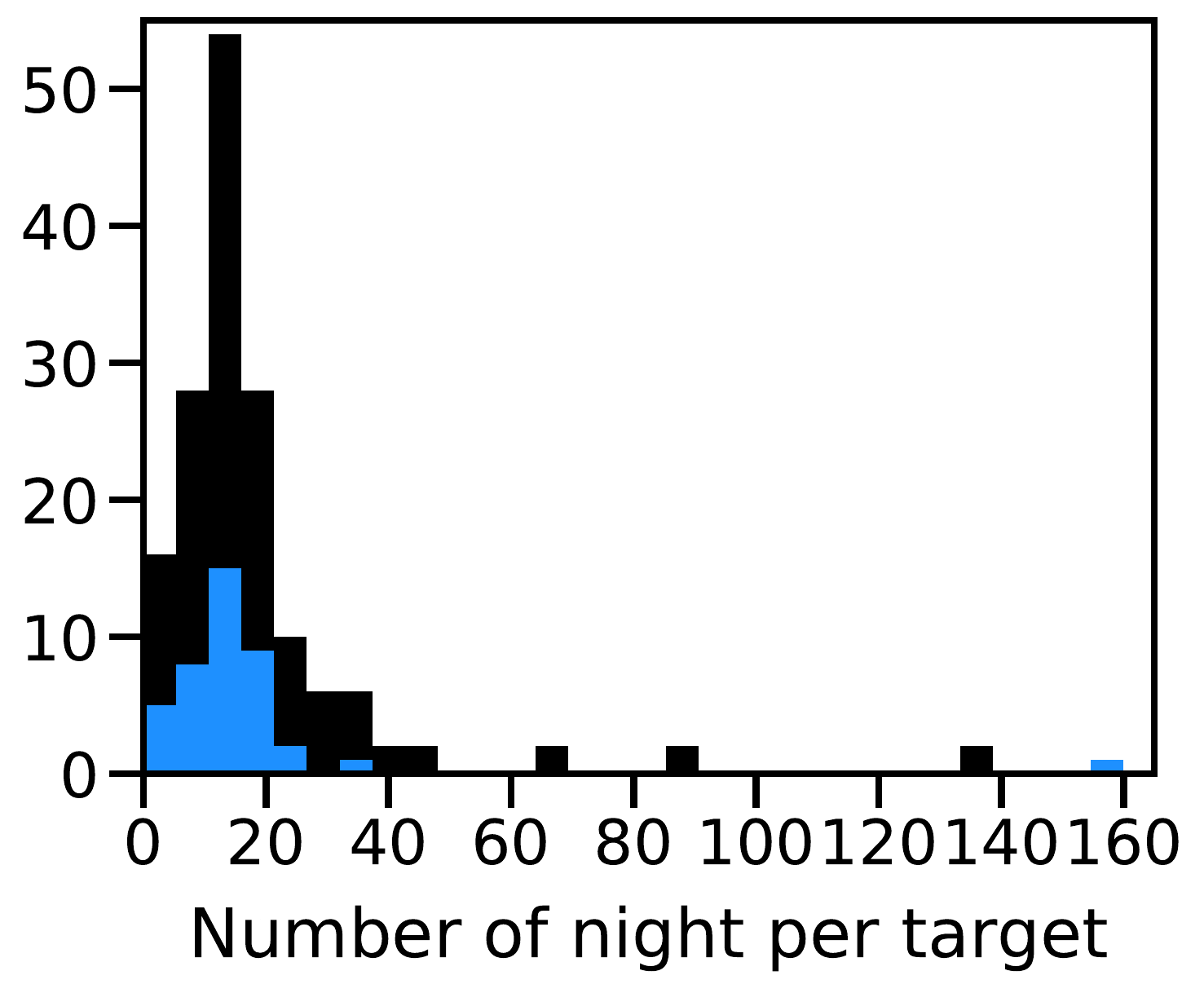}
\caption{\label{Nb_day}}
\end{subfigure}
\begin{subfigure}[t]{0.32\textwidth}
\includegraphics[width=1\hsize]{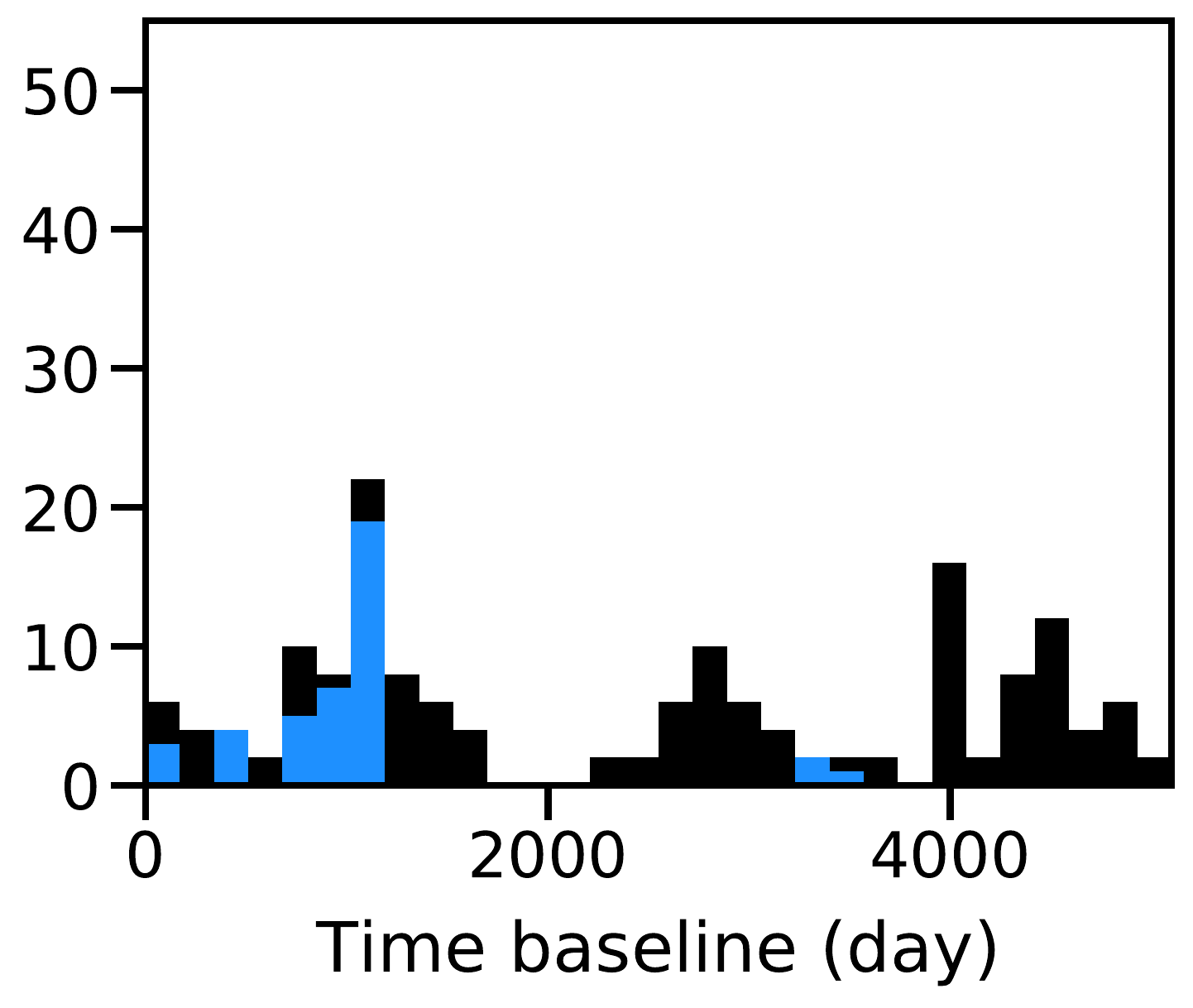}
\caption{\label{time_bsl}}
\end{subfigure}
\caption{Observation summary of the combined  \harps \ and \sophie \  YNS sample.
 \subref{Nmes}) Histogram of the number of spectra per target; HD 216956 (Fomalhaut, 834 spectra) and HD 039060 ($\beta$ Pic, 5108 spectra) are not displayed.
 \subref{Nb_day}) Histogram of the  number of nights per target. 
\subref{time_bsl}) Histogram of the time baselines.
The  \harps \ (black) histogram and the \sophie \  (blue) histogram are stacked.} 
       \label{survey_carac_2_hs}
\end{figure*}

The median time baseline is \SI{2621}{\day} (mean time baseline of  \SI{2493}{\day}), with a  median number of spectra per target of $25$ ($92$ on average) spaced over a median number of $\SI{13}{\night}$  ($18$ on average, \Cref{survey_carac_2_hs}).

Details can be found in  Tables \ref{tab_carac_s}, \ref{tab_carac_h}, \ref{tab_result_s}, and \ref{tab_result_h} (\cref{tab_carac_h} and  \ref{tab_result_h}  being updated versions of the Tables A.1 and A.2 presented in \citep{Grandjean_HARPS}.

\subsection{Stellar intrinsic variability}

The jitter observed in the RV time series of our combined sample is mainly caused by pulsations for early-type stars (from A to F5V), and by spots and faculae for late-type stars ($>F5V$). 
Those two regimes can be distinguished from each other as stars with pulsations show a vertically spread (BVS, RV) diagram, whereas stars with spots present a correlation between BVS and RV \citep{Lagrange_2009}.
The main origin of RV jitter is reported in Tables  \ref{tab_result_s} and  \ref{tab_result_h} for each target of the \harps \ YNS and \sophie \  YNS surveys, respectively. The distinction between stars with pulsations (labeled ``P'' in   these tables) and stars with spots   (labeled ``A'') was based on the stellar spectral type, the shape of the bisectors \citep{Lagrange_2009}, and the  (BVS, RV) diagram shape as exposed above.

\begin{figure*}[t!]
  \centering
\begin{subfigure}[m]{0.32\textwidth}
\includegraphics[width=1\hsize]{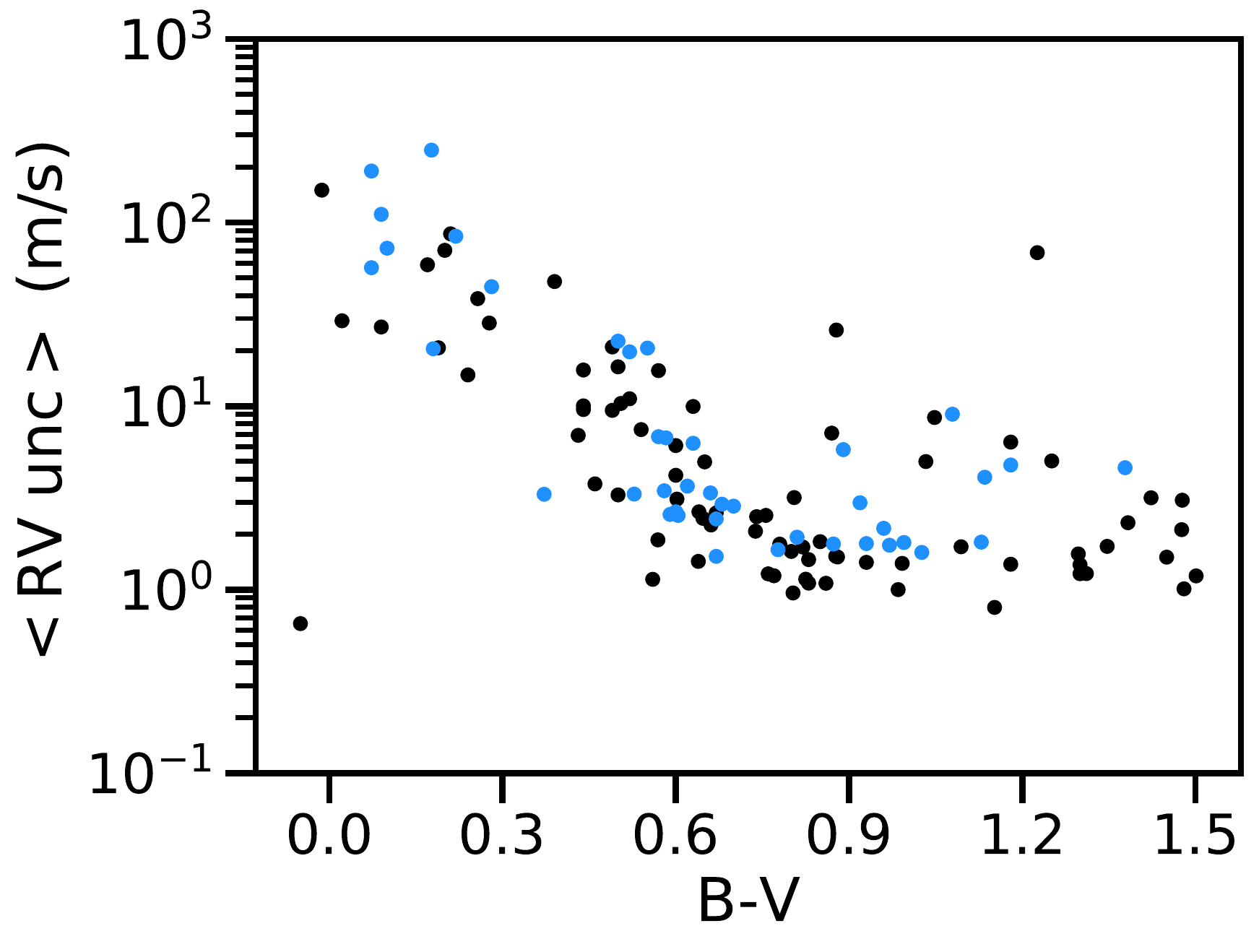}
\caption{\label{HS_unc_BV}}
\end{subfigure}
\begin{subfigure}[m]{0.32\textwidth}
\includegraphics[width=1\hsize]{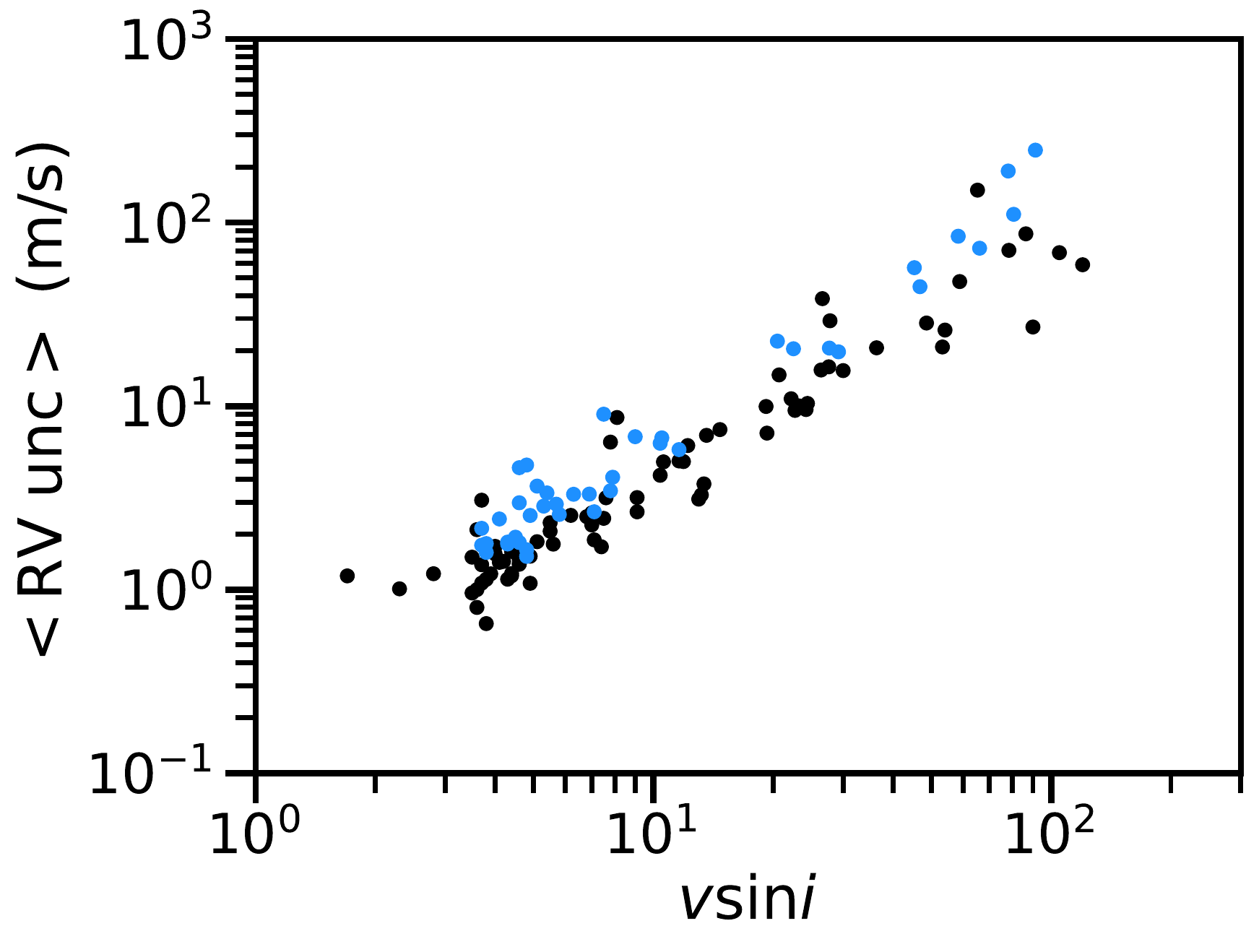}
\caption{\label{HS_unc_vsini}}
\end{subfigure}
\begin{subfigure}[m]{0.32\textwidth}
\includegraphics[width=1\hsize]{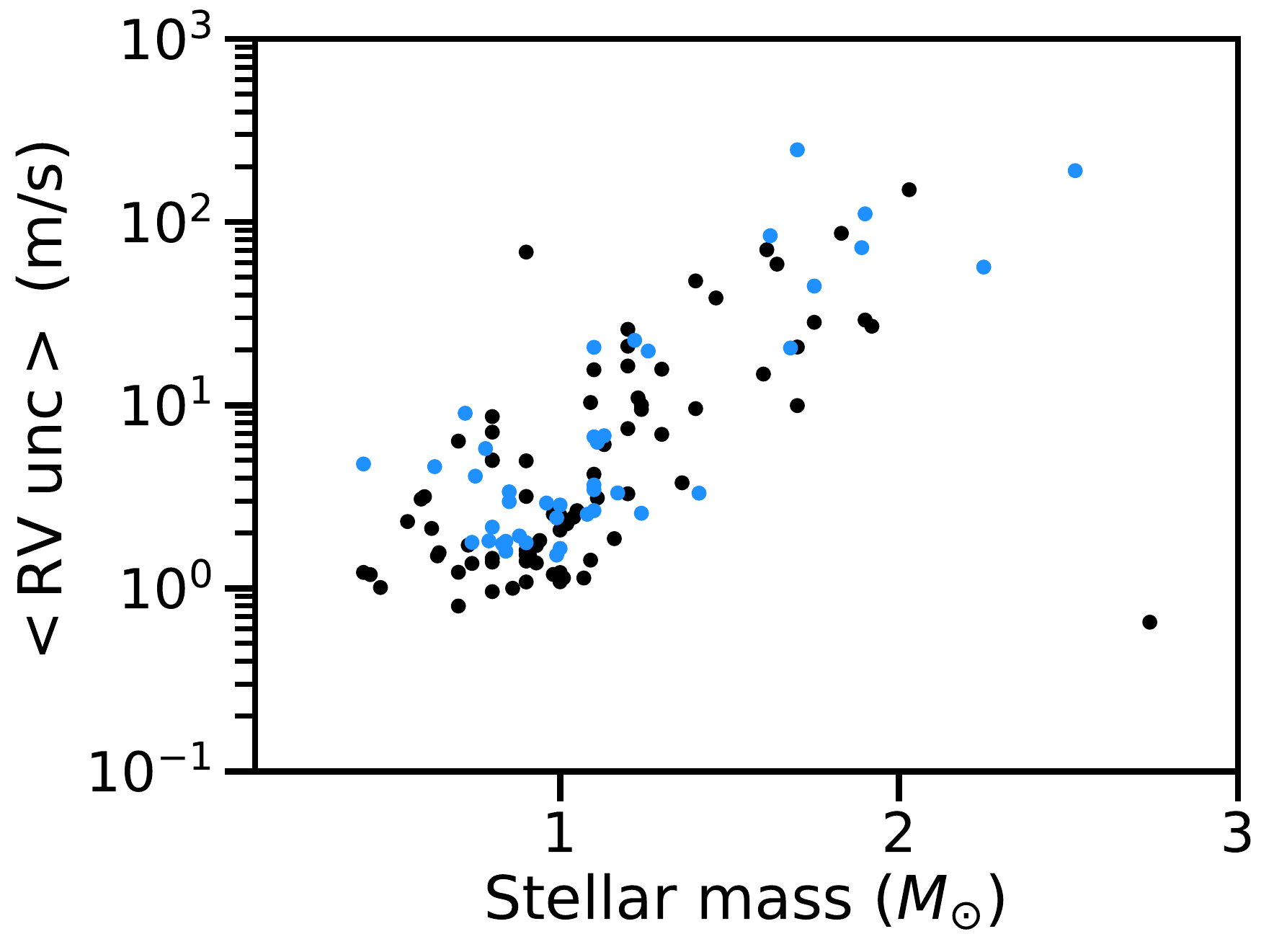}
\caption{\label{HS_unc_M}}
\end{subfigure}
\caption{Summary of the RV uncertainties of the combined survey.  Mean RV uncertainty (accounting for the photon noise only) vs \bv (\subref{HS_unc_BV}), vs \vsini \ (\subref{HS_unc_vsini}) and vs \Mstar \ (in \Msun, \subref{HS_unc_M}). \harps \ targets related data  are presented in black and \sophie \   related data are presented in blue. }
       \label{stell_var_hs_1}
\end{figure*}

\begin{figure*}[t!]
  \centering
\begin{subfigure}[m]{0.32\textwidth}
\includegraphics[width=1\hsize]{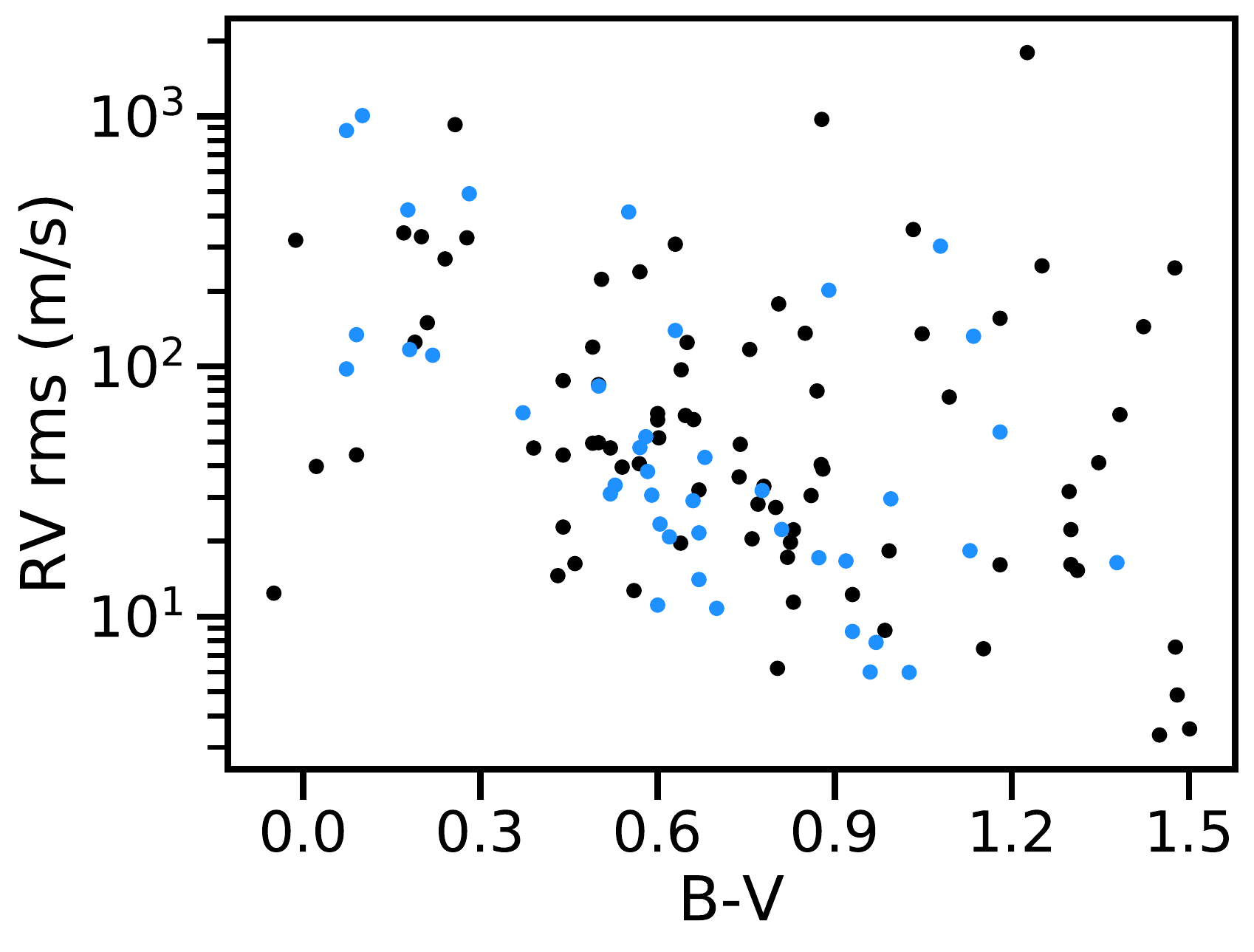}
\caption{\label{HS_rms_BV}}
\end{subfigure}
\begin{subfigure}[m]{0.32\textwidth}
\includegraphics[width=1\hsize]{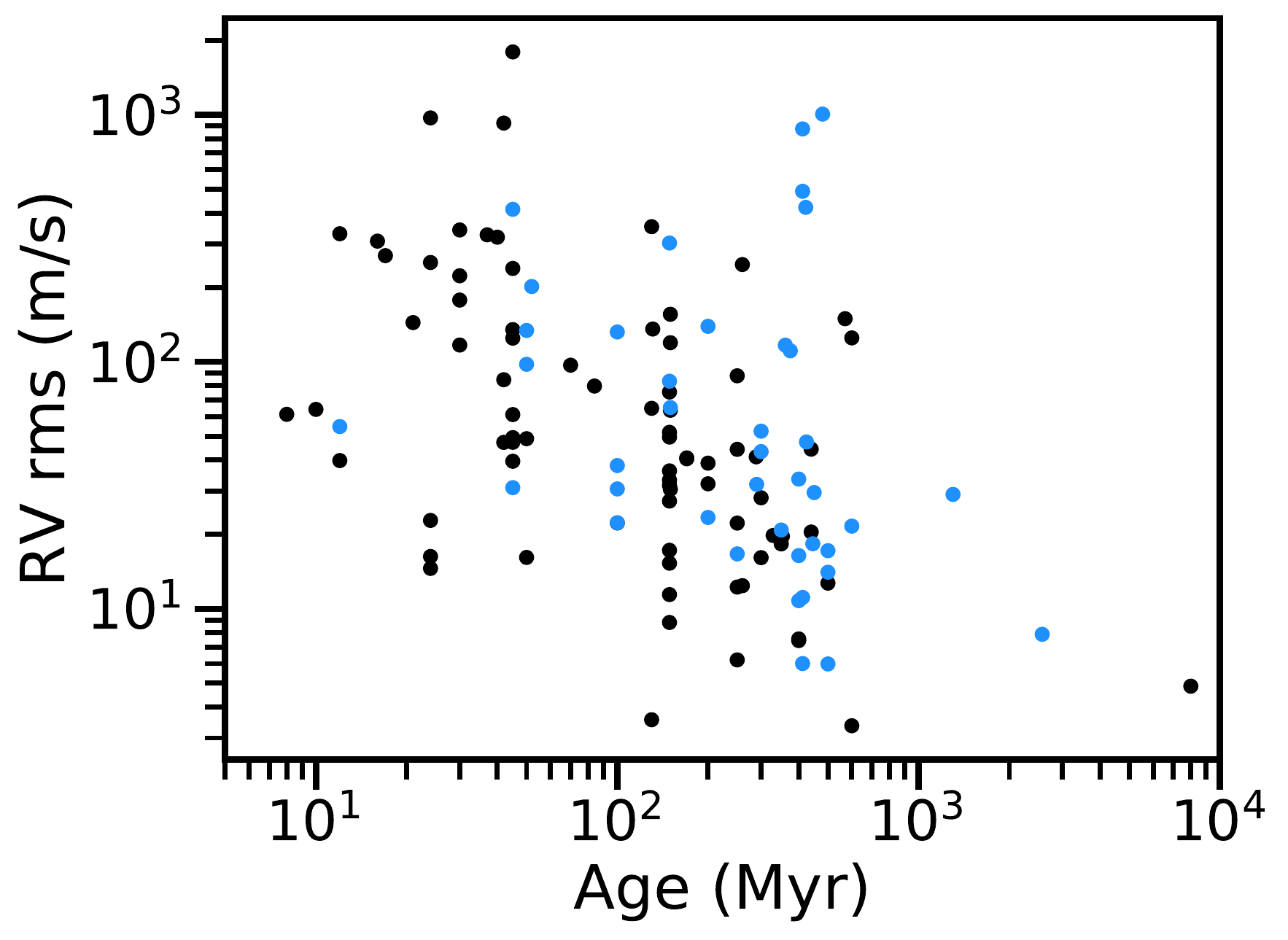}
\caption{\label{HS_rms_A}}
\end{subfigure}
\caption{Summary of the RV rms of the combined survey. RV rms vs \bv \ (\subref{HS_rms_BV}), and vs age  (\subref{HS_rms_A}). \harps \ targets related data  are presented in black and \sophie \   related data are presented in blue. }
       \label{stell_var_hs_2}
\end{figure*}

The stars of the combined survey present a strong jitter. After the removal of the companion's signal (see \Cref{comp}) and  the HD 217987 secular drift (see \cite{Grandjean_HARPS}), the ratio of the RV rms to the mean RV uncertainty is between $120$ and $1$, with a median at $12$.
The median of the RV rms is \SI{44}{\meter\per\second} (\SI{129}{\meter\per\second} on average).
We display in \Cref{stell_var_hs_1}  the mean RV uncertainty versus   \bv, versus   \vsini,  and versus $M_*$, of the combined sample.
We also display the RV rms versus \bv \ and versus age in \Cref{stell_var_hs_2}.
The mean RV uncertainty is strongly correlated with   \vsini \ ($Pearson=0.78$, $p_{value}< \num{6e-26}\%$). The correlation of the combined survey is stronger than for the  \harps \ YNS survey alone ($Pearson=0.69$).

\begin{figure}[h!]
  \centering

\includegraphics[width=0.75\hsize]{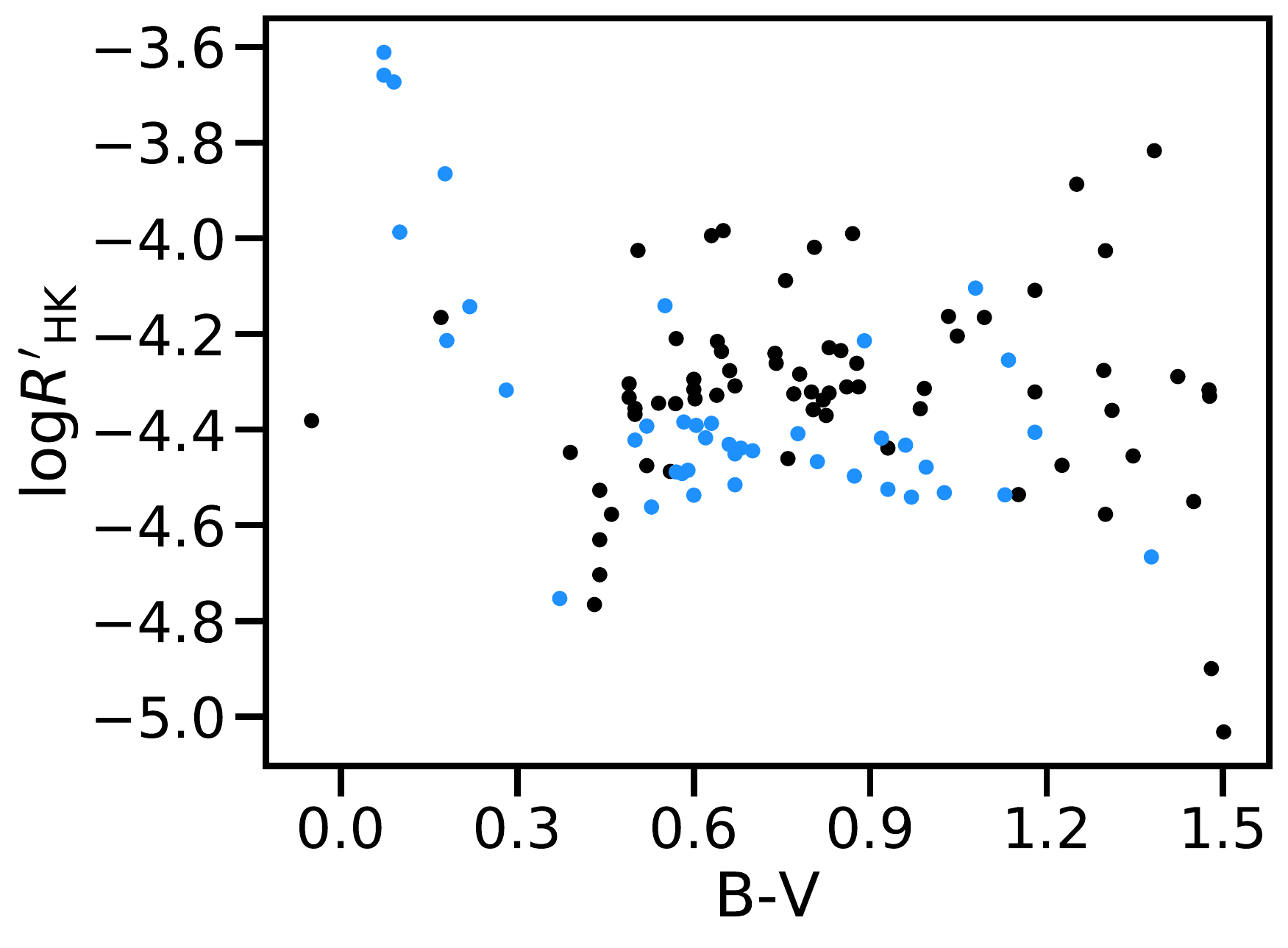}

\caption{Combined survey $<$\rhk$>$ \ vs  \bv. The data from \harps \ targets   are shown in black and from \sophie \ targets    in blue. } 
       \label{stell_rhk_hs}
\end{figure}

Out of $120$ stars in our sample, $110$  present variations in their Ca lines, which confirms the presence of stellar activity for a large number of targets. The median $<$\rhk$>$ \ of our sample is $-4.3$ with a standard deviation of $0.2$. 
Of these targets, $4$ present signs of low activity ($<$\rhk$>$ $<-4.75$), $85$ are active ($-4.75<$ $<$\rhk$>$ $< -4.2$), and $21$ present signs of high activity ($<$\rhk$>$ $>-4.2$). The median of the standard deviations of the $<$\rhk$>$ of the stars of the sample is  $0.03$ (mean of $0.04$).
We present in \Cref{stell_rhk_hs} $<$\rhk$>$ \ versus \bv.

\subsection{RV correction for further analysis}

We used the method presented in \cite{Grandjean_HARPS} to correct the RVs of the \sophie \  stars from their jitter and from their companion signal: for  the stars for which the RVs are dominated by spots (labeled A in  \cref{tab_result_s}), we corrected the RVs from the (BVS, RV) correlation using the  \cite{Melo_BVS_RV_corr} method.
For the stars that present a trend (see \Cref{trend_bin}), we applied a linear regression on their RVs. If the residuals presented a correlation between the BVS and the RVs, we corrected them for this correlation \citep{Melo_BVS_RV_corr}.
For the binary star for which it was possible to fit the companion signal  (see \cref{trend_bin}), we worked on the residuals. If the residuals presented a correlation between the BVS and the RVs we corrected them for this correlation \citep{Melo_BVS_RV_corr}.

For the  \harps \ stars we used the corrected data presented in \cite{Grandjean_HARPS}.

\subsection{Detection limits}
\label{detlim}

We used the local power analysis (LPA) \citep{Meunier,Simon_IX} to compute the $m_p\sin{i}$ detection limits   for periods between $1$ and \SI{1000}{\day} in the GP domain (between $1$ and $\SI{13}{\MJ}$), and in the BD domain (between $13$ and  $\SI{80}{\MJ}$). The LPA method  determines, for all periods $P$, the minimum $m_p\sin{i}$ for which a companion on a circular orbit\footnote{Assuming a circular orbit in the computation of the detection limits is common in the field of large RV surveys \citep{cumming, Lagrange_2009,Simon_IX,Grandjean_HARPS} despite the slight  underestimation of the detection limits it implies for the whole survey.} with a period $P$   leads to a signal consistent with the data;   this is done by comparing the synthetic companion maximum power of its periodogram to the maximum power of the data periodogram within a small period range around the period $P$. For a given star the detection limit is infinite for periods greater than its time baseline. We made this choice because the high jitter and the moderate number of spectra per target do not provide a strong constraint on a companion signal  that has a period greater than the time baseline.

\begin{figure}[t!]
  \centering
\includegraphics[width=1\hsize]{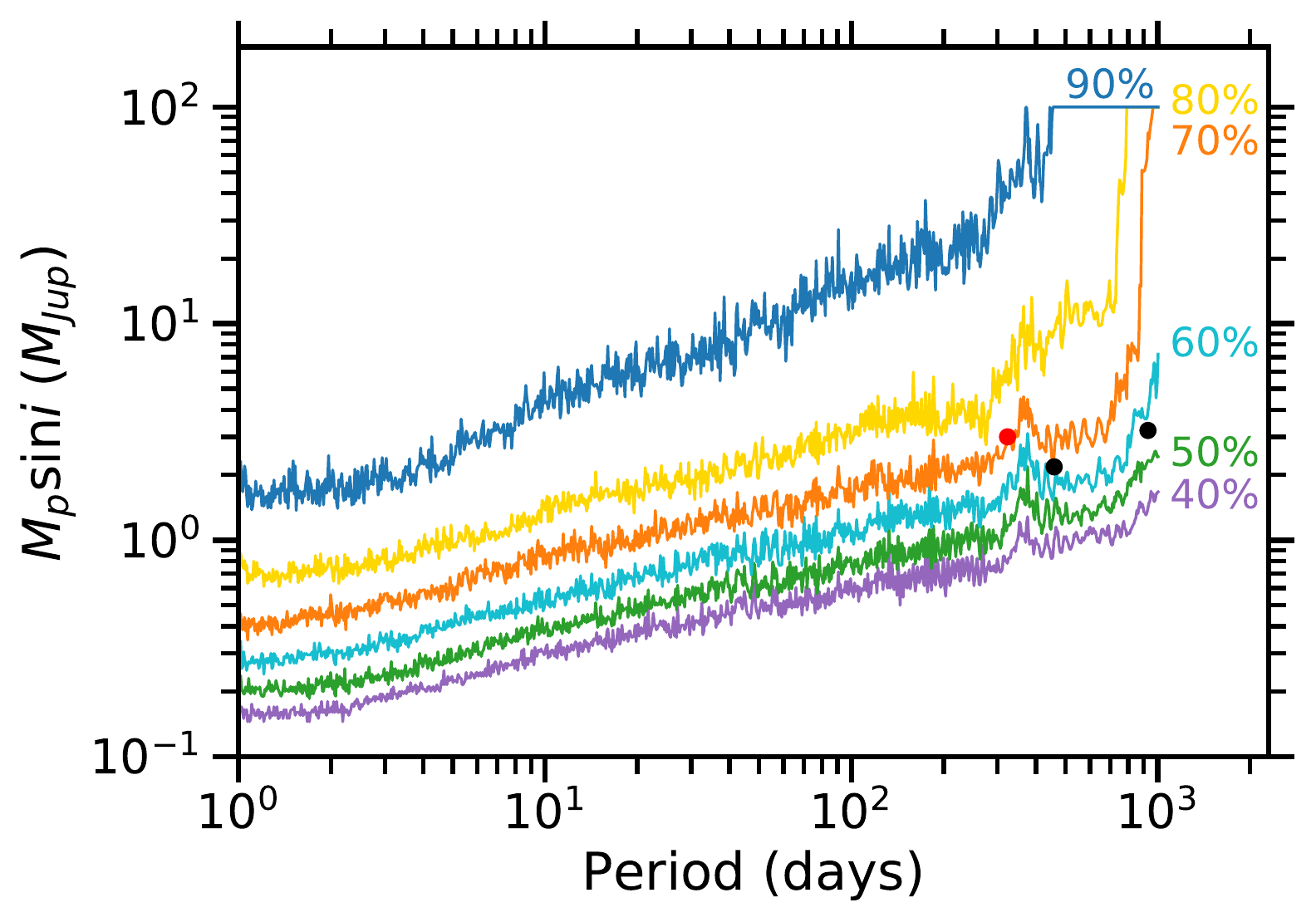}
\caption{Search completeness of our survey, corresponding to the lower $m_p \sin{i}$ for which $X\%$ of the stars in the survey have detection limits below this $m_p \sin{i}$ at a given period $P$; from bottom to top: $40\%$ to $90\%$. Our detected planet HD 113337 b is shown as  a red dot, while the non-detected planets in the survey, HD 128311 b and c, are shown as black dots. \label{Completeness}} 
\end{figure}

We then computed the completeness function $C(m_p\sin{i},P)$ of the sample which corresponds, for a given couple $(m_p\sin{i},P)$, to the fraction of stars in the sample for which a companion with this mass and period is excluded by the detection limits \citep{Simon_IX}. For the computation of the completeness we excluded HD113337, for which companions were detected during the survey.
The $40$ to $90\%$ search completeness values are presented in \Cref{Completeness}. 
We also computed the sample search completeness function $C_D$ in this period and mass ranges (see \cref{tab_occur}). It is over $75 \ \%$ for the AF and FK sub-samples.

\subsection{Companion occurrence rates}

From our combined \harps \ and \sophie \  sample of $119$ stars, we computed the occurrence rates of  GPs ($1$ to $\SI{13}{\MJ}$) and BDs ($13$ to $\SI{80}{\MJ}$) around AF-type ($B-V \in [-0.05:0.52[$), FK-type ($B-V \in [0.52:1.33[$), and KM-type ($B-V \geq 1.33$)   stars, and for different ranges of periods: $1$-$10$, $10$-$100$, $100$-$1000$, and $1$-\SI{1000}{\day}.
We used the method described in \cite{Simon_IX} to compute the occurrence rates and to correct them from the estimated number of missed companions $n_{miss}$ derived from the search completeness.
For the range where no companions were detected in the survey, only the upper limits of the occurrence rates are available.

\begin{figure}[h!]
  \centering
\includegraphics[width=1\hsize]{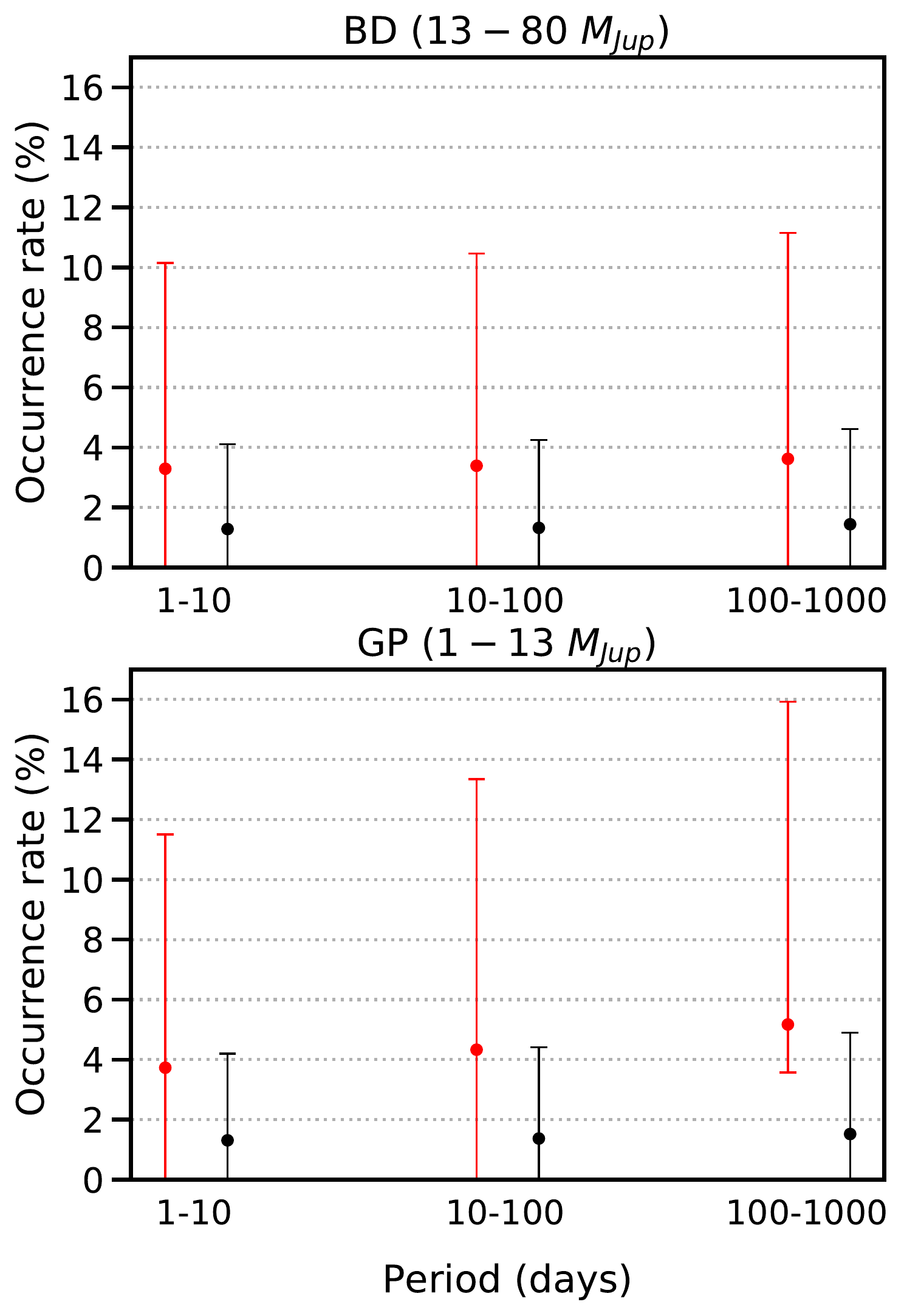}
\caption{Occurrence rates  and their $1 \sigma$ ranges  for the period ranges of $1$-$10$, $10$-$100$, and $100$- \SI{1000}{\day} in the GP domain ($1-13$ \Mjup,  \emph{Top}) and BD domain ($13-80$ \Mjup,  \emph{Bottom}), for the AF sub-sample (\emph{red}) and the FK sub-sample (\emph{black}).} 
       \label{sample}
\end{figure}

The inclusion of the SOPHIE YNS survey targets in the combined sample provided a better constraint on the GP and  BD  occurrence rates around young stars in comparison to our previous study based on the HARPS YNS survey alone  \citep{Grandjean_HARPS}. As an example, the upper limits on the GP and BD occurrence rates are now $10$ to $40\%$ (depending on the period and mass range) lower than our previous estimate. 
Moreover, one GP system with $P<\SI{1000}{\day}$ companions  was discovered in the SOPHIE YNS survey, which permit us to derive the occurrence rates of these objects in this period range  instead of an upper limit.  We computed an occurrence rate of $1_{-0.3}^{+2.2} \ \%$ for GP with periods under $ \SI{1000}{\day}$. The  BD occurrence rate is below $0.9_{-0.9}^{+2} \ \%$ in this period range.
We present these occurrence rates for AF, FK, M, and all stars in \cref{tab_occur}, and we present the AF and FK sub-samples occurrences rates  in \Cref{sample}.

\renewcommand{\arraystretch}{1.25}
\begin{table*}[t!]
\caption{GP ($m_p\sin{i} \in [1,13] \ \si{\MJ}$) and BD  ($m_p\sin{i} \in [13,80] \ \si{\MJ}$) occurrence rates around young nearby stars computed from our combined HARPS and SOPHIE sample. The parameters are displayed in normal, bold, italic, or bold  italic fonts when considering the full star sample, the  AF sub-sample, the FK sub-sample, or the M sub-sample, respectively.}
\label{tab_occur}
\begin{center}
\resizebox{1\textwidth}{!}{ 
\begin{tabular}{l c c c c c c l l}\\
\hline
\hline
\msini    & Orbital period  & \bv  & Search       & Detected    &  Missed       & GP occurrence rate    & \multicolumn{2}{c}{Confidence intervals}\\
interval  & interval        && completeness & GP systems  & GP systems    &                & $1\sigma$ & $2\sigma$        \\
(\Mjup)   & (day)           & & $C_D$  (\%)   &             &     upper limit          &                (\%)         & (\%)      &  (\%)            \\

\hline 
1-13 & 1-10 & all & 95 & 0 & 0.1 & <0.9 & 0-2.9 & 0-4.9 \\ 
(GP) &  & $[-0.05:0.52[$ & {\bf 87} & {\bf 0} & {\bf 0.2} & {\bf <3.7} & {\bf 0-11.5} & {\bf 0-19.1} \\ 
 &  & $[0.52:1.33[$ & {\it 97} & {\it 0} & {\it 0.0} & {\it <1.3} & {\it 0-4.2} & {\it 0-7.1} \\ 
 &  & $\geq 1.33$ &  \textbf{\textit{99}} &  \textbf{\textit{0}} & \textbf{\textit{0.0}} & \textbf{\textit{<11.2}} & \textbf{\textit{0-29.7}} & \textbf{\textit{0-45.6}} \\ 
\hline 
1-13 & 1-100 & all & 91 & 0 & 0.1 & <0.9 & 0-3.0 & 0-5.1 \\ 
&  & $[-0.05:0.52[$ & {\bf 81} & {\bf 0} & {\bf 0.2} & {\bf <4.0} & {\bf 0-12.4} & {\bf 0-20.5} \\ 
 &  & $[0.52:1.33[$ & {\it 95} & {\it 0} & {\it 0.1} & {\it <1.3} & {\it 0-4.3} & {\it 0-7.3} \\ 
 &  & $\geq 1.33$ &  \textbf{\textit{97}} &  \textbf{\textit{0}} & \textbf{\textit{0.0}} & \textbf{\textit{<11.5}} & \textbf{\textit{0-30.4}} & \textbf{\textit{0-46.6}} \\ 
\hline 
1-13 & 1-1000 & all & 87 & 1 & 0.2 & 1.0 & 0.7-3.1 & 0.2-5.4 \\ 
 &  & $[-0.05:0.52[$ & {\bf 75} & {\bf 1} & {\bf 0.3} & {\bf 4.3} & {\bf 3.0-13.4} & {\bf 1.0-22.1} \\ 
 &  & $[0.52:1.33[$ & {\it 91} & {\it 0} & {\it 0.1} & {\it <1.4} & {\it 0-4.5} & {\it 0-7.6} \\ 
 &  & $\geq 1.33$ &  \textbf{\textit{89}} &  \textbf{\textit{0}} & \textbf{\textit{0.1}} & \textbf{\textit{<12.5}} & \textbf{\textit{0-33.0}} & \textbf{\textit{0-50.7}} \\ 
\hline 
\hline 
13-80 & 1-10 & all & 99 & 0 & 0.0 & <0.8 & 0-2.8 & 0-4.7 \\ 
(BD) &  & $[-0.05:0.52[$ & {\bf 98} & {\bf 0} & {\bf 0.0} & {\bf <3.3} & {\bf 0-10.2} & {\bf 0-16.8} \\ 
 &  & $[0.52:1.33[$ & {\it 97} & {\it 0} & {\it 0.0} & {\it <1.3} & {\it 0-4.2} & {\it 0-7.1} \\ 
 &  & $\geq 1.33$ &  \textbf{\textit{100}} &  \textbf{\textit{0}} & \textbf{\textit{0.0}} & \textbf{\textit{<11.1}} & \textbf{\textit{0-29.4}} & \textbf{\textit{0-45.2}} \\ 
\hline 
13-80 & 1-100 & all & 97 & 0 & 0.0 & <0.9 & 0-2.8 & 0-4.8 \\ 
&  & $[-0.05:0.52[$ & {\bf 97} & {\bf 0} & {\bf 0.0} & {\bf <3.3} & {\bf 0-10.3} & {\bf 0-17.1} \\ 
 &  & $[0.52:1.33[$ & {\it 98} & {\it 0} & {\it 0.0} & {\it <1.3} & {\it 0-4.2} & {\it 0-7.1} \\ 
 &  & $\geq 1.33$ &  \textbf{\textit{100}} &  \textbf{\textit{0}} & \textbf{\textit{0.0}} & \textbf{\textit{<11.2}} & \textbf{\textit{0-29.5}} & \textbf{\textit{0-45.3}} \\ 
\hline 
13-80 & 1-1000 & all & 94 & 0 & 0.1 & <0.9 & 0-2.9 & 0-4.9 \\ 
 &  & $[-0.05:0.52[$ & {\bf 94} & {\bf 0} & {\bf 0.1} & {\bf <3.4} & {\bf 0-10.6} & {\bf 0-17.5} \\ 
 &  & $[0.52:1.33[$ & {\it 95} & {\it 0} & {\it 0.1} & {\it <1.3} & {\it 0-4.3} & {\it 0-7.3} \\ 
 &  & $\geq 1.33$ &  \textbf{\textit{94}} &  \textbf{\textit{0}} & \textbf{\textit{0.1}} & \textbf{\textit{<11.8}} & \textbf{\textit{0-31.4}} & \textbf{\textit{0-48.1}} \\ 
\hline 
\hline

\end{tabular}}
\end{center}
\end{table*}
\renewcommand{\arraystretch}{1.}

\subsection{Comparison to surveys on main sequence stars}

 In our survey two companions with periods  between $100$ and $\SI{1000}{\day}$ were detected on the same stars belonging to our AF sub-sample (HD 113337 b and c; \cite{Simon_VIII,Simon_X}). However, we may have missed some planets with low masses and long periods as only $40\%$ of the stars in the survey have detection limits lower than $\SI{2}{\MJ}$ between $100$ and $\SI{1000}{\day}$ (see \Cref{Completeness}).

No HJ was detected in the survey. This non-detection is robust as $70 \ \%$ of our stars present limits of detection lower than $\SI{1}{\MJ}$ for period lower than $\SI{10}{\day}$. In consequence only upper limits on the occurrence rates of HJ can be computed. 
To compare our results to previous surveys we adopted the same p-value formalism as in \cite{Grandjean_HARPS}. If the p-value of our non-detection of HJ around young stars  is below $10 \%$ for a given occurrence rate on MS stars found in the literature, it   indicates that the occurrence rate might be different between young and MS stars with a confidence level of $90\%$.

For the AF stars we computed  an occurrence rate of  $4.3_{-1.3}^{+9.0} \ \%$ for GP with periods lower $\SI{1000}{\day}$, which is in agreement with the  $3.7_{-1.1}^{+2.8} \ \%$ occurrence rate derived by \cite{Simon_X} on all age AF MS stars.
For the FK stars we computed an upper limit on the occurrence rate of  $1.4_{-1.4}^{+3.1} \ \%$ for GP with periods lower $\SI{1000}{\day}$. It is compatible at  $1\sigma$ with the GP occurrence rate of $4.3 \pm 1 \ \%$ obtained by \cite{cumming} on FK MS stars in the same period range, but it may be lower. The p-value test is validated with a confidence level of $90\%$ ($p_{value} =  3_{-2}^{+4}\%$), which indicates that the occurrence rates of these objects might be different between young and old FK stars. However, the level of confidence of this test is not strict, and the probability of  observing such a difference is still about $10\%$. Moreover, it is puzzling that we do not observe a similar lack of close GPs around young AF stars, as our AF and FK sub-samples are not significantly different in metallicity and age.
A statistical analysis on a larger number of FK  young  targets is needed to determine whether the GP occurrence rate is significantly lower for young FK stars than for MS FK stars.

Our upper limit on the occurrence rate of HJs ($P<\SI{10}{\day}$) around FK-type stars is    $1.3_{-1.3}^{+2.9} \ \%$, which is compatible with the HJ occurrence rate around FK MS stars       estimated at $0.46^{+0.3}_{-0.3} \ \%$ by  \cite{cumming}. We detected $0$ companion out of $32$ stars; the corresponding  p-value is  $86^{+9}_{-8} \ \%$.

We note that our upper limit on the GP occurrence rates around FK stars for periods lower than $\SI{1000}{\day}$ is lower, but compatible at $1\sigma$, with the occurrence rate we derived for AF stars on the same period range. If the occurrence rate of these planets is identical between AF and FK stars, then the p-value of our non-detection of such planets around our $32$ AF stars will be   $25_{-24}^{+13}\%$. The apparent difference  in occurrence rates between young AF and FK stars  is not likely to   be significant, but a possibility that it is significant remains.
The metallicity of our AF and FK sub-samples are not statically different. On the other hand, the stellar mass of these two sub-sample is significantly different. The occurrence rates of GP around young stars might then depend on the host star mass in favor of the high mass. This would be in agreement with the predictions of core accretion models \citep{Kennedy}.
It would also be in agreement with the host star mass-GP occurrence rate positive correlation observed for evolved stars in RV \citep{Bowler_2009,Johnson,Jones} and for wide orbit planets around young stars in direct imaging \citep{Baron_2019}. 
A statistical analysis on larger and similarly sized young AF and FK samples is needed to determine whether the GP occurrence rate for FK young stars is significantly lower than for AF young stars.

Close BDs ($P<\SI{1000}{\day}$)  are known to be rare \citep{Grether, Sahlmann,Grieves,Jones_BD,Simon_X}. Our upper limit of  $0.9_{-0.9}^{+2} \ \%$ on the BD occurrence rate for periods lower than $\SI{1000}{\day}$ is consistent with the literature.
A statistical analysis on a larger number of young targets is needed to study the difference in the BD occurrence rates between young and MS stars.

\section{Conclusion}
\label{conc}

We  carried out a three-year \sophie \  survey on $63$ young  A- to M-type  stars in the search for close GP and BD companions.
This survey allowed the discovery of a multiplanet system around HD113337 \citep{Simon_VIII,Simon_X}; two binary companions, HD 112097 B and  HD 105693 B; and  a long-term trend on HD 109647.
We confirmed numerous binary companions and  we constrained for the first time the orbital parameters of HD 195943 B.
No HJ or short-period ($P<\SI{10}{\day}$) BD was discovered in this survey.

We then combined our \sophie \  survey with the \harps \ YNS survey that was presented in \cite{Grandjean_HARPS}, leading to a statistical analysis on $120$ young stars.
We obtained a  GP occurrence rate of  $0.9_{-0.3}^{+2.2} \ \%$ for periods lower than  $\SI{1000}{\day}$ and an upper limit on the BD occurrence rate  of $0.9_{-0.9}^{+2}\ \%$ in the same period range.
We observed a barely significant difference of close GP occurrence rate between AF-type and FK-type stars.
We also observed a significant difference in GP occurrence rates between young and MS F- to K-type stars with a confidence level of $90\%$.
An analysis of a larger number of young stars is needed to determine whether these differences are actually  significant.

The  forthcoming analysis of our \harps \ survey on Sco-Cen stars will add $50$ stars to our analysis. This will permit us to reduce the uncertainties on the derived occurrence rates for young stars, and will also help in the search for the possible impact of system ages on occurrence rates. Moreover, the Sco-Cen survey is mainly composed of early-type stars, which will balance our AF and FK sub-samples, allowing a better comparison of the two sub-sample occurrence rates.

\begin{acknowledgements}
   We acknowledge support from the French CNRS and from the Agence Nationale de la Recherche (ANR grant GIPSE ANR-14-CE33-0018). 
This work has been supported by a grant from Labex OSUG@2020 (Investissements d’avenir – ANR10 LABX56).
These results have made use of the SIMBAD database, operated at the CDS, Strasbourg, France.  ESO
SD acknowledges the support by INAF/Frontiera through the "Progetti
Premiali" funding scheme of the Italian Ministry of Education,
University, and Research.
Based on observations collected at the European Southern Observatory under ESO programme(s) 060.A-9036(A),072.C-0488(E),072.C-0636(A),072.C-0636(B),073.C-0733(A),073.C-0733(C),073.C-0733(D),073.C-0733(E),074.C-0037(A),074.C-0364(A),075.C-0202(A),075.C-0234(A),075.C-0234(B),075.C-0689(A),075.C-0689(B),076.C-0010(A),076.C-0073(A),076.C-0279(A),076.C-0279(B),076.C-0279(C),077.C-0012(A),077.C-0295(A),077.C-0295(B),077.C-0295(C),077.C-0295(D),078.C-0209(A),078.C-0209(B),078.C-0751(A),078.C-0751(B),079.C-0170(A),079.C-0170(B),079.C-0657(C),080.C-0032(A),080.C-0032(B),080.C-0664(A),080.C-0712(A),081.C-0034(B),081.C-0802(C),082.C-0308(A),082.C-0357(A),082.C-0412(A),082.C-0427(A),082.C-0427(C),082.C-0718(B),083.C-0794(A),083.C-0794(B),083.C-0794(C),083.C-0794(D),084.C-1039(A),085.C-0019(A),087.C-0831(A),089.C-0732(A),089.C-0739(A),090.C-0421(A),091.C-0034(A),094.C-0946(A),098.C-0739(A),099.C-0205(A),0104.C-0418(A),1101.C-0557(A),183.C-0437(A),183.C-0972(A),184.C-0815(A),184.C-0815(B),184.C-0815(C),184.C-0815(E),184.C-0815(F),191.C-0873(A),192.C-0224(B),192.C-0224(C),192.C-0224(G),192.C-0224(H).
Based on observations collected at the Observatoire de Haute Provence under the programme(s)
06B.PNP.CONS,07A.PNP.CONS,07B.PNP.CONS, 08A.PNP.CONS,13A.PNP.DELF,13A.PNP.LAGR,14A.PNP.LAGR.
\end{acknowledgements}

\section{References}

\bibliography{sophie_yns_sco_cen}

\bibliographystyle{aa}

%
%


\begin{appendix}

\onecolumn
\section{Combined sample}

\subsection{\sophie \  sample}

\setcounter{footnote}{0}
\begingroup
\renewcommand\arraystretch{1.3}
\begin{longtable}{lcccccccc}
\caption{Star characteristics for the $54$ stars in  the \sophie \  YNS RV survey for which RVs could be computed  ($V\sin{i} <\SI{300}{\kilo\meter\per\second}$). The stars excluded from our analysis (too old stars, binary stars for which the companion signal could not be fitted) are flagged  with a dagger($^\dagger$).
Spectral type (ST) were taken from the CDS database at the beginning of the survey, and the  \bv~values are    from the CDS database.
The \vsini~ are computed with \safir~based on the CCF width.
The IR/D column reports whether   an IR excess is reported ($y$) or not ($n$), and whether a disk has been resolved ($y$) or not ($n$) in the literature.}
\label{tab_carac_s}
 \\ \hline
Name & HIP & ST & \bv & Mass & Age  &\vsini          &  IR/D\\ 
  HD/BD/CD & &&&(\Msun) & (\si{\mega\year}) &(\kms) &  & \\
\hline \hline
\endfirsthead
\caption{Continued.}\\
 \\ \hline
Name & HIP & ST & \bv & Mass & Age  &\vsini         &  IR/D\\ 
  HD/BD/CD & &&&(\Msun) & (\si{\mega\year})  &(\kms)   & \\
 \\ \hline
  \hline
\endhead
\hline
\endfoot
HD377 & 682 &  G2V & 0.630 & 1.11\footnote{\label{Sylvano_mass_s} This work (see \Cref{age_masse})} & $200_{-200}^{+200}$\footnote{\label{Chavero}\cite{Chavero}}  & 10.4 & y\footnote{\label{Meyer_excess}\cite{Meyer_excess}}/y\footnote{\label{Cotten}\cite{Cotten}} \\ 
HD2454$^\dagger$ & 2235 &  F6V & 0.430 & 1.21\footnoteref{Sylvano_mass_s} & $800_{-300}^{+300}$\footnote{\label{Gomez}\cite{Gomez}}  & 6.7 & -/- \\ 
HD7804  & 6061 & A3V &  0.073 & 2.52\footnote{\label{Stone}\cite{Stone}} & $412$\footnoteref{Stone} &  78.0 & y\footnote{\label{McDonald}\cite{McDonald}}/- \\ 
HD13507 & 10321 &  G0V & 0.670 & 0.99\footnoteref{Sylvano_mass_s} & $500_{-100}^{+100}$\footnoteref{Gomez}  & 4.1 & y\footnoteref{McDonald}/- \\ 
HD13531  & 10339 & G0V &  0.700 & 1\footnoteref{Sylvano_mass_s} & $400$\footnote{\label{Gaspar}\cite{Gaspar}} &  5.3 & y\footnoteref{McDonald}/- \\ 
HD17250  & 12925 & F8V &  0.520 & 1.26\footnoteref{Sylvano_mass_s} & $45$\footnote{\label{Vigan}\cite{Vigan}} &  29.2 & y\footnoteref{McDonald}/- \\ 
HD20630 & 15457 &  G5V & 0.670 & $0.99^{+0.02}_{-0.03}$\footnoteref{Gomez} & $600_{-200}^{+200}$\footnoteref{Gomez}  & 4.8 & n\footnoteref{Gaspar}/- \\ 
HD25457 & 18859 &  F5V & 0.500 & 1.2\footnoteref{Vigan} & $149_{-49}^{+31}$\footnoteref{Vigan}  & 12.4 & y\footnote{\label{Zuckerman}\cite{Zuckerman}}/- \\ 
HD25953 & 19183 &  F5V & 0.500 & 1.22\footnoteref{Sylvano_mass_s} & $149_{-49}^{+31}$\footnoteref{Vigan}  & 20.5 & y\footnoteref{McDonald}/- \\ 
HD26913 & 19855 &  G8V & 0.660 & 0.85\footnote{\label{Luck}\cite{Luck}} & $1300_{-1200}^{+1200}$\footnoteref{Chavero}  & 5.4 & y\footnoteref{McDonald}/- \\ 
HD26923 & 19859 &  G0IV & 0.560 & 1.07\footnote{\label{Ammler}\cite{Ammler}} & $500_{-100}^{+50}$\footnoteref{Sylvano_mass_s}  & 3.8 & n\footnoteref{McDonald}/- \\ 
HD28495$^\dagger$  & 21276 & G0V &  0.778 & 0.82\footnoteref{Sylvano_mass_s} & $100$\footnote{\label{Ballering}\cite{Ballering}} &  4.8 & y\footnoteref{McDonald}/- \\ 
HD29697 & 21818 &  K3V & 1.135 & 0.75\footnoteref{Sylvano_mass_s} & $100_{-70}^{+100}$\footnoteref{Sylvano_mass_s}  & 7.9 & -/- \\ 
HD32977 & 23871 &  A5V & 0.100 & 1.89\footnoteref{Sylvano_mass_s} & $480_{-193}^{+193}$\footnoteref{Sylvano_mass_s}  & 66.1 & y\footnoteref{McDonald}/- \\ 
HD35171 & 25220 &  K2V & 1.129 & 0.79\footnoteref{Sylvano_mass_s} & $445_{-222}^{+445}$\footnoteref{Sylvano_mass_s}  & 4.3 & -/- \\ 
HD39587  & 27913 & G0V &  0.600 & 1.1\footnoteref{Stone} & $412$\footnoteref{Stone} &  7.1 & y\footnoteref{McDonald}/- \\ 
HD41593 & 28954 &  K0V & 0.825 & 1.01\footnoteref{Ammler} & $329_{-93}^{+93}$\footnoteref{Plavchan}  & 4.6 & y\footnoteref{McDonald}/- \\ 
HD48299$^\dagger$ & 32262 &  F5IV & 0.520 & $1.18^{+0.05}_{-0.04}$\footnote{\label{Casagrande}\cite{Casagrande}} & $2940_{-1660}^{+560}$\footnoteref{Casagrande}  & 4.3 & n\footnoteref{McDonald}/- \\ 
HD63433 & 38228 &  G5IV & 0.680 & $0.96^{+0.02}_{-0.02}$\footnoteref{Gomez} & $300_{-100}^{+100}$\footnoteref{Gomez}  & 5.7 & y\footnoteref{McDonald}/- \\ 
HD75332 & 43410 &  F7V & 0.528 & 1.17\footnoteref{Sylvano_mass_s} & $400_{-100}^{+100}$\footnoteref{Gomez}  & 6.9 & y\footnoteref{McDonald}/- \\ 
HD82106 & 46580 &  K3V & 1.026 & 0.84\footnoteref{Sylvano_mass_s} & $500_{-100}^{+100}$\footnoteref{Sylvano_mass_s}  & 3.8 & -/- \\ 
HD89449$^\dagger$  & 50564 & F6IV &  0.440 & 1.44\footnote{\label{Simon_X}\cite{Simon_X}} & $3100$\footnoteref{Gaspar} &  11.9 & -/- \\ 
HD90905 & 51386 &  F5V & 0.569 & 1.16\footnote{\label{Lagrange}\cite{Lagrange}} & $170_{-70}^{+180}$\footnoteref{Vigan}  & 7.0 & y\footnote{\label{Carpenter}\cite{Carpenter}}/y\footnote{\label{Morales}\cite{Morales}} \\ 
HD94765 & 53486 &  K0V & 0.873 & 0.9\footnoteref{Vigan} & $500_{-200}^{+200}$\footnoteref{Vigan}  & 4.3 & -/- \\ 
HD97244 & 54688 &  A5V & 0.219 & 1.62\footnoteref{Sylvano_mass_s} & $375_{-337}^{+337}$\footnoteref{Sylvano_mass_s}  & 58.4 & y\footnoteref{McDonald}/- \\ 
HD102647  & 57632 & A3V &  0.090 & 1.90\footnoteref{Sylvano_mass_s} & $50$\footnoteref{Vigan} &  80.5 & y\footnote{\label{Stencel}\cite{Stencel}}/y\footnote{\label{Matthews}\cite{Matthews}} \\ 
HD105963$^\dagger$ & 59432 &  K0V & 0.210 & 0.75\footnoteref{Sylvano_mass_s} & $266_{-11}^{+11}$\footnote{\label{Plavchan}\cite{Plavchan}}  & 8.7 & -/- \\ 
HD107146 & 60074 &  G2V & 0.604 & 1.08\footnoteref{Sylvano_mass_s} & $200_{-80}^{+100}$\footnoteref{Vigan}  & 4.9 & -/y\footnote{\label{Corder}\cite{Corder}} \\ 
HD109647  & 61481 & K0V &  0.960 & 0.8\footnoteref{Stone} & $412$\footnoteref{Stone} &  3.7 & y\footnoteref{McDonald}/- \\ 
HD110463  & 61946 & K3V &  0.970 & 0.83\footnote{\label{Luck_17}\cite{Luck_17}} & $2570$\footnoteref{Luck_17} &  3.7 & n\footnoteref{McDonald}/- \\ 
HD112097  & 62933 & A7III &  0.281 & 1.75\footnoteref{Stone} & $412$\footnoteref{Stone} &  46.8 & n\footnoteref{McDonald}/- \\ 
HD113337 & 63584 &  F6V & 0.372 & 1.41\footnoteref{Sylvano_mass_s} & $150_{-50}^{+50}$\footnote{\label{Simon_VIII}\cite{Simon_VIII}}  & 6.3 & y\footnote{\label{Rhee}\cite{Rhee}}/y\footnote{\label{Su}\cite{Su}} \\
\pagebreak 
HD115383 & 64792 &  G0V & 0.590 & 1.24\footnoteref{Sylvano_mass_s} & $100_{-100}^{+100}$\footnoteref{Gomez}  & 5.8 & -/- \\ 
HD128311 & 71395 &  K0V & 0.995 & 0.84\footnoteref{Sylvano_mass_s} & $450_{-100}^{+100}$\footnoteref{Sylvano_mass_s}  & 4.6 & y\footnote{\label{Beichman_2006}\cite{Beichman_2006}}\footnote{\label{Saffe}\cite{Saffe}}/- \\ 
HD131156  & 72659 & G8V &  0.777 & 1.00\footnoteref{Stone} & $290$\footnoteref{Stone} &  4.8 & y\footnoteref{McDonald}/- \\ 
HD135559 & 74689 &  A4V & 0.177 & 1.70\footnoteref{Sylvano_mass_s} & $422_{-322}^{+322}$\footnoteref{Sylvano_mass_s}  & 91.3 & y\footnoteref{McDonald}/- \\ 
HD142229 & 77810 &  G5V & 0.620 & 1.10\footnoteref{Sylvano_mass_s} & $350_{-150}^{+100}$\footnoteref{Sylvano_mass_s}  & 5.1 & y\footnoteref{McDonald}/- \\ 
HD148387$^\dagger$ & 80331 &  G8III & 0.910 & $2.45^{+0.1}_{-0.1}$\footnote{\label{Baines}\cite{Baines}} & $650_{-100}^{+100}$\footnoteref{Baines}  & 4.1 & y\footnoteref{McDonald}/- \\ 
HD161284  & 86456 & K0V & 0.930 & 0.74\footnoteref{Sylvano_mass_s} & - &  3.8 & y\footnoteref{McDonald}/- \\ 
HD166435$^\dagger$ & 88945 &  G0V & 0.628 & $1.01^{+0.02}_{-0.04}$\footnoteref{Gomez} & $3320_{-1590}^{+2290}$\footnoteref{Gomez}  & 6.2 & y\footnoteref{McDonald}/- \\ 
HD171488 & 91043 &  G0V & 0.551 & 1.1\footnoteref{Vigan} & $45_{-25}^{+35}$\footnoteref{Vigan}  & 27.7 & y\footnoteref{McDonald}/- \\ 
HD175726 & 92984 &  G0V & 0.570 & 1.13\footnoteref{Sylvano_mass_s} & $424_{-212}^{+424}$\footnoteref{Sylvano_mass_s}  & 9.0 & y\footnoteref{McDonald}/- \\ 
HD186689 & 97229 &  A3IV & 0.180 & 1.68\footnoteref{Sylvano_mass_s} & $361_{-308}^{+308}$\footnoteref{Sylvano_mass_s}  & 22.5 & y\footnoteref{McDonald}/- \\ 
HD186704  & 97255 & G0V &  0.583 & 1.10\footnote{\label{Bonavita}\cite{Bonavita}} & $100$\footnote{\label{Z_13}\cite{Zuckerman_2013}} &  10.5 & y\footnoteref{McDonald}/- \\ 
HD195943  & 101483 & A3IV &  0.073 & 2.25\footnoteref{Sylvano_mass_s} & $50$\footnoteref{Vigan} &  45.3 & n\footnoteref{McDonald}/- \\ 
HD206860 & 107350 &  G0V & 0.580 & 1.1\footnoteref{Vigan} & $300_{-100}^{+700}$\footnoteref{Vigan}  & 7.8 & y\footnoteref{Beichman_2006}\footnote{\label{Moro-Martin}\cite{Moro-Martin}}/- \\ 
HD211472  & 109926 & K1V &  0.810 & $0.88^{+0.01}_{-0.05}$\footnoteref{Casagrande} & $100$\footnoteref{Casagrande} &  4.5 & -/- \\ 
HD218396  & 114189 & A5V &  0.257 & 1.46\footnoteref{Vigan} & $42$\footnoteref{Vigan} &  29.8 & y\footnote{\label{Zuckerman_8799}\cite{Zuckerman_8799}}/y\footnote{\label{Su_8799}\cite{Su_8799}} \\ 
HD218738$^\dagger$ & 114379 &  G5V & 0.929 & $0.79^{+0.02}_{-0.03}$\footnoteref{Casagrande} & $100_{-70}^{+200}$\footnote{\label{Galicher}\cite{Galicher}}  & 9.4 & -/- \\ 
HD220140 & 115147 &  G9V & 0.890 & $0.78^{+0.02}_{-0.01}$\footnoteref{Casagrande} & $52_{-22}^{+28}$\footnoteref{Vigan}  & 11.6 & -/- \\ 
HD245409 & 26335 &  K7V & 1.378 & 0.63\footnoteref{Sylvano_mass_s} & $400_{-250}^{+400}$\footnoteref{Sylvano_mass_s}  & 4.6 & y\footnoteref{McDonald}/- \\ 
-  & 11437 & K7V &  1.180 & 0.42\footnote{\label{Baron}\cite{Baron}} & $12$\footnote{\label{Paulson}\cite{Paulson}} &  4.8 & y\footnoteref{McDonald}/- \\ 
-  & 40774 & G5V &  0.919 & 0.85\footnoteref{Sylvano_mass_s} & $250$\footnoteref{Vigan} &  4.6 & -/- \\ 
BD$+$20 1790 & - &  K5V & 1.079 & 0.72\footnoteref{Sylvano_mass_s} & $149_{-49}^{+31}$\footnoteref{Vigan}  & 7.5 & -/- \\ 
\end{longtable}
\endgroup

\setcounter{footnote}{0}
\begin{landscape}
\begingroup
\begin{longtable}{lcc|cccccccccccccc}
\caption{Results, before any correction, for the $54$ stars in our  \sophie \  YNS RV survey  for which RVs could be computed  ($V\sin{i} <\SI{300}{\kilo\meter\per\second}$). The stars excluded from our analysis (too old stars, binary stars for which the companion signal could not be fitted) are flagged  with a dagger($^\dagger$). 
Spectral type (ST)  were taken from the CDS database at the beginning of the survey.
The table provides the time baseline (TBL); the number of computed spectra $N_{\rm m}$; the amplitude corresponding to the difference between the maximum and the minimum of the RVs({\it A}), rms, and mean uncertainty $<${\it U}$>$ on the RV and BVS measurements; the RV-BVS correlation factor (slope of the best linear fit); the mean FWHM ($<$FWHM$>$); and the mean \rhk \ ($<$\rhk$>$) and its standard deviation $\sigma_{logR'_{\rm HK}}$.
The V column presents, if identified,  the source of stellar jitter: A stands for stellar activity (spots) and P stands for pulsations.
The T column presents the stars with a long-term trend induced by a companion.
The B column presents the nature of the companion:  C stands for sub-stellar companion, SB1 for stands for single-line spectroscopic binary, and SB2 stands for double-line spectroscopic binary. }
\label{tab_result_s}\\
\\ \hline
\multicolumn{3}{c|}{Stellar characteristics} & \multicolumn{12}{c}{Survey results.}
 \\ \hline
Name & HIP & ST &  TBL & $N_{\rm m}$   & \multicolumn{3}{c}{RV}        & \multicolumn{3}{c}{BVS}      & RV-    & $<$FWHM$>$ & $<$\rhk$>$  & V   & T & B\\ 
 HD/BD/CD & &  & & & \multicolumn{3}{c}{\raisebox{.5\baselineskip}{$\overbrace{\hspace{2.7cm}}$}} & \multicolumn{3}{c}{\raisebox{.5\baselineskip}{$\overbrace{\hspace{2.7cm}}$}} & BVS & &($\sigma_{logR'_{\rm HK}}$)  & & &\\
&& &  &  & {\it A} &rms  & $<${\it U}$>$ & {\it A} & rms &$<${\it U}$>$ & corr.    &                 &  & & &  \\
&&&&    & \multicolumn{3}{c}{\raisebox{.5\baselineskip}{$\underbrace{\hspace{2.7cm}}$}} & \multicolumn{3}{c}{\raisebox{.5\baselineskip}{$\underbrace{\hspace{2.7cm}}$}} & &  & & &\\
&& &     (day)       && \multicolumn{3}{c}{($\si{\meter\per\second}$)}      & \multicolumn{3}{c}{($\si{\meter\per\second}$)}     &   &      ($\si{\kilo\meter\per\second}$)  &    & & &\\ \hline
\endfirsthead
\caption{Continued.}\\
 \\ \hline
\multicolumn{3}{c|}{Stellar characteristics} & \multicolumn{12}{c}{Survey results.}
 \\ \hline
Name & HIP & ST &  TBL & $N_{\rm m}$   & \multicolumn{3}{c}{RV}        & \multicolumn{3}{c}{BVS}      & RV-    & $<$FWHM$>$ & $<$\rhk$>$ & V  & T & V \\ 
 HD/BD/CD & &  & & & \multicolumn{3}{c}{\raisebox{.5\baselineskip}{$\overbrace{\hspace{2.7cm}}$}} & \multicolumn{3}{c}{\raisebox{.5\baselineskip}{$\overbrace{\hspace{2.7cm}}$}} & BVS & &($\sigma_{logR'_{\rm HK}}$)  & & &\\
&& &  &  & {\it A} &rms  & $<${\it U}$>$ & {\it A} & rms &$<${\it U}$>$ & corr.    &                 &    & & &\\
&&&&    & \multicolumn{3}{c}{\raisebox{.5\baselineskip}{$\underbrace{\hspace{2.7cm}}$}} & \multicolumn{3}{c}{\raisebox{.5\baselineskip}{$\underbrace{\hspace{2.7cm}}$}} & & & & & \\
&& & (day)           && \multicolumn{3}{c}{($\si{\meter\per\second}$)}      & \multicolumn{3}{c}{($\si{\meter\per\second}$)}     &   &       ($\si{\kilo\meter\per\second}$) &  & & &\\ \hline
\endhead
\hline
\endfoot
HD377 & 682 & G2V & 747 & 17 & 652.3 & 139.1 & 6.3 & 302.6 & 15.0 & 73.6 & -0.84 & 22.8 & -4.39(0.04) & A & - &- \\ 
HD2454$^\dagger$ & 2235 & F6V & 805 & 29 & 545.6 & 154.9 & 4.2 & 105.4 & 11.0 & 28.4 & - & 14.6 & -4.89(0.08) & A & T &SB1 \\ 
HD7804 & 6061 & A3V & 740 & 92 & 8767.7 & 875.7 & 190.9 & 161539.7 & 447.1 & 15742.1 & - & 175.9 & -3.61(0.03) & P & - &- \\ 
HD13507 & 10321 & G0V & 886 & 24 & 100.0 & 24.5 & 2.2 & 42.7 & 6.5 & 10.8 & - & 9.9 & -4.45(0.04) & A & T &SB1 \\ 
HD13531 & 10339 & G0V & 882 & 33 & 103.1 & 28.6 & 2.7 & 52.9 & 7.7 & 11.6 & - & 11.9 & -4.44(0.04) & A & T &SB1 \\ 
HD17250 & 12925 & F8V & 5 & 5 & 1157.2 & 420.0 & 19.7 & 579.6 & 49.9 & 199.5 & - & 63.9 & -4.393(0.01) & - & T &SB1 \\ 
HD20630 & 15457 & G5V & 740 & 29 & 80.1 & 21.6 & 1.5 & 48.2 & 4.4 & 11.9 & -1.61 & 11.0 & -4.52(0.04) & A & - &- \\ 
HD25457 & 18859 & F5V & 498 & 10 & 121.0 & 39.7 & 6.3 & 158.5 & 14.8 & 54.8 & -0.59 & 27.1 & -4.48(0.02) & A & - &- \\ 
HD25953 & 19183 & F5V & 883 & 22 & 339.7 & 83.4 & 22.6 & 508.7 & 56.1 & 133.1 & 0.05 & 45.1 & -4.42(0.08) & A & - &- \\ 
HD26913 & 19855 & G8V & 883 & 18 & 112.5 & 29.0 & 3.4 & 84.0 & 8.5 & 20.0 & -1.21 & 12.2 & -4.43(0.05) & A & - &- \\ 
HD26923 & 19859 & G0IV & 501 & 22 & 23.6 & 6.7 & 1.9 & 20.3 & 5.2 & 5.0 & -0.21 & 9.4 & -4.56(0.05) & A & - &- \\ 
HD28495$^\dagger$ & 21276 & G0V & 500 & 12 & 5293.0 & 1867.9 & 2.3 & 849.5 & 6.3 & 307.6 & - & 11.1 & -4.35(0.02) & - & - &SB1 \\ 
HD29697 & 21818 & K3V & 498 & 11 & 461.4 & 131.9 & 4.1 & 221.6 & 10.6 & 56.9 & -1.93 & 17.1 & -4.25(0.04) & A & - &- \\ 
HD32977 & 23871 & A5V & 881 & 58 & 7983.5 & 1005.2 & 72.5 & 8043.9 & 180.4 & 1139.2 & - & 141.2 & -3.99(0.05) & P & - &- \\ 
HD35171 & 25220 & K2V & 882 & 14 & 56.9 & 18.3 & 1.8 & 25.1 & 5.1 & 8.0 & -0.48 & 10.1 & -4.54(0.03) & A & - &- \\ 
HD39587 & 27913 & G0V & 457 & 21 & 1342.4 & 569.7 & 2.4 & 55.3 & 6.2 & 15.9 & - & 15.6 & -4.54(0.03) & - & - &SB1 \\ 
HD41593 & 28954 & K0V & 42 & 9 & 32.5 & 10.1 & 2.2 & 20.5 & 6.1 & 8.5 & -0.66 & 10.7 & -4.43(0.04) & A & - &- \\ 
HD48299$^\dagger$ & 32262 & F5IV & 499 & 11 & 19.0 & 5.3 & 3.0 & 42.0 & 7.9 & 11.7 & -0.25 & 10.2 & -4.9(0.1) & A & - &- \\ 
HD63433 & 38228 & G5IV & 1109 & 22 & 135.1 & 43.3 & 2.9 & 81.1 & 7.8 & 24.7 & -1.62 & 12.9 & -4.44(0.02) & A & - &- \\ 
HD75332 & 43410 & F7V & 3413 & 69 & 125.4 & 33.5 & 3.3 & 120.7 & 8.8 & 22.2 & -0.7 & 15.2 & -4.56(0.06) & A & - &- \\ 
HD82106 & 46580 & K3V & 1106 & 33 & 33.4 & 6.0 & 1.6 & 18.4 & 4.7 & 4.6 & 0.15 & 9.3 & -4.53(0.03) & A & - &- \\ 
HD89449$^\dagger$ & 50564 & F6IV & 3332 & 43 & 112.2 & 24.4 & 5.2 & 206.0 & 12.2 & 51.1 & -0.13 & 25.4 & -4.87(0.07) & A & - &- \\ 
HD90905 & 51386 & F5V & 1109 & 19 & 116.6 & 28.9 & 4.4 & 110.3 & 11.5 & 33.2 & -0.53 & 15.3 & -4.43(0.04) & A & - &- \\ 
HD94765 & 53486 & K0V & 1148 & 32 & 65.6 & 17.2 & 1.8 & 49.7 & 5.1 & 10.5 & -0.75 & 10.1 & -4.5(0.03) & A & - &- \\ 
HD97244 & 54688 & A5V & 1150 & 61 & 446.3 & 110.8 & 84.2 & 78962.9 & 207.6 & 9458.8 & - & 122.5 & -4.14(0.04) & P & - &- \\ 
HD102647 & 57632 & A3V & 1146 & 102 & 702.7 & 133.8 & 110.9 & 23552.8 & 250.4 & 2136.4 & - & 180.3 & -3.67(0.02) & P & - &- \\ 
HD105963$^\dagger$ & 59432 & K0V & 1105 & 10 & 7239.8 & 2756.8 & 3.5 & - & - & - & - & 5.6 & - & - & - &SB2 \\ 
HD107146 & 60074 & G2V & 1150 & 30 & 89.9 & 23.4 & 2.5 & 49.2 & 6.5 & 13.7 & -1.18 & 11.2 & -4.39(0.03) & A & - &- \\ 
HD109647 & 61481 & K0V & 1147 & 21 & 62.1 & 19.4 & 1.9 & 39.4 & 5.8 & 9.7 & - & 9.2 & -4.43(0.08) & A & T &- \\ 
HD110463 & 61946 & K3V & 1150 & 21 & 26.6 & 7.9 & 1.7 & 27.5 & 5.0 & 6.8 & -0.76 & 9.2 & -4.54(0.03) & A & - &- \\ 
HD112097 & 62933 & A7III & 3 & 6 & 8778.4 & 3352.6 & 44.7 & 7892.0 & 109.3 & 2480.6 & - & 100.9 & -4.318(0.01) & P & T &SB1 \\ 
HD113337 & 63584 & F6V & 3367 & 304 & 327.4 & 65.2 & 3.3 & 132.8 & 8.2 & 23.6 & - & 13.6 & -4.75(0.06) & - & - &C\footnote{\label{Simon_VIII_2}\cite{Simon_VIII}}\footnote{\label{Simon_X}\cite{Simon_X}} \\ 
HD115383 & 64792 & G0V & 1149 & 38 & 136.4 & 30.6 & 2.6 & 117.5 & 6.0 & 22.7 & -1.21 & 13.2 & -4.49(0.05) & A & - &- \\ 
HD128311 & 71395 & K0V & 1153 & 41 & 107.5 & 29.5 & 1.8 & 55.5 & 5.1 & 10.9 & 0.26 & 10.7 & -4.48(0.04) & - & - &- \\ 
HD131156 & 72659 & G8V & 1154 & 37 & 210.9 & 52.9 & 1.3 & 72.8 & 4.7 & 18.4 & - & 10.9 & -4.41(0.04) & A & T &SB1 \\ 
HD135559 & 74689 & A4V & 1152 & 70 & 2104.0 & 421.5 & 248.1 & 9971236.0 & 626.8 & 1174299.1 & - & 203.0 & -3.87(0.07) & P & - &- \\ 
HD142229 & 77810 & G5V & 1111 & 20 & 234.4 & 65.6 & 3.5 & 57.1 & 9.5 & 16.2 & - & 11.6 & -4.42(0.09) & A & T &SB1 \\ 
HD148387$^\dagger$ & 80331 & G8III & 851 & 87 & 41.1 & 10.2 & 0.9 & 19.8 & 3.4 & 4.3 & -0.44 & 9.9 & -5.17(0.07) & A & - &- \\ 
HD161284 & 86456 & K0V & 1151 & 28 & 35.8 & 8.7 & 1.8 & 36.9 & 5.1 & 9.6 & -0.42 & 9.3 & -4.52(0.04) & A & - &- \\ 
HD166435$^\dagger$ & 88945 & G0V & 1153 & 33 & 185.9 & 49.4 & 2.9 & 125.8 & 7.7 & 33.4 & -1.32 & 13.9 & -4.41(0.03) & A & - &- \\ 
HD171488 & 91043 & G0V & 1151 & 25 & 1104.2 & 413.9 & 20.7 & 1217.2 & 50.8 & 421.6 & -0.87 & 59.3 & -4.14(0.04) & A & - &- \\ 
HD175726 & 92984 & G0V & 1044 & 31 & 154.5 & 47.3 & 6.8 & 151.7 & 17.4 & 37.5 & -1.05 & 19.7 & -4.49(0.04) & A & - &- \\ 
HD186689 & 97229 & A3IV & 3234 & 114 & 603.4 & 116.7 & 20.5 & 2026.0 & 49.4 & 384.4 & - & 48.6 & -4.21(0.07) & - & - &- \\ 
HD186704 & 97255 & G0V & 1042 & 22 & 191.8 & 38.0 & 6.7 & 256.6 & 15.4 & 57.6 & - & 22.8 & -4.38(0.06) & A & T &SB1 \\ 
HD195943 & 101483 & A3IV & 1043 & 82 & 12156.0 & 4363.6 & 56.8 & 7775.5 & 133.2 & 1261.6 & - & 98.3 & -3.66(0.04) & P & - &SB1 \\ 
HD206860 & 107350 & G0V & 1044 & 38 & 182.7 & 52.4 & 3.5 & 184.2 & 8.5 & 51.6 & -0.93 & 17.1 & -4.49(0.04) & A & - &- \\ 
HD211472 & 109926 & K1V & 975 & 30 & 77.2 & 22.3 & 1.9 & 53.3 & 5.4 & 13.7 & -1.12 & 10.5 & -4.47(0.02) & A & - &- \\ 
HD218396 & 114189 & A5V & 809 & 112 & 4510.0 & 1070.1 & 36.3 & 4233.3 & 85.7 & 1034.4 & - & 62.7 & -4.19(0.03) & P & - &- \\ 
HD218738$^\dagger$ & 114379 & G5V & 115 & 6 & 1618.3 & 676.0 & 7.6 & - & - & - & - & -9.3 & - & - & - &SB2 \\ 
HD220140 & 115147 & G9V & 735 & 17 & 629.4 & 201.5 & 5.8 & 501.6 & 14.7 & 138.9 & -1.33 & 25.3 & -4.21(0.03) & A & - &- \\ 
HD245409 & 26335 & K7V & 454 & 5 & 44.8 & 16.4 & 4.6 & 46.2 & 12.3 & 14.2 & 0.57 & 10.6 & -4.67(0.05) & A & - &- \\ 
- & 11437 & K7V & 67 & 11 & 189.2 & 54.6 & 4.8 & 66.5 & 11.1 & 21.3 & -1.82 & 11.1 & -4.41(0.08) & A & - &- \\ 
- & 40774 & G5V & 715 & 10 & 46.5 & 16.7 & 3.0 & 55.1 & 8.0 & 17.5 & -0.02 & 10.6 & -4.42(0.06) & A & - &- \\ 
BD$+$20 1790 & - & K5V & 345 & 10 & 910.6 & 302.3 & 9.0 & 512.0 & 22.1 & 187.1 & -1.54 & 16.3 & -4.1(0.1) & A & - &- \\ 
\end{longtable}
\endgroup
\end{landscape}

\clearpage

\subsection{\harps \ sample}

\setcounter{footnote}{0}
\begingroup
\renewcommand\arraystretch{1.3}
\begin{longtable}{lcccccccc}
\caption{Star characteristics of the $89$ stars in the initial sample of our \harps~YNS RVs survey. The stars excluded from our analysis (see \cite{Grandjean_HARPS}) are flagged  with a dagger($^\dagger$).
Spectral type (ST) were taken from the CDS database at the beginning of the survey, and the  \bv~values are   from the CDS database.
The \vsini~ were computed with \safir~based on the CCF width.
The IR/D column reports whether  an IR excess is reported ($y$) or not ($n$), and whether  a disk has been resolved ($y$) or not ($n$) in the literature.
\label{tab_carac_h} }
 \\ \hline
Name & HIP & ST & \bv & Mass & Age  &\vsini &Rotation          &  IR/D\\ 
  HD/BD/CD & &&&(\Msun) & (\si{\mega\year}) &(\kms) & period ($\si{\day}$)   & \\
\hline \hline
\endfirsthead
\caption{Continued.}\\
 \\ \hline
Name & HIP & ST & \bv & Mass & Age  &\vsini &Rotation          &  IR/D\\ 
  HD/BD/CD & &&&(\Msun) & (\si{\mega\year})  &(\kms) & period ($\si{\day}$)   & \\
 \\ \hline
  \hline
\endhead
\hline
\endfoot
HD105 & 490 &  G0V & 0.600 & 1.1\footnote{\label{Vigan_h}\cite{Vigan}} & $45_{-10}^{+5}$\footnoteref{Vigan_h}  & 10.4 & - & y\footnote{\label{Meyer}\cite{Meyer_105}}/y\footnote{\label{Donaldson}\cite{Donaldson}} \\ 
HD984 & 1134 &  F7V & 0.500 & 1.2\footnoteref{Vigan_h} & $42_{-7}^{+8}$\footnoteref{Vigan_h}  & 27.6 & - & n\footnote{\label{Carpenter_h}\cite{Carpenter}}/- \\ 
HD987 & 1113 &  G8V & 0.756 & 0.98\footnote{\label{Lagrange_h}\cite{Lagrange}} & $30_{-15}^{+15}$\footnote{\label{Weise}\cite{Weise}}  & 6.2 & $3.72\pm0.01$\footnote{\label{Messina}\cite{Messina}} & n\footnote{\label{Rebull}\cite{Rebull}}/- \\ 
HD1466 & 1481 &  F8V & 0.540 & 1.2\footnoteref{Vigan_h} & $45_{-10}^{+5}$\footnoteref{Vigan_h}  & 14.7 & - & y\footnote{\label{Mamajek_1466}\cite{Mamajek_1466}}/- \\ 
HD3221 & 2729 &  K4V & 1.226 & 0.9\footnoteref{Messina} & $45_{-10}^{+5}$\footnoteref{Vigan_h}  & 104.9 & $0.370\pm0.002$\footnoteref{Messina} & -/- \\ 
HD6569 & 5191 &  K1V & 0.830 & 0.8\footnoteref{Vigan_h} & $149_{-49}^{+31}$\footnoteref{Vigan_h}  & 4.6 & $7.13\pm0.05$\footnoteref{Messina} & -/- \\ 
HD7661 & 5938 &  K0V & 0.770 & 0.98\footnote{\label{Sylvano_mass} This work (see \Cref{age_masse})} & $300_{-50}^{+50}$\footnoteref{Weise}  & 4.4 & 7.46\footnote{\label{Wright_prot}\cite{Wright_prot}} & n\footnote{\label{Lawler}\cite{Lawler}}/- \\ 
HD10008 & 7576 &  K0V & 0.803 & 0.8\footnoteref{Vigan_h} & $250_{-50}^{+50}$\footnoteref{Vigan_h}  & 3.5 & $7.15\pm0.10$\footnote{\label{Folsom}\cite{Folsom}} & y\footnote{\label{Plavchan_h}\cite{Plavchan}}/- \\ 
HD16765 & 12530 &  F71V & 0.520 & 1.23\footnoteref{Sylvano_mass} & $45_{-10}^{+5}$\footnoteref{Sylvano_mass}  & 22.2 & - & -/- \\ 
HD17925 & 13402 &  K1V & 0.860 & 0.9\footnoteref{Vigan_h} & $150_{-80}^{+150}$\footnoteref{Vigan_h}  & 4.9 & 6.76\footnoteref{Wright_prot} & y\footnote{\label{Hallenbrand}\cite{Hallenbrand}}/y\footnote{\label{Eiroa}\cite{Eiroa}} \\ 
HD18599 & 13754 &  K2V & 0.880 & 0.91\footnoteref{Sylvano_mass} & $200_{-75}^{+200}$\footnoteref{Sylvano_mass}  & 4.6 & - & -/- \\ 
HD19668 & 14684 &  K0V & 0.780 & 0.9\footnoteref{Vigan_h} & $149_{-49}^{+31}$\footnoteref{Vigan_h}  & 5.6 & $5.46\pm0.08$\footnoteref{Messina} & y\footnoteref{Carpenter_h}/y\footnoteref{Weise} \\ 
HD24916 & 18512 &  K4V & 1.152 & 0.7\footnote{\label{Ammler_h}\cite{Ammler}} & $400_{-200}^{+400}$\footnoteref{Sylvano_mass}  & 3.6 & - & -/- \\ 
HD25457 & 18859 &  F6V & 0.500 & 1.2\footnoteref{Vigan_h} & $149_{-49}^{+31}$\footnoteref{Vigan_h}  & 13.2 & 3.13\footnoteref{Wright_prot} & y\footnote{\label{Zuckerman_h}\cite{Zuckerman}}/- \\ 
HD26923 & 19859 &  G0IV & 0.560 & 1.07\footnoteref{Ammler_h} & $500_{-100}^{+50}$\footnoteref{Sylvano_mass}  & 3.8 & - & n\footnote{\label{McDonald_h}\cite{McDonald}}/- \\ 
HD29391 & 21547 &  F0IV & 0.277 & $1.75^{+0.05}_{-0.05}$\footnote{\label{Simon}\cite{Simon}} & $37_{-9}^{+9}$\footnoteref{Montet}  & 48.6 & - & y\footnote{\label{Patel}\cite{Patel}}/- \\ 
HD30447 & 22226 &  F3V & 0.390 & 1.4\footnoteref{Vigan_h} & $42_{-7}^{+8}$\footnoteref{Vigan_h}  & 58.9 & - & y\footnote{\label{Chen}\cite{Chen}}/y\footnote{\label{Soummer}\cite{Soummer}} \\ 
HD35650 & 25283 &  K6V & 1.311 & 0.7\footnoteref{Messina} & $149_{-49}^{+31}$\footnoteref{Sylvano_mass}  & 4.4 & $9.34\pm0.08$\footnoteref{Messina} & y\footnoteref{Zuckerman_h}/y\footnote{\label{Choquet}\cite{Choquet}} \\ 
HD37572 & 26373 &  K0V & 1.094 & 0.93\footnoteref{Lagrange_h} & $149_{-49}^{+31}$\footnoteref{Vigan_h}  & 7.4 & $4.52\pm0.02$\footnoteref{Messina} & y\footnoteref{Zuckerman_h}/- \\ 
HD39060 & 27321 & A6V & 0.170 & 1.64\footnoteref{Lagrange_h}  & $30$\footnoteref{Lagrange_h} &  130.0 & - & y\footnote{\label{Beta_pic_IR}\cite{Beta_pic_IR}}/y\footnote{\label{Beta_pic_disc}\cite{Beta_pic_disc}} \\ 
HD41593 & 28954 &  K0V & 0.825 & 1.01\footnoteref{Ammler_h} & $329_{-93}^{+93}$\footnoteref{Plavchan_h}  & 4.3 & - & y\footnoteref{McDonald_h}/- \\ 
HD43989 & 30030 &  G0V & 0.570 & 1.1\footnoteref{Vigan_h} & $45_{-10}^{+5}$\footnoteref{Vigan_h}  & 30.0 & 1.15\footnoteref{Wright_prot} & y\footnoteref{Carpenter_h}/- \\ 
HD44627 & 30034 &  K1V & 0.805 & 0.9\footnoteref{Messina} & $30_{-15}^{+15}$\footnoteref{Weise}  & 9.1 & $3.85\pm0.01$\footnoteref{Messina} & -/n\footnoteref{Weise} \\ 
HD45081 & 29964 &  K4V & 1.251 & 0.8\footnoteref{Messina} & $24_{-5}^{+5}$\footnoteref{Sylvano_mass}  & 11.6 & $2.67\pm0.01$\footnoteref{Messina} & n\footnoteref{McDonald_h}\footnoteref{Weise}/- \\ 
HD45270 & 30314 &  G1V & 0.602 & 1.11\footnoteref{Lagrange_h} & $149_{-49}^{+31}$\footnoteref{Vigan_h}  & 13.0 & - & y\footnoteref{Carpenter_h}/- \\ 
HD59967 & 36515 &  G3V & 0.639 & 1.09\footnoteref{Plavchan_h} & $353_{-58}^{+58}$\footnoteref{Plavchan_h}  & 4.2 & - & y\footnoteref{Plavchan_h}/- \\ 
HD61005 & 36948 &  G8V & 0.740 & 1.0\footnoteref{Vigan_h} & $50_{-10}^{+20}$\footnoteref{Vigan_h}  & 6.8 & $5.04\pm0.03$\footnote{\label{Folsom}\cite{Folsom}} & y\footnoteref{Meyer_excess}/y\footnote{\label{Hines}\cite{Hines}} \\ 
HD63608 & 37923 &  K0V & 0.830 & 1.0\footnoteref{Vigan_h} & $250_{-50}^{+50}$\footnoteref{Vigan_h}  & 3.7 & - & -/- \\ 
HD77825 & 44526 &  K2V & 0.992 & 0.8\footnoteref{Vigan_h} & $350_{-150}^{+150}$\footnoteref{Vigan_h}  & 4.6 & 8.64\footnote{\label{Kiraga}\cite{Kiraga}} & -/- \\ 
HD82558 & 46816 &  K1V & 0.870 & 0.8\footnote{\label{Kovari}\cite{Kovari}} & $84_{-59}^{+64}$\footnote{\label{Brandt_2014}\cite{Brandt_2014}}  & 19.3 & 1.70\footnoteref{Wright_prot} & -/- \\ 
HD89449$^\dagger$  & 50564 & F6IV &  0.440 & 1.44\footnote{\label{Simon_X_h}\cite{Simon_X}} & $3100$\footnote{\label{Gaspar_h}\cite{Gaspar}} &  12.0 & - & -/- \\ 
HD90905 & 51386 &  F5V & 0.569 & 1.16\footnoteref{Lagrange_h} & $170_{-70}^{+180}$\footnoteref{Vigan_h}  & 7.1 & 2.60\footnoteref{Wright_prot} & y\footnoteref{Carpenter_h}/y\footnote{\label{Morales}\cite{Morales}} \\ 
HD92945 & 52462 &  K1V & 0.877 & 0.9\footnoteref{Vigan_h} & $170_{-70}^{+130}$\footnoteref{Vigan_h}  & 4.9 & - & n\footnote{\label{Chen_92945}\cite{Chen_92945}}/y\footnote{\label{Golimowski}\cite{Golimowski}} \\ 
\pagebreak 
HD95086 & 53524 &  A8III & 0.240 & 1.6\footnote{\label{Rameau_95086}\cite{Rameau_95086}} & $17_{-4}^{+4}$\footnote{\label{Meshkat_2013}\cite{Meshkat_2013}}  & 20.7 & - & y\footnote{\label{Rhee}\cite{Rhee}}/y\footnote{\label{Moor}\cite{Moor}} \\ 
HD95650 & 53985 &  M2V & 1.477 & 0.59\footnote{\label{Montet}\cite{Montet_14}} & $400_{-200}^{+150}$\footnoteref{Sylvano_mass}  & 3.7 & 14.80\footnote{\label{Kiraga}\cite{Kiraga}} & -/- \\ 
HD99211  & 55705 & A7V &  0.210 & 1.83\footnoteref{Gaspar_h} & $570$\footnoteref{Gaspar_h} &  86.4 & - & y\footnote{\label{Mannings}\cite{Mannings}}/- \\ 
HD102458 & 57524 &  G4V & 0.630 & 1.70\footnote{\label{Reza}\cite{reza}} & $16_{-4}^{+4}$\footnoteref{Sylvano_mass}  & 19.2 & - & y\footnoteref{Rebull}/- \\ 
HD103743 & 58241 &  G4V & 0.670 & 1.05\footnoteref{Sylvano_mass} & $200_{-50}^{+100}$\footnoteref{Sylvano_mass}  & 7.0 & - & -/- \\ 
HD105690 & 59315 &  G5V & 0.661 & 1.02\footnoteref{Lagrange_h} & $8_{-8}^{+15}$\footnoteref{Weise}  & 7.0 & - & -/- \\ 
HD106906$^\dagger$ & 59960 &  F5V & 0.400 & 1.5\footnote{\label{Pecaut_2012}\cite{Pecaut_2012}} & $13_{-2}^{+2}$\footnoteref{Pecaut_2012}  & 34.0 & - & y\footnote{\label{Sierchio}\cite{Sierchio}}/y\footnote{\label{Kalas_2015}\cite{Kalas_2015}} \\ 
HD108767  & 60965 & K0V &  -0.050 & $2.74^{+0.07}_{-0.06}$\footnote{\label{Montesinos}\cite{Montesinos}} & $260$\footnoteref{Sylvano_mass} &  3.8 & - & -/- \\ 
HD116434$^\dagger$ & 65426 & A2V &  0.100 & $1.96^{+0.04}_{-0.04}$\footnote{\label{Chauvin}\cite{Chauvin}} & $14_{-4}^{+4}$\footnoteref{Chauvin} &  300.0 & - & n\footnoteref{Chauvin}/- \\ 
HD118100 & 66252 &  K5V & 1.180 & 0.7\footnoteref{Vigan_h} & $150_{-50}^{+50}$\footnoteref{Vigan_h}  & 7.8 & 3.96\footnoteref{Wright_prot} & -/- \\ 
HD131399$^\dagger$ & 72940 &  A1V & 0.110 & $2.08^{+0.12}_{-0.11}$\footnote{\label{Nielsen}\cite{Nielsen}} & $21_{-3}^{+4}$\footnoteref{Nielsen}  & 19.3 & - & -/- \\ 
HD141943 & - &  G2V & 0.505 & 1.09\footnoteref{Lagrange_h} & $30_{-15}^{+15}$\footnoteref{Weise}  & 24.4 & 2.2\footnoteref{Kiraga} & -/- \\ 
HD146464 & 79958 &  K3V & 1.033 & 0.80\footnoteref{Sylvano_mass} & $130_{-50}^{+170}$\footnoteref{Sylvano_mass}  & 11.9 & 2.329\footnote{\label{Koen}\cite{Koen}} & y\footnote{\label{Cotten}\cite{Cotten}}/y\footnoteref{Soummer} \\ 
HD146624  & 79881 & A0V &  0.022 & 1.9\footnoteref{Lagrange_h} & $12$\footnoteref{Lagrange_h} &  27.8 & - & -/n\footnoteref{Lagrange_h} \\ 
HD152555 & 82688 &  F8V & 0.600 & 1.13\footnoteref{Sylvano_mass} & $130_{-19}^{+18}$\footnoteref{Brandt_2014}  & 12.2 & 2.77\footnoteref{Wright_prot} & n\footnoteref{Zuckerman_h}/- \\ 
HD159492 & 86305 &  A5IV & 0.189 & 1.70\footnoteref{Sylvano_mass} & $600_{-220}^{+220}$\footnoteref{Sylvano_mass}  & 36.4 & - & -/y\footnote{\label{Morales}\cite{Morales}} \\ 
HD164249 & 88399 &  F6V & 0.431 & 1.3\footnoteref{Vigan_h} & $24_{-5}^{+5}$\footnoteref{Vigan_h}  & 13.6 & - & y\footnoteref{Chen}/- \\ 
HD169178 & - &  K0V & 0.850 & 0.94\footnoteref{Sylvano_mass} & $131_{-65}^{+131}$\footnoteref{Sylvano_mass}  & 5.1 & - & -/- \\ 
HD171488 & 91043 &  G2V & 0.551 & 1.1\footnoteref{Vigan_h} & $45_{-25}^{+35}$\footnoteref{Vigan_h}  & 26.5 & $1.3371\pm0.0002$\footnote{\label{Strassmeier}\cite{Strassmeier}} & y\footnoteref{McDonald_h}/- \\ 
HD172555  & 92024 & A7V &  0.200 & 1.61\footnoteref{Lagrange_h} & $12$\footnote{\label{Messina_2017}\cite{Messina_2017}} &  78.3 & - & y\footnoteref{Lagrange_h}/y\footnote{\label{Smith}\cite{Smith}} \\ 
HD174429 & 92680 &  G9IV & 0.878 & 1.2\footnoteref{Vigan_h} & $24_{-5}^{+5}$\footnoteref{Vigan_h}  & 54.1 & $0.944\pm0.001$\footnoteref{Messina_2017} & y\footnoteref{Chen}/- \\ 
HD177171$^\dagger$ & 93815 & F6V & 0.526 & 1.25\footnoteref{Reza} & $30$\footnoteref{Lagrange_h} &  300.0 & 4.737\footnoteref{Koen} & -/- \\ 
HD181321$^\dagger$ & 95149 &  G2V & 0.630 & 1.01\footnote{\label{Fuhrmann}\cite{Fuhrmann}} & $320_{-120}^{+180}$\footnoteref{Sylvano_mass}  & 9.1 & 5.7\footnote{\label{Olmedo}\cite{Olmedo}} & -/- \\ 
HD181327 & 95270 &  F6V & 0.460 & 1.36\footnoteref{Lagrange_h} & $24_{-5}^{+5}$\footnoteref{Vigan_h}  & 13.4 & - & y\footnoteref{Mannings}/y\footnote{\label{Schneider}\cite{Schneider}} \\ 
HD183414 & 96334 &  G3V & 0.647 & 1.04\footnoteref{Lagrange_h} & $150_{-80}^{+70}$\footnoteref{Vigan_h}  & 7.5 & 3.924\footnoteref{Koen} & -/- \\ 
HD186704  & 97255 & G0V &  0.583 & 1.10\footnote{\label{Bonavita_h}\cite{Bonavita}} & $100$\footnote{\label{Zuckerman_13_h}\cite{Zuckerman_2013}} &  10.3 & 3.511\footnoteref{Kiraga} & y\footnoteref{McDonald_h}/- \\ 
HD188228  & 98495 & A0V &  -0.013 & 2.03\footnoteref{Lagrange_h} & $40$\footnoteref{Lagrange_h} &  65.3 & - & n\footnoteref{McDonald_h}/y\footnote{\label{Booth}\cite{Booth}} \\ 
HD189245 & 98470 &  F7V & 0.490 & 1.2\footnoteref{Vigan_h} & $150_{-50}^{+150}$\footnoteref{Vigan_h}  & 53.3 & $1.88\pm0.01$\footnote{\label{Desidera}\cite{Desidera}} & -/- \\ 
HD191089 & 99273 &  F5V & 0.440 & 1.3\footnoteref{Vigan_h} & $24_{-5}^{+5}$\footnoteref{Weise}  & 26.4 & $0.488\pm0.005$\footnoteref{Desidera} & y\footnoteref{Mannings}/y\footnote{\label{Churcher}\cite{Churcher}} \\ 
HD197481 & 102409 &  M1V & 1.423 & 0.6\footnoteref{Messina} & $21_{-5}^{+7}$\footnoteref{Brandt_2014}  & 7.6 & $4.84\pm0.02$\footnoteref{Messina} & y\footnote{\label{Rameau}\cite{Rameau}}/y\footnote{\label{Kalas}\cite{Kalas}} \\ 
HD197890$^\dagger$ & 102626 &  K3V & 1.053 & 1.0\footnoteref{Vigan_h} & $45_{-35}^{+55}$\footnoteref{Vigan_h}  & 107.6 & 0.3804\footnoteref{Kiraga} & y\footnoteref{McDonald_h}/- \\ 
HD202917 & 105388 &  G7V & 0.650 & 0.9\footnoteref{Vigan_h} & $45_{-10}^{+5}$\footnoteref{Vigan_h}  & 10.6 & $3.36\pm0.01$\footnoteref{Messina} & y\footnoteref{Patel}/y\footnoteref{Soummer} \\ 
HD206860 & 107350 &  G0V & 0.580 & 1.1\footnoteref{Vigan_h} & $300_{-100}^{+700}$\footnoteref{Vigan_h}  & 7.5 & 4.86\footnoteref{Wright_prot} & y\footnote{\label{Beichman_2006_h}\cite{Beichman_2006}}\footnote{\label{Moro-Martin}\cite{Moro-Martin}}/- \\ 
HD206893 & 107412 &  F5V & 0.440 & 1.24\footnote{\label{Milli}\cite{Milli}} & $250_{-200}^{+450}$\footnote{\label{Delorme}\cite{Delorme}}  & 23.2 & - & y\footnoteref{Sierchio}/y\footnoteref{Milli} \\ 
HD207575 & 107947 &  F6V & 0.490 & 1.24\footnoteref{Lagrange_h} & $45_{-15}^{+5}$\footnoteref{Vigan_h}  & 22.7 & - & y\footnoteref{Zuckerman_h}/- \\ 
HD213845 & 111449 &  F7V & 0.440 & 1.4\footnoteref{Vigan_h} & $250_{-50}^{+750}$\footnoteref{Vigan_h}  & 24.2 & - & y\footnote{\label{Beichman}\cite{Beichman}}/- \\ 
\pagebreak 
HD215641 & 112491 &  G8V & 0.760 & 1.00\footnoteref{Sylvano_mass} & $440_{-40}^{+40}$\footnoteref{Brandt_2014}  & 3.9 & - & -/- \\ 
HD216956 & 113368 & A3V & 0.090 & $1.92^{+0.02}_{-0.02}$\footnote{\label{Mamajek}\cite{Mamajek}} & $440_{-40}^{+40}$\footnoteref{Mamajek}  & 90.0 & - & y\footnote{\label{Backman}\cite{Backman}}/y\footnote{\label{Holland}\cite{Holland}} \\ 
HD217343  & 113579 & G5V &  0.640 & 1.05\footnoteref{Vigan_h} & $70$\footnoteref{Lagrange_h} &  9.1 & - & -/- \\ 
HD217987 & 114046 & M2V & 1.480 & 0.47\footnoteref{Montet}& $100-10000$\footnote{\label{Delorme_2012}\cite{Delorme_2012}} &   2.3 & - & -/- \\ 
HD218396  & 114189 & A5V &  0.257 & 1.46\footnoteref{Vigan_h} & $42$\footnoteref{Vigan_h} &  26.6 & - & y\footnote{\label{Zuckerman_8799}\cite{Zuckerman_8799}}/y\footnote{\label{Su_8799}\cite{Su_8799}} \\ 
HD218860 & 114530 &  G8V & 0.738 & 1.0\footnoteref{Messina} & $149_{-49}^{+31}$\footnoteref{Sylvano_mass}  & 5.5 & $5.17\pm0.02$\footnoteref{Messina} & y\footnoteref{Zuckerman_h}/- \\ 
HD221575 & 116258 &  K2V & 0.930 & 0.90\footnoteref{Sylvano_mass} & $250_{-100}^{+150}$\footnoteref{Sylvano_mass}  & 4.1 & - & -/- \\ 
HD223340 & - &  K1V & 0.820 & 0.91\footnoteref{Sylvano_mass} & $149_{-49}^{+31}$\footnoteref{Sylvano_mass}  & 4.6 & - & -/- \\ 
HD224228 & 118008 &  K2V & 0.985 & 0.86\footnoteref{Lagrange_h} & $149_{-49}^{+31}$\footnoteref{Vigan_h}  & 3.6 & - & y\footnoteref{Zuckerman_h}/- \\ 
- & 6276 &  G9V & 0.800 & 0.9\footnoteref{Vigan_h} & $149_{-49}^{+31}$\footnoteref{Vigan_h}  & 4.4 & 6.40\footnoteref{Wright_prot} & y\footnoteref{Sierchio}/- \\ 
- & 116384 &  K7V & 1.347 & 0.729\footnote{\label{Newton}\cite{Newton}} & $289_{-260}^{+971}$\footnoteref{Brandt_2014}  & 4.0 & - & -/- \\ 
- & 17157 &  K7V & 1.300 & 0.74\footnoteref{Sylvano_mass} & $100_{-30}^{+50}$\footnoteref{Sylvano_mass}  & 3.7 & - & -/- \\ 
- & 23309 &  M0V & 1.383 & 0.55\footnoteref{Reza} & $10_{-3}^{+3}$\footnoteref{Weise}  & 5.5 & $8.60\pm0.07$\footnoteref{Messina} & n\footnoteref{Rebull}/- \\ 
- & 31878 &  K7V & 1.297 & 0.643\footnote{\label{Lindgren}\cite{Lindgren}} & $149_{-49}^{+31}$\footnoteref{Sylvano_mass}  & 4.0 & $9.06\pm0.08$\footnoteref{Messina} & -/- \\ 
- & 36985 &  M2V & 1.476 & 0.621\footnote{\label{Carmenes}\cite{Carmenes}} & $260_{-260}^{+420}$\footnote{\label{Poveda}\cite{Poveda}}  & 3.6 & 12.16\footnoteref{Kiraga} & -/- \\ 
- & 44722 &  K7V & 1.450 & 0.638\footnoteref{Lindgren} & $600_{-500}^{+200}$\footnoteref{Sylvano_mass}  & 3.5 & - & -/- \\ 
- & 46634 &  G5V & 1.180 & 0.93\footnoteref{Sylvano_mass} & $300_{-150}^{+300}$\footnoteref{Sylvano_mass}  & 4.6 & $3.05\pm0.03$\footnoteref{Messina} & -/- \\ 
- & 51317 &  M2V & 1.501 & 0.44\footnoteref{Sylvano_mass} & $130_{-20}^{+40}$\footnoteref{Brandt_2014}  & 1.7 & - & -/- \\ 
BD$+$20 2465$^\dagger$ & - &  M5V & 1.300 & $0.42^{+0.01}_{-0.01}$\footnote{\label{Morin}\cite{Morin}} & $50_{-30}^{+150}$\footnoteref{Sylvano_mass}  & 2.8 & 2.60\footnoteref{Wright_prot} & -/- \\ 
CD$-$46 1064 & - &  K3V & 1.048 & 0.8\footnoteref{Vigan_h} & $45_{-10}^{+5}$\footnoteref{Vigan_h}  & 8.1 & $3.74\pm0.04$\footnoteref{Messina} & -/- \\ 
\end{longtable}
\endgroup

\setcounter{footnote}{0}
\begin{landscape}
\begingroup
\begin{longtable}{lcc|cccccccccccccc}
\caption{Results for the $89$ stars in the initial sample of our  \harps~YNS RV survey before any correction. The stars excluded from our analysis (see \cite{Grandjean_HARPS}) are flagged  with a dagger($^\dagger$).
Spectral type (ST)  were taken from the CDS database at the beginning of the survey.
The table provides the time baseline (TBL); the number of computed spectra $N_{\rm m}$; the amplitude corresponding to the difference between the maximum and the minimum of the RVs({\it A}), rms, and mean uncertainty $<${\it U}$>$ on the RV and BVS measurements; the RV-BVS correlation factor (slope of the best linear fit); the mean FWHM ($<$FWHM$>$);  the mean \rhk \ ($<$\rhk$>$); and its standard deviation $\sigma_{logR'_{\rm HK}}$.
The V column presents, if identified,   the source of stellar jitter: A stands for stellar activity (spots) and P stands  for pulsations.
The T column presents the stars with a long-term trend induced by a companion.
The B column presents the nature of the companion:  C stands for sub-stellar companion, SB1 for stands for single-line spectroscopic binary, and SB2 stands for double-line spectroscopic binary.
\label{tab_result_h}}\\ 
\\ \hline
\multicolumn{3}{c|}{Stellar characteristics} & \multicolumn{12}{c}{Survey results.}
 \\ \hline
Name & HIP & ST &  TBL & $N_{\rm m}$   & \multicolumn{3}{c}{RV}        & \multicolumn{3}{c}{BVS}      & RV-    & $<$FWHM$>$ & $<$\rhk$>$  & V   & T & B\\ 
 HD/BD/CD & &  & & & \multicolumn{3}{c}{\raisebox{.5\baselineskip}{$\overbrace{\hspace{2.7cm}}$}} & \multicolumn{3}{c}{\raisebox{.5\baselineskip}{$\overbrace{\hspace{2.7cm}}$}} & BVS & & ($\sigma_{logR'_{\rm HK}}$) & & &\\
&& &  &  & {\it A} &rms  & $<${\it U}$>$ & {\it A} & rms &$<${\it U}$>$ & corr.    &                 &  & & &  \\
&&&&    & \multicolumn{3}{c}{\raisebox{.5\baselineskip}{$\underbrace{\hspace{2.7cm}}$}} & \multicolumn{3}{c}{\raisebox{.5\baselineskip}{$\underbrace{\hspace{2.7cm}}$}} & &  & & &\\
&& (day)&            && \multicolumn{3}{c}{($\si{\meter\per\second}$)}      & \multicolumn{3}{c}{($\si{\meter\per\second}$)}     &   &      ($\si{\kilo\meter\per\second}$) &    & & &\\ \hline
\endfirsthead
\caption{Continued.}\\
 \\ \hline
\multicolumn{3}{c|}{Stellar characteristics} & \multicolumn{12}{c}{Survey results.}
 \\ \hline
Name & HIP & ST &  TBL & $N_{\rm m}$   & \multicolumn{3}{c}{RV}        & \multicolumn{3}{c}{BVS}      & RV-    & $<$FWHM$>$ & $<$\rhk$>$ & V  & T & V \\ 
 HD/BD/CD & &  & & & \multicolumn{3}{c}{\raisebox{.5\baselineskip}{$\overbrace{\hspace{2.7cm}}$}} & \multicolumn{3}{c}{\raisebox{.5\baselineskip}{$\overbrace{\hspace{2.7cm}}$}} & BVS & & ($\sigma_{logR'_{\rm HK}}$) & & &\\
&& &  &  & {\it A} &rms  & $<${\it U}$>$ & {\it A} & rms &$<${\it U}$>$ & corr.    &                 &    & & &\\
&&&&    & \multicolumn{3}{c}{\raisebox{.5\baselineskip}{$\underbrace{\hspace{2.7cm}}$}} & \multicolumn{3}{c}{\raisebox{.5\baselineskip}{$\underbrace{\hspace{2.7cm}}$}} & & & & & \\
&& (day)&            && \multicolumn{3}{c}{($\si{\meter\per\second}$)}      & \multicolumn{3}{c}{($\si{\meter\per\second}$)}     &   &      ($\si{\kilo\meter\per\second}$) &   & & &\\ \hline
\endhead
\hline
\endfoot
HD105 & 490 & G0V & 4606 & 36 & 236.8 & 61.1 & 4.2 & 310.7 & 9.8 & 72.9 & -0.67 & 22.6 & -4.32(0.02) & A & - &- \\ 
HD984 & 1134 & F7V & 867 & 21 & 301.9 & 84.6 & 16.4 & 571.8 & 38.2 & 137.4 & -0.46 & 59.7 & -4.37(0.02) & A & - &- \\ 
HD987 & 1113 & G8V & 2621 & 19 & 502.6 & 116.8 & 2.5 & 393.6 & 6.4 & 103.1 & -1.08 & 13.0 & -4.09(0.04) & A & - &- \\ 
HD1466 & 1481 & F8V & 4400 & 19 & 135.8 & 39.6 & 7.4 & 189.8 & 17.6 & 55.9 & -0.66 & 32.3 & -4.34(0.02) & A & - &- \\ 
HD3221 & 2729 & K4V & 4014 & 5 & 4793.4 & 1794.3 & 68.6 & 3115.1 & 172.5 & 1447.8 & -0.54 & 203.6 & -4.47(0.01) & - & - &- \\ 
HD6569 & 5191 & K1V & 8 & 4 & 24.0 & 11.4 & 1.5 & 17.8 & 3.8 & 7.5 & -1.49 & 9.6 & -4.229(0.003) & A & - &- \\ 
HD7661 & 5938 & K0V & 1525 & 29 & 96.0 & 28.1 & 1.2 & 50.5 & 3.3 & 12.5 & -1.42 & 9.2 & -4.33(0.05) & A & - &- \\ 
HD10008 & 7576 & K0V & 4021 & 17 & 20.5 & 6.2 & 1.0 & 17.5 & 2.7 & 5.2 & 0.44 & 7.5 & -4.36(0.03) & A & - &- \\ 
HD16765 & 12530 & F71V & 926 & 27 & 173.0 & 47.2 & 11.0 & 402.9 & 27.9 & 93.1 & -0.42 & 48.5 & -4.48(0.02) & A & - &- \\ 
HD17925 & 13402 & K1V & 1470 & 40 & 113.6 & 30.5 & 1.1 & 58.7 & 3.1 & 17.4 & -0.87 & 10.0 & -4.31(0.02) & A & - &- \\ 
HD18599 & 13754 & K2V & 1055 & 16 & 115.9 & 38.9 & 1.5 & 54.3 & 3.9 & 17.6 & -1.46 & 9.8 & -4.31(0.02) & A & - &- \\ 
HD19668 & 14684 & K0V & 4402 & 20 & 143.1 & 33.2 & 1.8 & 91.9 & 4.7 & 26.0 & -0.95 & 11.8 & -4.28(0.03) & A & - &- \\ 
HD24916 & 18512 & K4V & 792 & 22 & 24.3 & 7.4 & 0.8 & 36.7 & 2.2 & 11.6 & 0.46 & 7.8 & -4.54(0.04) & A & - &- \\ 
HD25457 & 18859 & F6V & 4089 & 78 & 187.2 & 49.5 & 3.3 & 345.4 & 7.7 & 60.2 & -0.66 & 28.2 & -4.36(0.04) & A & - &- \\ 
HD26923 & 19859 & G0IV & 4010 & 47 & 50.2 & 12.7 & 1.1 & 24.7 & 3.1 & 5.6 & 1.12 & 8.0 & -4.49(0.05) & A & - &- \\ 
HD29391 & 21547 & F0IV & 3996 & 81 & 1476.1 & 326.2 & 28.4 & 2564.3 & 72.1 & 487.8 & - & 101.4 & -4.36(0.01) & P & - &- \\ 
HD30447 & 22226 & F3V & 792 & 19 & 194.1 & 47.2 & 47.7 & 7048.7 & 106.6 & 1728.3 & - & 120.7 & -4.45(0.02) & P & - &- \\ 
HD35650 & 25283 & K6V & 1195 & 13 & 48.3 & 15.3 & 1.2 & 43.6 & 3.0 & 12.3 & -0.67 & 9.4 & -4.36(0.04) & A & - &- \\ 
HD37572 & 26373 & K0V & 2826 & 34 & 347.1 & 75.4 & 1.7 & 223.6 & 4.4 & 52.8 & -1.28 & 15.6 & -4.17(0.02) & A & - &- \\ 
HD39060 & 27321 & A6V & 3702 & 5108 & 755.5 & 341.4 & 58.9 & - & - & - & - & - & - & P & - & - \\ 
HD41593 & 28954 & K0V & 1194 & 15 & 69.5 & 19.8 & 1.1 & 59.4 & 2.9 & 16.7 & -0.66 & 9.1 & -4.37(0.03) & A & - &- \\ 
HD43989 & 30030 & G0V & 4433 & 17 & 819.0 & 238.6 & 15.6 & 743.7 & 34.3 & 197.6 & -0.83 & 64.2 & -4.21(0.06) & A & - &- \\ 
HD44627 & 30034 & K1V & 4892 & 23 & 670.6 & 177.7 & 3.2 & 428.6 & 8.1 & 110.9 & - & 19.1 & -4.02(0.03) & A & - &- \\ 
HD45081 & 29964 & K4V & 4020 & 17 & 852.4 & 252.1 & 5.0 & 911.0 & 13.0 & 246.9 & - & 24.8 & -3.89(0.07) & A & - &- \\ 
HD45270 & 30314 & G1V & 2827 & 19 & 137.4 & 51.8 & 3.1 & 193.3 & 8.4 & 59.2 & -0.76 & 27.6 & -4.34(0.03) & A & - &- \\ 
HD59967 & 36515 & G3V & 1065 & 25 & 65.3 & 19.7 & 1.4 & 41.6 & 3.6 & 10.5 & -0.99 & 9.0 & -4.33(0.03) & A & - &- \\ 
HD61005 & 36948 & G8V & 2369 & 33 & 186.5 & 48.8 & 2.5 & 127.4 & 6.3 & 32.7 & -1.16 & 14.5 & -4.26(0.04) & A & - &- \\ 
HD63608 & 37923 & K0V & 1069 & 36 & 80.7 & 22.2 & 1.1 & 49.1 & 3.0 & 11.7 & -0.63 & 8.0 & -4.32(0.05) & A & - &- \\ 
HD77825 & 44526 & K2V & 283 & 6 & 47.0 & 18.3 & 1.4 & 46.5 & 3.4 & 14.9 & -1.09 & 9.6 & -4.31(0.03) & A & - &- \\ 
HD82558 & 46816 & K1V & 1092 & 24 & 296.0 & 79.7 & 7.1 & 475.6 & 17.5 & 124.1 & -0.43 & 42.0 & -3.99(0.06) & A & - &- \\ 
HD89449$^\dagger$ & 50564 & F6IV & 1227 & 28 & 129.9 & 41.4 & 4.3 & 351.6 & 9.1 & 113.6 & -0.26 & 25.3 & -4.97(0.11) & A & - &- \\ 
HD90905 & 51386 & F5V & 2551 & 24 & 125.6 & 40.8 & 1.9 & 187.7 & 5.0 & 49.0 & -0.74 & 15.0 & -4.35(0.04) & A & - &- \\ 
HD92945 & 52462 & K1V & 4533 & 38 & 182.3 & 40.5 & 1.5 & 86.2 & 3.9 & 23.8 & -1.37 & 10.3 & -4.26(0.03) & A & - &- \\ 
HD95086 & 53524 & A8III & 1532 & 103 & 1279.0 & 268.8 & 14.8 & 2594.0 & 36.3 & 511.1 & - & 45.4 & -4.47(0.03) & P & - &- \\ 
HD95650 & 53985 & M2V & 4046 & 12 & 23.8 & 7.6 & 3.1 & 46.1 & 7.3 & 13.6 & - & 8.0 & -4.33(0.04) & - & - &- \\ 
HD99211 & 55705 & A7V & 3068 & 112 & 818.0 & 149.4 & 86.8 & 49704.0 & 214.7 & 5222.3 & - & 183.4 & -4.243(0.008) & P & - &- \\ 
HD102458 & 57524 & G4V & 2964 & 26 & 988.5 & 307.7 & 10.0 & 1358.4 & 23.8 & 431.7 & -0.79 & 43.6 & -3.99(0.02) & A & - &- \\ 
HD103743 & 58241 & G4V & 1065 & 30 & 102.2 & 32.1 & 2.6 & 141.6 & 6.6 & 35.1 & -0.79 & 14.9 & -4.31(0.03) & A & - &- \\ 
HD105690 & 59315 & G5V & 2975 & 133 & 269.4 & 61.3 & 2.3 & 193.7 & 5.4 & 44.0 & -0.91 & 15.1 & -4.28(0.03) & A & - &- \\ 
HD106906$^\dagger$ & 59960 & F5V & 1230 & 46 & 5290.4 & 1285.7 & 32.9 & 9949.8 & 83.3 & 2272.4 & - & 74.0 & -4.57(0.02) & A & - &SB1 \\ 
HD108767 & 60965 & K0V & 1201 & 18 & 47.9 & 12.4 & 0.7 & 25.3 & 2.0 & 7.7 & 0.21 & 8.2 & -4.38(0.01) & A & - &- \\ 
HD116434$^\dagger$ & 65426 & A2V & 439 & 58 & 6273.2 & 1223.2 & 1235.6 & - & - & - & - & - & - & P & - & - \\ 
HD118100 & 66252 & K5V & 720 & 10 & 534.7 & 155.6 & 6.4 & 139.5 & 15.6 & 43.0 & -3.48 & 16.5 & -4.11(0.03) & A & - &- \\ 
HD131399$^\dagger$ & 72940 & A1V & 189 & 87 & 19478.2 & 6381.7 & 16.4 & 1287.6 & 39.3 & 237.0 & - & 43.1 & -3.85(0.01) & - & - &SB1 \\ 
HD141943 & - & G2V & 2648 & 58 & 861.4 & 222.8 & 10.3 & 1169.7 & 25.3 & 269.6 & -0.66 & 53.0 & -4.03(0.02) & A & - &- \\ 
HD146464 & 79958 & K3V & 51 & 5 & 872.1 & 352.1 & 5.0 & 546.2 & 12.3 & 220.6 & -1.55 & 25.8 & -4.16(0.1) & A & - &- \\ 
HD146624 & 79881 & A0V & 4762 & 335 & 232.7 & 39.8 & 29.2 & 70472.6 & 72.9 & 6163.4 & - & 65.1 & -3.429(0.004) & P & - &- \\ 
HD152555 & 82688 & F8V & 1135 & 22 & 225.0 & 64.8 & 6.1 & 283.2 & 15.9 & 69.9 & -0.74 & 26.3 & -4.29(0.02) & A & - &- \\ 
HD159492 & 86305 & A5IV & 4751 & 90 & 612.4 & 124.9 & 20.8 & 2413.2 & 52.1 & 359.7 & - & 79.4 & -4.45(0.01) & P & - &- \\ 
HD164249 & 88399 & F6V & 1113 & 25 & 56.9 & 14.6 & 6.9 & 134.2 & 17.0 & 36.7 & -0.24 & 29.8 & -4.77(0.07) & A & - &- \\ 
HD169178 & - & K0V & 1123 & 19 & 354.7 & 135.7 & 1.8 & 85.6 & 4.7 & 25.0 & - & 10.7 & -4.23(0.03) & A & - &- \\ 
HD171488 & 91043 & G2V & 1111 & 18 & 2190.4 & 717.7 & 13.1 & 2249.5 & 33.2 & 680.4 & -1.0 & 58.2 & -3.99(0.03) & A & - &- \\ 
HD172555 & 92024 & A7V & 2975 & 262 & 1883.8 & 329.6 & 70.6 & 94415.3 & 185.8 & 6148.5 & - & 172.4 & -4.167(0.007) & P & - &- \\ 
HD174429 & 92680 & G9IV & 4572 & 42 & 3362.3 & 970.2 & 26.0 & 1167.7 & 60.0 & 320.7 & -1.77 & 109.9 & - & A & - &- \\ 
HD177171$^\dagger$ & 93815 & F6V & 144 & 21 & 21233.2 & 5931.1 & 53.4 & - & - & - & - & - & -4.05(0.04) & - & - & SB1 \\ 
HD181321$^\dagger$ & 95149 & G2V & 3757 & 28 & 2610.2 & 683.2 & 3.9 & 244.0 & 9.6 & 59.1 & - & 19.5 & -4.24(0.02) & A & - &SB1 \\ 
HD181327 & 95270 & F6V & 3496 & 56 & 63.1 & 16.3 & 3.8 & 156.8 & 9.0 & 30.6 & -0.26 & 28.8 & -4.58(0.03) & A & - &- \\ 
HD183414 & 96334 & G3V & 3097 & 68 & 247.9 & 63.6 & 2.4 & 214.0 & 6.2 & 58.3 & -1.01 & 16.0 & -4.24(0.02) & A & - &- \\ 
HD186704 & 97255 & G0V & 451 & 4 & 345.8 & 168.8 & 4.7 & 77.1 & 11.9 & 29.3 & - & 22.3 & -4.309(0.008) & - & T &SB1 \\ 
HD188228 & 98495 & A0V & 4315 & 194 & 1398.7 & 319.2 & 150.2 & 211414.9 & 368.1 & 18294.0 & - & 148.0 & -2.928(0.003) & P & - &- \\ 
HD189245 & 98470 & F7V & 4285 & 46 & 646.2 & 119.4 & 21.0 & 5149.4 & 43.1 & 757.5 & - & 104.2 & -4.3(0.02) & P & - &- \\ 
HD191089 & 99273 & F5V & 1025 & 26 & 107.1 & 22.8 & 15.7 & 300.0 & 37.3 & 78.4 & - & 57.0 & -4.53(0.01) & P & - &- \\ 
HD197481 & 102409 & M1V & 5619 & 55 & 667.6 & 144.0 & 3.2 & 470.1 & 8.0 & 101.2 & -1.37 & 16.2 & -4.29(0.23) & A & - &- \\ 
HD197890$^\dagger$ & 102626 & K3V & 13 & 3 & 922.1 & 376.6 & 85.2 & 3975.6 & 217.1 & 1713.8 & - & 216.2 & -4.28(0.03) & - & - &- \\ 
HD202917 & 105388 & G7V & 4438 & 20 & 436.1 & 124.6 & 5.0 & 438.1 & 12.4 & 134.8 & -0.84 & 23.0 & -3.98(0.04) & A & - &- \\ 
HD206860 & 107350 & G0V & 1332 & 22 & 122.1 & 35.8 & 2.8 & 180.9 & 6.7 & 51.4 & - & 16.1 & -4.39(0.03) & A & - &- \\ 
\pagebreak
HD206893 & 107412 & F5V & 941 & 78 & 297.4 & 87.7 & 10.0 & 357.5 & 27.6 & 85.4 & - & 50.7 & -4.63(0.02) & P & T &C\footnote{\label{Grandjean}\cite{Grandjean}} \\ 
HD207575 & 107947 & F6V & 2391 & 39 & 227.6 & 49.3 & 9.5 & 389.5 & 19.9 & 97.6 & - & 49.1 & -4.33(0.02) & P & - &- \\ 
HD213845 & 111449 & F7V & 4436 & 79 & 220.4 & 44.2 & 9.6 & 454.3 & 25.8 & 82.6 & - & 52.3 & -4.7(0.02) & P & - &- \\ 
HD215641 & 112491 & G8V & 1639 & 78 & 101.5 & 20.5 & 1.2 & 84.4 & 3.4 & 12.2 & 0.22 & 8.3 & -4.46(0.03) & A & - &- \\ 
HD216956 & 113368 & A3V & 4929 & 834 & 382.4 & 44.3 & 27.0 & - & - & - & - & - & -3.965 & P & - & - \\ 
HD217343 & 113579 & G5V & 2727 & 26 & 276.6 & 96.9 & 2.7 & 298.7 & 6.6 & 85.6 & -1.08 & 19.3 & -4.22(0.03) & A & - &- \\ 
HD217987 & 114046 & M2V & 5109 & 130 & 63.8 & 18.5 & 0.9 & 39.5 & 2.3 & 8.2 & 0.95 & 5.8 & -4.9(0.06) & A & - &- \\ 
HD218396 & 114189 & A5V & 2727 & 124 & 3616.3 & 924.5 & 38.5 & 7376.0 & 95.5 & 2191.3 & - & 62.3 & -4.3(0.02) & P & - &- \\ 
HD218860 & 114530 & G8V & 1137 & 18 & 137.1 & 36.2 & 2.1 & 82.8 & 5.4 & 18.6 & -1.3 & 11.7 & -4.24(0.03) & A & - &- \\ 
HD221575 & 116258 & K2V & 869 & 16 & 45.7 & 12.2 & 1.4 & 33.4 & 3.7 & 12.0 & -0.54 & 8.8 & -4.44(0.02) & A & - &- \\ 
HD223340 & - & K1V & 868 & 10 & 54.4 & 17.3 & 1.7 & 42.2 & 4.4 & 10.7 & -1.34 & 9.8 & -4.34(0.03) & A & - &- \\ 
HD224228 & 118008 & K2V & 2815 & 31 & 35.4 & 8.8 & 1.0 & 31.5 & 2.8 & 7.5 & 0.32 & 7.8 & -4.36(0.03) & A & - &- \\ 
- & 6276 & G9V & 1201 & 20 & 95.6 & 27.3 & 1.6 & 91.0 & 4.2 & 27.6 & -0.65 & 9.4 & -4.32(0.03) & A & - &- \\ 
- & 116384 & K7V & 733 & 8 & 104.3 & 41.2 & 1.7 & 60.2 & 4.5 & 19.2 & -1.18 & 8.6 & -4.46(0.18) & A & - &- \\ 
- & 17157 & K7V & 1457 & 6 & 67.5 & 22.3 & 1.4 & 37.1 & 3.6 & 14.9 & - & 8.1 & -4.58(0.03) & A & - &- \\ 
- & 23309 & M0V & 4030 & 16 & 245.4 & 64.1 & 2.3 & 116.1 & 5.9 & 33.0 & -1.83 & 11.7 & -3.82(0.03) & A & - &- \\ 
- & 31878 & K7V & 435 & 11 & 84.3 & 31.6 & 1.6 & 48.3 & 4.1 & 16.1 & -1.86 & 8.5 & -4.28(0.02) & A & - &- \\ 
- & 36985 & M2V & 789 & 20 & 670.1 & 247.4 & 2.1 & 58.8 & 5.4 & 14.7 & - & 8.1 & -4.32(0.03) & A & - &SB1 \\ 
- & 44722 & K7V & 90 & 4 & 8.0 & 3.4 & 1.5 & 10.7 & 3.8 & 4.0 & -0.49 & 7.6 & -4.55(0.04) & A & - &- \\ 
- & 46634 & G5V & 281 & 3 & 38.2 & 16.1 & 1.4 & 32.6 & 3.5 & 13.9 & -1.0 & 9.8 & -4.32(0.02) & A & - &- \\ 
- & 51317 & M2V & 4541 & 139 & 17.7 & 3.6 & 1.2 & 24.5 & 3.2 & 3.5 & -0.38 & 4.8 & -5.03(0.06) & A & - &- \\ 
BD$+$20 2465$^\dagger$ & - & M5V & 3983 & 40 & 63.5 & 16.1 & 1.2 & 33.6 & 3.1 & 6.5 & -0.61 & 6.7 & -4.03(0.05) & A & - &- \\ 
CD$-$46 1064 & - & K3V & 1185 & 12 & 477.8 & 134.9 & 8.7 & 211.2 & 21.6 & 68.7 & -1.53 & 17.4 & -4.2(0.03) & A & - &- \\ 
\end{longtable}
\endgroup
\end{landscape}

\clearpage

\twocolumn

\section{Age and mass estimations}

\label{age_masse}
When available, the ages and masses of the targets of the \harps \ YNS and the \sophie \ YNS surveys were taken from the literature. 
When it was possible, the ages and masses were derived using the methods described in \cite{Desidera} and in an upcoming paper, then on a homogeneous scale with the above works. Briefly, we considered membership to groups and associations (adopting the group ages from \cite{Bonavita}, also discussed in the upcoming paper); indirect indicators such as rotation, chromospheric, and coronal activity; and Li \SI{6708}{\angstrom} equivalent width, complemented by isochrone fitting. Preference was given to the moving group criterion whenever possible (confirmed members). For field objects  the weight assigned to the various methods depends on color and/or spectral type and age range (e.g., saturation of chromospheric activity and coronal emission versus age below $100-\SI{150}{\mega\year}$, and  high sensitivity of lithium to age for K dwarfs younger than $300-\SI{500}{\mega\year}$, with  limits only at older ages). Masses were derived using the PARAM interpolation code\footnote{\url{http://stev.oapd.inaf.it/cgi-bin/param_1.3} } \citep{Da_Silva}, as in \cite{Desidera}.
For the remaining targets the  masses were estimated from the spectral type by using an empirical $M_*=f(\bv)$ relation (see page 564 of \cite{Lang}  and page 209 of \cite{Allen}).

\end{appendix}

\end{document}